\documentstyle[12pt]{article}
\input{psfig}

\newcommand{\braket}[2]{\langle #1 | #2 \rangle}

\newcommand{\beq}{\begin{equation}}
\newcommand{\eeq}{\end{equation}}
\newcommand{\half}{\mbox{$\textstyle \frac{1}{2}$} }

\newcommand{\ket}[1]{| #1 \rangle}
\newcommand{\bra}[1]{\langle #1 |}
\newcommand{\proj}[1]{\ket{#1}\! \bra{#1}}

\newcommand{\outerprod}[2]{| #1 \rangle\!\langle #2 |}
\newcommand{\Tr}{{\rm Tr}}
\renewcommand{\choose}[2]{{{#1}\atopwithdelims(){#2}}}
\renewcommand{\paragraph}{\section}

\begin{document}

{\large Mixed State Entanglement and Quantum Error Correction}
 
\bigskip
\begin{center}
Charles H. Bennett$^{(1)}$, David P. DiVincenzo$^{(1)}$, John A.~Smolin$^{(2)}$,\\
and William K.~Wootters$^{(3)}$
\end{center}
 
\bigskip
(1) IBM Research Division, Yorktown Heights, NY 10598;
(2) Physics Department, University of California at Los Angeles, Los Angeles,
CA 90024; (3) Physics Department, Williams College, Williamstown, MA 01267
 
\date{\today}
 
\begin{abstract}
Entanglement purification protocols (EPP) and quantum error-correcting
codes (QECC) provide two ways of protecting quantum states from
interaction with the environment.  In an EPP, perfectly entangled pure
states are extracted, with some yield $D$, from a mixed
state $M$ shared by two parties; with a QECC, an arbitrary quantum 
state $|\xi\rangle$
can be transmitted at some rate $Q$ through a noisy channel $\chi$
without degradation.  We prove that an EPP involving one-way classical
communication and acting on mixed state $\hat{M}(\chi)$
(obtained by sharing halves of EPR pairs through a channel $\chi$)
yields a QECC on $\chi$ with
rate $Q=D$, and vice versa.  We compare the amount of entanglement
$E(M)$ required to prepare a mixed state $M$ by local actions with the
amounts $D_1(M)$ and $D_2(M)$ that can be locally distilled from it by
EPPs using one- and two-way classical communication respectively, and give
an exact expression for $E(M)$
when $M$ is Bell-diagonal.  While EPPs require classical
communication, QECCs do not, and we prove $Q$ is not
increased by adding one-way classical communication.  However, both
$D$ and $Q$ can be increased by adding two-way communication.  We show
that certain noisy quantum channels, for example a 50\% depolarizing
channel, can be used for reliable transmission of quantum states if
two-way communication is available, but cannot be used if only one-way
communication is available.  We exhibit a family of codes based on
universal hashing able to achieve an asymptotic $Q$ (or $D$) of $1-S$
for simple noise models, where $S$ is the error entropy.  We also
obtain a specific, simple 5-bit single-error-correcting quantum block
code.  We prove that {\em iff} a QECC results in high fidelity for 
the case of no error the QECC can be recast into a form where the encoder 
is the matrix inverse of the decoder.

\end{abstract}
 
 

\section{Introduction}
\label{sec:intro}
 
\subsection{Entanglement and nonlocality in quantum physics}
\label{subsec:E}
 
Among the most celebrated features of quantum mechanics is the
Einstein-Podolsky-Rosen~\cite{EPR} (EPR) effect, in which anomalously
strong correlations are observed between presently noninteracting
particles that have interacted in the past.  These nonlocal correlations
occur only when the quantum state of the entire system is {\em
entangled\/}, i.e., not representable as a tensor product of states of
the parts.  In Bohm's version of the EPR paradox, a pair
of spin-1/2 particles, prepared in the singlet state
\begin{equation}
\Psi^-=\frac{1}{\sqrt{2}}(|\!\uparrow\downarrow\rangle-
|\!\downarrow\uparrow\rangle),\label{sing}
\end{equation}
and then separated, exhibit perfectly anticorrelated spin components
when locally measured along any axis.  Bell~\cite{Bell} and Clauser
{\it et al.}~\cite{CHSH} showed that these statistics violate
inequalities that must be satisfied by any classical local hidden
variable model of the particles' behavior.  Repeated experimental
confirmation~\cite{bellex}
of the nonlocal correlations predicted by quantum
mechanics is regarded as strong evidence in its favor.
 
Besides helping to confirm the validity of quantum mechanics,
entanglement has assumed an important role in quantum information
theory, a role in many ways complementary to the role of classical
information. Much recent work in quantum information theory has aimed
at characterizing the channel resources necessary and sufficient to
transmit unknown quantum states, rather than classical data, from a
sender to a receiver. To avoid violations of physical law, the intact
transmission of a general quantum state requires both a quantum
resource, which cannot be cloned, and a directed resource, which
cannot propagate superluminally.  The sharing of entanglement requires
only the former, while purely classical communication requires only
the latter.  In quantum teleportation~\cite{teleportation} the two
requirements are met by two separate systems, while in the direct,
unimpeded transmission of a quantum particle, they are met by the same
system.  Quantum data compression~\cite{Schu} optimizes the use of
quantum channels, allowing redundant quantum data, such as a random
sequence of two non-orthogonal states, to be compressed to a bulk
approximating its von Neumann entropy, then recovered at the receiving
end with negligible distortion.  On the other hand, quantum superdense
coding~\cite{superdense} uses previously shared entanglement to double
a quantum channel's capacity for carrying classical information.
 
Probably the most important achievement of classical information theory
is the ability, using error-correcting codes, to transmit data reliably
through a noisy channel.  Quantum error-correcting codes
(QECC)~\cite{ChuangLaf,shellgame,CS,Steane,Laflamme,EM,Sam,balance,newest} use
coherent generalizations of classical error-correction techniques to
protect quantum states from noise and decoherence during transmission
through a noisy channel or storage in a noisy environment.  Entanglement
purification protocols (EPP)~\cite{purification} achieve a similar
result indirectly, by distilling pure entangled states (e.g. singlets)
from a larger number of impure entangled states (e.g. singlets shared
through a noisy channel).  The purified entangled states can then be
used for reliable teleportation, thereby achieving the same effect as if
a noiseless storage or transmission channel had been available.  The
present paper develops the quantitative theory of mixed state
entanglement and its relation to reliable transmission of quantum
information.
 
\begin{figure}[htbp]
\centerline{\psfig{figure=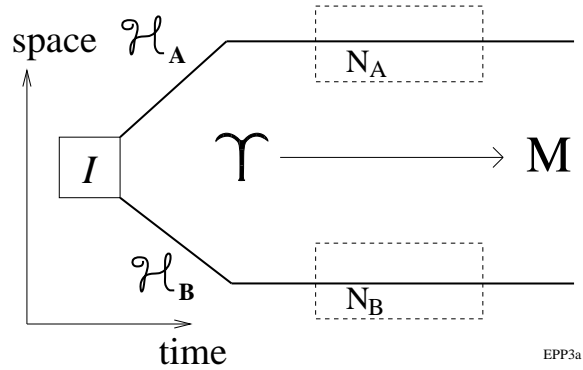,width=3in}}
\caption[Typical scenario for creation of entangled quantum states]
{Typical scenario for creation of entangled quantum states.
At some early time and at location $I$, two quantum
systems $A$ and $B$ interact~\protect\cite{foot1}, then become 
spatially separated, one
going to Alice and the other to Bob.  The joint system's state lies in
a Hilbert space ${\cal H}= {\cal H}_A\otimes{\cal H}_B$ that is the
tensor product of the spaces of the subsystems, but the state itself
is not expressible as a product of states of the subsystems:
$\Upsilon\neq \Upsilon_A\otimes\Upsilon_B$. 
State $\Upsilon$, its pieces acted upon
separately by noise processes $N_A$ and $N_B$, evolves into mixed
state $M$.} 
\label{f1}
\end{figure}
 
Entanglement is a property of bipartite systems---systems
consisting of two parts $A$ and $B$ that are too far apart to interact,
and whose state, pure or mixed, lies in a Hilbert space ${\cal H}= {\cal
H}_A\otimes{\cal H}_B$ that is the tensor product of Hilbert spaces of
these parts.   Our goal is to develop a general theory of state
transformations that can be performed on a bipartite system without
bringing the parts together.  We consider these transformations to be
performed by two observers, ``Alice'' and ``Bob,'' each having access to
one of the subsystems.  We allow Alice and Bob to perform local actions,
e.g. unitary transformations and measurements, on their respective
subsystems along with whatever ancillary systems they might create in
their own labs.  Sometimes we will also  allow them to coordinate their
actions through one-way or two-way classical communication; however, we
do not allow them to perform nonlocal quantum operations on the entire
system nor to transmit fresh quantum states from one observer to the
other.  Of course two-way or even one-way classical communication is
itself an element of nonlocality that would not be permitted, say, in a
local hidden variable model, but we find that giving Alice and Bob the
extra power of classical communication considerably enhances their power
to manipulate bipartite states, without giving them so much power as to
make all state transformations trivially possible, as would be the case
if nonlocal quantum operations were allowed.  We will usually assume that
${\cal H}_A$ and ${\cal H}_B$ have equal dimension $N$ (no generality is
lost, since either subsystem's Hilbert space can be embedded in a larger
one by local actions).
 
\subsection{Pure-state entanglement}
\label{subsec:pse}
 
For pure states, a sharp distinction can be drawn between entangled and
unentangled states: a pure state is entangled or nonlocal if and only
if its state vector $\Upsilon$ cannot be expressed as a product
$\Upsilon_A\otimes\Upsilon_B$ of pure states of its parts. It has been shown
that every entangled pure state violates some Bell-type
inequality~\cite{Gisin}, while no product state does.  Entangled states
cannot be prepared from unentangled states by any sequence of local
actions of Alice and Bob, even with the help of classical communication.
 
Quantitatively, a pure state's entanglement is conveniently measured
by its entropy of entanglement,
\beq
E(\Upsilon)=S(\rho_A)=S(\rho_B),\label{tangy}
\eeq
the apparent entropy of either subsystem considered
alone.  Here $S(\rho)=-\Tr\rho\log_2\rho$ is the von Neumann entropy
and $\rho_A=\Tr_B\proj{\Upsilon}$ is the reduced density matrix obtained
by tracing the whole system's pure-state density matrix
$\proj{\Upsilon}$ over Bob's degrees of freedom.  Similarly
$\rho_B=\Tr_A\proj{\Upsilon}$ is the partial trace over Alice's
degrees of freedom.
 
The quantity $E$, which we shall henceforth often call simply {\em
entanglement}, ranges from zero for a product state to $\log_2 N$ for
a maximally-entangled state of two $N$-state particles.  $E=1$ for the
singlet state $\Psi^-$ of Eq. (\ref{sing}), either of whose spins,
considered alone, appears to be in a maximally-mixed state with 1 bit
of entropy.  Paralleling the term {\em qubit\/} for any
two-state quantum system (e.g. a spin-$\half$ particle),
we define an {\em ebit\/} as the amount of entanglement in a maximally
entangled state of two qubits, or any other pure bipartite state
for which $E=1$.
 
Properties of $E$ that make it a natural entanglement measure for pure
states include:
\begin{itemize}
\item
The entanglement of independent systems is additive, $n$ shared
singlets for example having $n$ ebits of entanglement.
\item
$E$ is conserved under local unitary operations, i.e., under any unitary
transformation $U$ that can be expressed as a product $U=U_A\otimes
U_B$ of unitary operators on the separate subsystems.
\item
The expectation of $E$ cannot be increased by local nonunitary operations:
if a bipartite
pure state $\Upsilon$ is subjected to a local nonunitary operation (e.g.
measurement by Alice) resulting in residual pure states $\Upsilon_j$
with respective probabilities $p_j$, then the expected entanglement of
the final states $\sum_j p_jE(\Upsilon_j)$ is no greater, but may be
less, than the original entanglement
$E(\Upsilon)$~\cite{Concentration}. In the present paper we generalize
this result to mixed states: see Sec.~\ref{subsec:EMa}.
 \item
Entanglement can be concentrated and diluted with unit asymptotic
efficiency~\cite{Concentration}, in the sense that for any two bipartite
pure states $\Upsilon$ and $\Upsilon'$, if Alice and Bob are given a
supply of $n$ identical systems in a state
$\mbox{\boldmath$\Upsilon$}=(\Upsilon)^n$, they can use local actions
and one-way classical communication to prepare $m$ identical systems in
state $\mbox{\boldmath$\Upsilon'$}\approx(\Upsilon')^m$, with the yield
$m/n$ approaching $E(\Upsilon)/E(\Upsilon')$, the fidelity
$|\braket{\mbox{\boldmath$\Upsilon'$}}{(\Upsilon')^m}|^2$ approaching 1,
and probability of failure approaching zero in the limit of large $n$.
 \end{itemize}
With regard to entanglement, a pure bipartite state $\Upsilon$ is thus
completely parameterized by $E(\Upsilon)$, with $E(\Upsilon)$ being
both the asymptotic number of standard singlets required to locally
prepare a system in state $\Upsilon$---its ``entanglement of
formation''---and the asymptotic number of standard singlets that can
be prepared from a system in state $\Upsilon$ by local
operations---its ``distillable entanglement''.
 
\subsection{Mixed-state entanglement}
\label{subsec:mse}
 
One aim of the present paper is to extend the quantitative theory of
entanglement to the more general situation in which Alice and Bob
share a {\it mixed} state $M$, rather than a pure state $\Upsilon$ as
discussed above.  Entangled mixed states may arise (cf. Fig.~\ref{f1}) 
when one or both parts of an initially pure entangled
state interact, intentionally or inadvertently, with other quantum
degrees of freedom (shown in the diagram as noise processes $N_A$ and
$N_B$ and shown explicity in quantum channel $\xi$ in Fig.~\ref{epp16}) 
resulting in a non-unitary evolution of the pure state
$\Upsilon$ into a mixed state $M$.  Another principal aim is to
elucidate the extent to which mixed entangled states, or the noisy
channels used to produce them, can nevertheless be used to transmit
quantum information reliably.  In this connection we develop a family
of one-way entanglement purification protocols~\cite{purification} and
corresponding quantum error-correcting codes, as well as two-way
entanglement purification protocols which can be used to transmit
quantum states reliably through channels too noisy to be used reliably
with any quantum error-correcting code.
 
The theory of mixed-state entanglement is more complicated and less
well understood than that of pure-state entanglement. Even the
qualitative distinction between local and nonlocal states is less
clear.  For example, Werner~\cite{Werner} has described mixed states
which violate no Bell inequality with regard to simple spin
measurements, yet appear to be nonlocal in other subtler ways.  These
include improving the fidelity of quantum teleportation
above what could be achieved by purely classical
communication~\cite{Popescu1}, and giving nonclassical statistics when
subjected to a sequence of measurements~\cite{Popescu2}.
 
Quantitatively, no single parameter completely characterizes mixed
state entanglement the way $E$ does for pure states.  For a generic
mixed state, we do not know how to distill out of the mixed state as
much pure entanglement (e.g. standard singlets) as was required to
prepare the state in the first place; moreover, for some mixed states,
entanglement can be distilled with the help of two-way communication
between Alice and Bob, but not with one-way communication.  In order
to deal with these complications, we introduce three entanglement
measures $D_1(M)\leq D_2(M)\leq E(M)$, each of which reduces to $E$
for pure states, but at least two of which ($D_1$ and $D_2$) are known
to be inequivalent for a generic mixed state.
 
Our fundamental measure of entanglement, for which we continue to use
the symbol $E$, will be a mixed state's {\em entanglement of
formation\/} $E(M)$, defined as the least expected entanglement of any
ensemble of pure states realizing $M$.  We show that local actions and
classical communication cannot increase the expectation of $E(M)$ and we
give exact expressions for the entanglement of formation of a simple
class of mixed states: states of two spin-$\half$ particles that are
diagonal in the so-called {\em Bell basis}.  This basis consists of four
maximally-entangled states --- the singlet state of Eq. (\ref{sing}),
and the three triplet states
 \begin{eqnarray}
\Psi^+=\frac{1}{\sqrt{2}}(|\!\uparrow\downarrow\rangle+
|\!\downarrow\uparrow\rangle)\ \label{bell1}\\
\Phi^\pm=\frac{1}{\sqrt{2}}
(|\!\uparrow\uparrow\rangle\pm |\!\downarrow\downarrow\rangle)\ .\label{bell2}
\end{eqnarray}
We also give lower bounds on the entanglement of formation of other,
more general mixed states. Nonzero $E(M)$ will again serve as our
qualitative criterion of nonlocality; thus, a mixed state will be
considered local if can be expressed as a mixture of product states,
and nonlocal if it cannot.
 
By {\em distillable entanglement\/} we will mean the asymptotic yield
of arbitrarily pure singlets that can be prepared locally from mixed state $M$
by entanglement purification protocols (EPP) involving one-way or
two-way communication between Alice and Bob.  Distillable entanglement
for one- and two-way communication will be denoted $D_1(M)$ and
$D_2(M)$, respectively.  Except in cases where we have been able to
prove that $D_1$ or $D_2$ is identically zero, we have no explicit
values for distillable entanglement, but we will exhibit various upper
bounds, as well as lower bounds given by the yield of particular
purification protocols.

\subsection{Entanglement purification and quantum error correction}
\label{subsec:epqec}
 
\begin{figure}[htbp]
\centerline{\psfig{figure=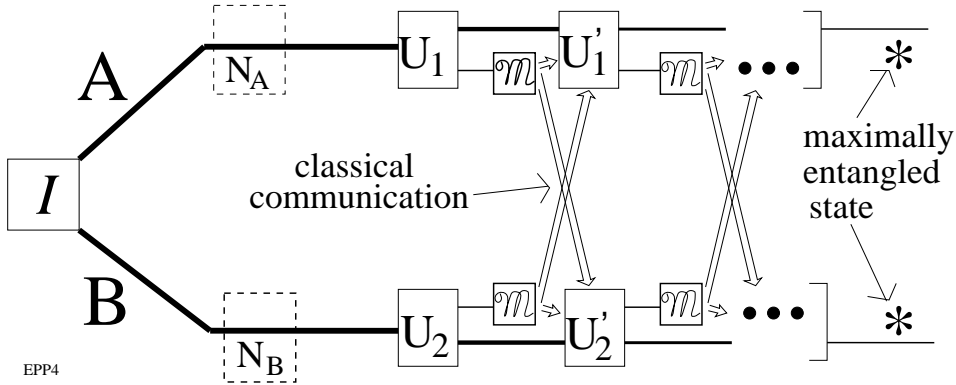,width=5in}}
\caption[Entanglement purification protocol involving two-way
classical communication (2-EPP)]
{Entanglement purification
protocol involving two-way
classical communication (2-EPP).  In the basic step of 2-EPP, Alice
and Bob subject the bipartite mixed state to two local unitary
transformations $U_1$ and $U_2$.  They then measure some of their
particles $\cal M$, and interchange the results of these
measurements (classical data transmission indicated by double lines).
After a number of stages, such a protocol can produce a pure,
near-maximally-entangled state (indicated by *'s).}
\label{2way}
\end{figure}

Entanglement purification protocols (EPP) will be the subject of a
large portion of this paper; we describe them briefly here. The most
powerful protocols, depicted in Fig.~\ref{2way}, involve two-way
communication.  Alice and Bob begin by sharing a bipartite mixed state
${\bf M}=(M)^n$ consisting of $n$ entangled pairs of particles each
described by the density matrix $M$, then proceed by repeated
application of three steps: 1) Alice and Bob perform unitary
transformations on their states; 2) They perform measurements on some
of the particles; and 3) They share the results of these measurements,
using this information to choose which unitary transformations to
perform in the next stage. The object is to sacrifice some of the
particles, while maneuvering the others into a close approximation of
a maximally entangled state such as
$\mbox{\boldmath$\Upsilon$}=(\Psi^-)^m$, the tensor product of $m$
singlets, where $0<m<n$.  No generality is lost by using only unitary
transformations and von Neumann measurements in steps 1) and 2), because
Alice and Bob are free at the outset to enlarge the Hilbert spaces
${\cal H}_A$ and ${\cal H}_A$ to include whatever
ancillas they might need to perform nonunitary
operations and generalized measurements on the original systems.
 
\begin{figure}[htbp]
\centerline{\psfig{figure=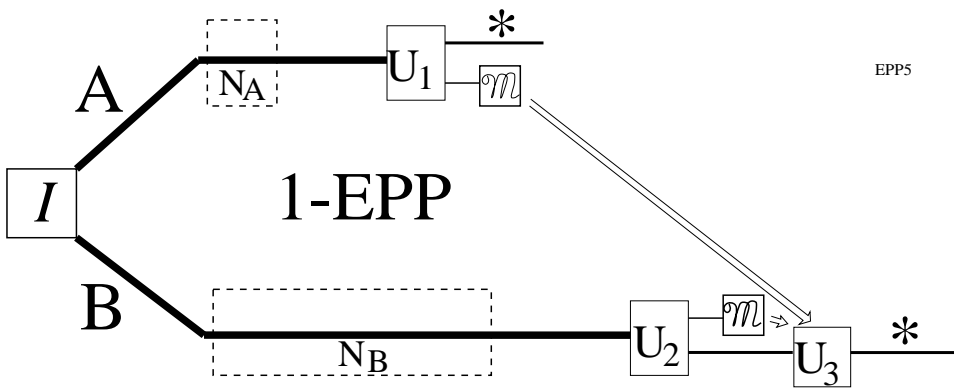,width=5in}}
\caption[One-way Entanglement Purification Protocol (1-EPP)]{One-way
Entanglement Purification Protocol (1-EPP).  In 1-EPP there is
only one stage; after unitary transformation $U_1$ and
measurement $\cal M$, Alice sends her classical result to Bob,
who uses it in combination with his measurement result to control a
final transformation $U_3$.  The unidirectionality of communication
allows the final, maximally-entangled state (*) to be separated both in
space and in time.}
\label{1way}
\end{figure}
 
A restricted version of the purification protocol involving only one-way
communication is illustrated in Fig. \ref{1way}.  Here, without loss of
generality, we permit only one stage of unitary operation and
measurement, followed by a  one-way classical communication.  The
principal advantage of such a protocol is that the components
of the resulting purified maximally entangled state indicated by (*) can
be separated both in space {\it and in time}. In
Secs.~\ref{CandM} and~\ref{simplecode}  we show that the 
time-separated EPR pairs
resulting from such a one-way protocol (1-EPP) always permit the
creation of a quantum error-correction code (QECC) whose rate and
fidelity are respectively no less than the yield $m/n$ and fidelity of
the purified states produced by the 1-EPP.
 
\begin{figure}[htbp]
\centerline{\psfig{figure=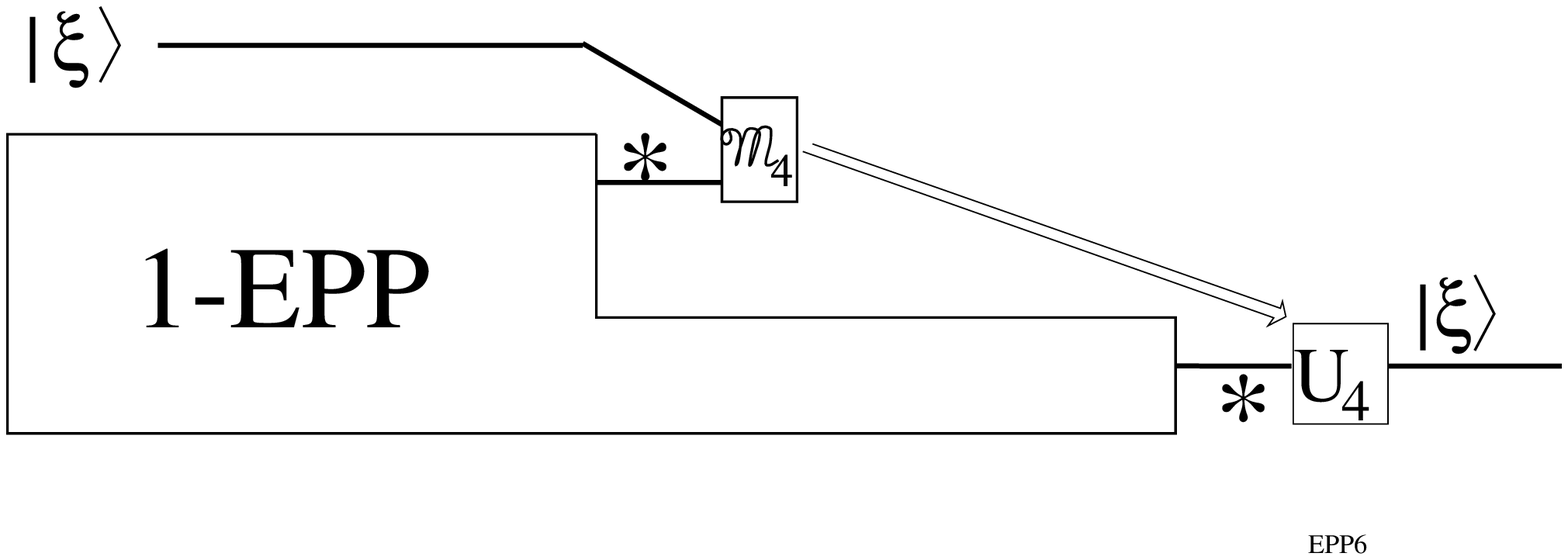,width=5in}}
\caption[EPP6.eps 1EPP used as module]
{If the 1-EPP of Fig.~\protect\ref{1way} is used as a module
for creating time-separated EPR pairs (*), then by using quantum
teleportation\protect\cite{teleportation}, an arbitrary quantum state
$|\xi\rangle$ may be recovered exactly after $U_4$, despite the presence
of intervening noise.  This is the desired effect of a quantum error
correcting code (QECC).} \label{qecc}
\end{figure}
 
The link between 1-EPP and QECC is provided by quantum
teleportation\cite{teleportation}. As Fig.~\ref{qecc} illustrates, the
availability of the time-separated EPR state (*) means that an
arbitrary quantum state $|\xi\rangle$ (in a Hilbert space no larger
than $2^m$) can be teleported forward in time: the teleportation is
initiated with Alice's Bell measurement ${\cal M}_4$, and is completed by
Bob's unitary transformation $U_4$.  The net effect is that an exact
replica of $|\xi\rangle$ reappears at the end, despite the presence of
noise ($N_{A,B}$) in the intervening quantum
environment.  Moreover, we will show in detail in
Sec.~\ref{simplecode} that the protocol of Fig.~\ref{qecc} can be
converted into a much simpler protocol with the same quantum
communication capacity but involving neither entanglement nor
classical communication, and having the topology of a quantum error
correcting code (Fig.
\ref{qeccf})~\cite{ChuangLaf,shellgame,CS,Steane,Laflamme,EM,Sam,
balance,newest}.
 
Many features of mixed-state entanglement, along with their consequences
for noisy-channel coding, are illustrated by a particular mixed state,
the Werner state~\cite{Werner}
\beq
W_{5/8} = \frac{5}{8}\proj{\Psi^-}
+ \frac{1}{8}(\proj{\Psi^+}+ \proj{\Phi^+}+\proj{\Phi^-}).\label{W58}
\eeq
This state, a 5/8 vs. 3/8 singlet-triplet mixture, can be produced by
mixing equal amounts of singlets and random uncorrelated spins, or
equivalently by sending one spin of an initially pure singlet through a
50\% depolarizing channel.  (A $x$-depolarizing channel is one in
which a state is transmitted unaltered with probability $1-x$ and
is replaced with a completely random qubit with probability $x$.) 
These recipes suggest that $E(W_{5/8})$, the
amount of pure entanglement required to prepare a Werner state, might
be 0.5, but we show~(Sec.~\ref{sec:EM}) that in fact that
$E(W_{5/8})\approx 0.117$.  The Werner state $W_{5/8}$ is also remarkable
in that pure entanglement can be distilled from it by two-way protocols
but not by any one-way protocol.  In terms of noisy-channel coding, this
means that a 50\% depolarizing channel, which has a positive capacity
for transmitting classical information, has zero capacity for
transmitting intact quantum states if used in a one-way fashion, even
with the help of quantum error-correcting codes.  This will be proved
in Sec.~\ref{sec:D1D2}.  If the same channel is
used in a two-way fashion, or with the help of two-way classical
communication, it has a positive capacity due to the non-zero
distillable entanglement $D_2(W_{5/8})$, which is known to lie between
0.00457 and 0.117 pure singlets out per impure pair in.  The lower bound
is from an explicit 2-EPP, while the upper bound comes from the known
entanglement of formation, which is always an upper bound on distillable
entanglement.
 
The remainder of this paper is organized as follows.  Section~\ref{sec:EM} 
contains our results on the entanglement of formation of
mixed states.  Section~\ref{sec:D} explains
purification of pure, maximally entangled states from mixed states.  
Section~\ref{sec:D1D2} exhibits a class of mixed states for 
which $D_1=0$ but $D_2>0$.  Section~\ref{CandM} shows the
relationship between mixed states and quantum channels.
Section~\ref{simplecode} shows how a class of quantum error 
correction codes may be derived from one-way purification protocols 
and contains our efficient 5 qubit code.  
Finally, Sec.~\ref{sec:theend} reviews several important remaining open 
questions.
 
\section{Entanglement of Formation}
\label{sec:EM}
 
\subsection{Justification of the Definition}
\label{subsec:EMa}
 
As noted above, we define the entanglement of formation $E(M)$ of a mixed
state $M$ as the least expected entanglement of any ensemble of pure states
realizing $M$.  The point of this subsection is to show that the 
designation ``entanglement of
formation" is justified: in order for Alice and Bob to create the state
$M$ without transferring quantum states between them, they must already
share the equivalent of $E(M)$ pure singlets; moreover, if they do
share this much entanglement already, then they will be able to create $M$.
(Both of these statements are to be taken in the asymptotic sense
explained in the Introduction.)  In this sense $E(M)$ is the amount of
entanglement needed to create $M$.
 
Consider any specific ensemble of pure states that realizes the mixed state
$M$.  By means of the asymptotically entanglement-conserving mapping between
arbitrary pure states and singlets~\cite{Concentration},
such an ensemble provides an asymptotic recipe for
locally preparing $M$ from a number of singlets equal to the mean
entanglement of the pure states in the ensemble. Clearly some ensembles
are more economical than others.  For example, the totally mixed state
of two qubits can be prepared at zero cost, as an equal mixture of four
product states, or at unit cost, as an equal mixture of the four Bell
states.   The quantity $E(M)$ is the minimum cost in this sense.
However, this fact does not yet justify calling $E(M)$ the entanglement
of formation, because one can imagine more complicated recipes for
preparing $M$: Alice and Bob could conceivably start with an initial
mixture whose expected entanglement is less than $E(M)$ and somehow,
by local actions and classical communication,
transform it into another mixture with greater expected entanglement.
We thus need to show that such entanglement-enhancing transformations 
are not possible.
 
We start by summarizing the definitions that lead to $E(M)$:
 
{\bf Definition:} The entanglement of formation of a
bipartite pure state $\Upsilon$ is the von Neumann entropy
$E(\Upsilon)=S(\Tr_A\proj{\Upsilon})$ of the reduced density matrix
as seen by Alice or Bob (see Eq.~\ref{tangy}).
 
{\bf Definition:} The entanglement of formation $E({\cal E})$ of an
ensemble of bipartite pure states ${\cal E}=\{p_i,\Upsilon_i\}$ is the
ensemble average $\sum_i p_i E(\Upsilon_i)$ of the entanglements of
formation of the pure states in the ensemble.
 
{\bf Definition:} The entanglement of formation $E(M)$ of a bipartite
mixed state $M$ is the minimum of $E(\cal E)$ over ensembles
${\cal E}=\{p_i, \Upsilon_i\}$ realizing the mixed state:
$M=\sum_i p_i \proj{\Upsilon_i}$
 
We now prove that $E(M)$ is nonincreasing under local operations and
classical communication.  First we prove two lemmas about the
entanglement of bipartite {\em pure} states under local operations by one
party, say Alice. Any such local action can be decomposed into
four basic kinds of operation: (i) appending an ancillary system
not entangled with Bob's part, (ii) performing a unitary transformation,
(iii) performing an orthogonal measurement, and (iv) throwing away,
{\em i.e.}, tracing out, part of the system.  (There is no need
to add generalized measurements as a separate category, since such
measurements can be constructed from operations of the above
kinds.)  It is clear that neither of the first two kinds of
operation can change the entanglement
of a pure state shared by Alice and Bob: the entanglement in
these cases remains equal to the von Neumann entropy of Bob's
part of the system.  However, for the last two kinds of operation,
the entanglement can change.  In the following two lemmas we
show that the expected entanglement in these cases cannot increase.
 
{\bf Lemma:} If a bipartite pure state $\Upsilon$ is subjected to a
measurement by Alice, giving outcomes $k$ with probabilities $p_k$,
and leaving residual bipartite pure states $\Upsilon_k$, then the
expected entanglement of formation $\sum_k p_k E(\Upsilon_k)$ of the
residual states is no greater than the entanglement of formation
$E(\Upsilon)$ of the original state.
\beq
\sum_k p_k E(\Upsilon_k)\leq E(\Upsilon)
\eeq
{\bf Proof.} Because the measurement is performed locally by Alice, it
cannot affect the reduced density matrix seen by Bob.  Therefore the
reduced density matrix seen by Bob before measurement,
$\rho=\Tr_A\proj{\Upsilon}$, must equal the ensemble average of the
reduced density matrices of the residual states after measurement:
$\rho_k=\Tr_A\proj{\Upsilon_k}$ after measurement.  It is well known
that von Neumann entropy, like classical Shannon entropy, is convex,
in the sense that the entropy of a weighted mean of several density
matrices is no less than the corresponding mean of their separate
entropies~\cite{Wehrl}.  Therefore
\beq
S(\rho) \geq \sum_k p_k S(\rho_k).
\eeq
But the left side of this expression is the original pure state's
entanglement before measurement, while the right side is the expected
entanglement of the residual pure states after measurement.\newline$\Box$
 
{\bf Lemma:} Consider a tripartite pure state $\Upsilon$, in which the
parts are labeled A, B, and C.  (We imagine Alice holding parts A and
C and Bob holding part B.)  Let $M = \Tr_C \proj{\Upsilon}$.  Then
$E(M) \leq E(\Upsilon)$, where the latter is understood to be the
entanglement between Bob's part B and Alice's part AC.  That is, Alice
cannot increase the minimum expected entanglement by throwing away
system C.
 
{\noindent}{\bf Proof.} Again, whatever pure-state ensemble one takes
as the realization of the mixed state $M$, the entropy at Bob's end of
the {\em average} of these states must equal $E(\Upsilon)$, because
the density matrix held by Bob has not changed.  By the above
argument, then, the average of the entropies of the reduced density
matrices associated with these pure states cannot exceed the entropy
of Bob's overall density matrix; that is, $E(M) \leq
E(\Upsilon)$.\newline$\Box$
 
We now prove a theorem that extends both of the
above results to mixed states:
 
{\bf Theorem:} If a bipartite mixed state $M$ is subjected to an
operation by Alice, giving outcomes $k$ with probabilities $p_k$,
and leaving residual bipartite mixed states $M_k$, then the expected
entanglement of formation $\sum_k p_k E(M_k)$ of the residual states
is no greater than the entanglement of formation $E(M)$ of the
original state.
\beq
\sum_k p_k E(M_k) \leq E(M)
\eeq
(If the operation is simply throwing away part of Alice's system,
then there will be only one value of $k$, with unit probability.)
 
{\noindent}{\bf Proof.} Given mixed state $M$ there will exist some
minimal-entanglement ensemble
\beq
{\cal E}=\{p_j,\Upsilon_j\}
\eeq
of pure states realizing $M$.
 
For any ensemble ${\cal E}'$ realizing $M$,
\beq
E(M) \leq E({\cal E}').
\label{nonopt}
\eeq
Applying the above lemmas to each pure state
in the minimal-entanglement ensemble ${\cal E}$, we get, for each $j$,
\beq
\sum_k p_{k|j} E(M_{jk}) \leq E(\Upsilon_j),\label{BBPSeach}
\eeq
where $M_{jk}$ is the residual state if pure state $\Upsilon_j$ is
subjected to Alice's operation and yields result $k$,
and $p_{k|j}$ is the conditional probability of obtaining this
outcome when the initial state is $\Upsilon_j$.
 
Note that when the outcome $k$ has occurred the residual mixed
state is described by the density matrix
 \beq
M_k = \sum_j p_{j|k} M_{jk}.
 \eeq
Multiplying Eq.~(\ref{BBPSeach}) by $p_j$ and summing over $j$
gives
 \beq
\sum_{j,k}p_j p_{k|j} E(M_{jk}) \leq \sum_j p_j E(\Upsilon_j)
= E(M).
\label{sumi} \eeq
By Bayes theorem,
\beq
p_{j,k}=p_j p_{k|j}=p_k p_{j|k},
\eeq
Eq.~(\ref{sumi}) becomes
\beq
\sum_{j,k}p_k p_{j|k} E(M_{jk}) \leq  E(M).
\label{voila} \eeq
Using the bound Eq.~(\ref{nonopt}), we get
\beq
\sum_kp_kE(M_k) \leq
\sum_k p_k \sum_j p_{j|k} E(M_{jk}) \leq E(M).
\label{finis}
\eeq
\newline$\Box$
 
Although the above theorem concerns a single operation by Alice, it
evidently applies to any finite preparation procedure, involving local
actions and one- or two-way classical communication, because any such
procedure can be expressed as sequence of operations of the above
type, performed alternately by Alice and Bob.  Each measurement-type
operation, for example,
generates a new classical result, and partitions the before-measurement
mixed state into residual after-measurement mixed states whose mean
entanglement of formation does not exceed the entanglement of formation
of the mixed state before measurement.  Hence we may summarize the
result of this section by saying that expected entanglement of
formation of a bipartite system's state does not increase under local
operations and classical communication.  As noted in
\cite{Concentration}, entanglement itself {\em can\/} increase under
local operations, even though its expectation cannot.  Thus it is
possible for Alice and Bob to gamble with entanglement, risking some of
their initial supply with a chance of winning more than they originally had.
 
\subsection{Entanglement of Formation for Mixtures of Bell States}
\label{mestates}
 
In the previous subsection it was shown that an ensemble of {\em pure
states} with minimum average pure-state entanglement realizing a given
density matrix defines a maximally economical way of creating that
density matrix.  In general it is not known how to find such an
ensemble of minimally entangled states for a given density matrix $M$.
We have, however, found such minimal ensembles for a particular class
of states of two spin-\half particles, namely, mixtures that are
diagonal when written in the Bell basis Eqs.~(\ref{sing}),
(\ref{bell1}), and (\ref{bell2}).  We have also found a lower bound on
$E(M)$ applicable to any mixed state of two spin-\half particles.  We
present these results in this subsection.
 
As a motivating example consider the Werner states of~\cite{Werner}.
A Werner state is a state drawn from an ensemble of $F$ parts
pure singlet, and $(1\!-\!F)/3$ parts of each of the other Bell
states --- that is, a generalization of Eq.~(\ref{W58}):
\beq
W_F = F\proj{\Psi^-}
+ \frac{1\!-\!F}{3}(\proj{\Psi^+}+ \proj{\Phi^+}+\proj{\Phi^-}).\label{WF}
\eeq
This is equivalent to saying it is drawn from an ensemble of
$x=(4F\!-\!1)/3$ parts pure singlet, and $1\!-\!x$ parts the totally mixed
``garbage'' density matrix (equal to the identity operator)
\beq
G=I=\frac{1}{4}
(\proj{\Psi^+} + \proj{\Psi^-} + \proj{\Phi^+} + \proj{\Phi^-}),
\label{garbage}
\eeq
which was Werner's original formulation.  We label these generalized
Werner states $W_F$, with their $F$ value, which is their fidelity or
purity $\bra{\Psi^-}W_F\ket{\Psi^-}$ relative to a perfect singlet
(even though this fidelity is defined nonlocally, it can be computed
from the results of local measurements, as $1\!-\!3P_\parallel/3$, where
$P_\parallel$ is the probability of obtaining parallel outcomes if the
two spins are measured along the same random axis).
 
It would take $x=(4F\!-\!1)/3$ pure singlets to create a mixed state
$W_F$ by directly implementing Werner's ensemble.  One
might assume that this prescription is the one requiring the least
entanglement, so that the $W_{5/8}$ state would cost 0.5 ebits to prepare.
However, through a numerical minimization technique we
found four pure states, each having only $0.117$ ebits of entanglement,
that when mixed with equal probabilities create the $W_{5/8}$ mixed
state much more economically.  Below we derive an explicit
minimally-entangled ensemble for any Bell-diagonal mixed state $W$,
including the Werner states $W_F$ as a special case, as well
as a giving a general lower bound for general mixed states $M$ of a pair of
spin-$\half$ particles.  For pure states and Bell-diagonal mixtures
$E(M)$ is simply equal to this bound.
 
The lower bound is expressed in terms of a quantity $f(M)$ which we
call the ``fully entangled fraction''of $M$ and define as
\beq
f(M)={\rm max} \langle e|M|e\rangle\ ,
\label{fef}
\eeq
where the maximum is over all completely entangled states
$|e\rangle$.  Specifically, we will see that for all states
of a pair of spin-\half particles,
$E(M) \ge h[f(M)]$, where the function $h$ is defined by
\begin{equation}
h(f) = \cases{H(\half + \sqrt{f(1\!-\!f)}) & for $f \ge \half$, \cr
              0 & for $f < \half$.   \cr} \label {h}
\end{equation}
Here $H(x)=-x\log_2x-(1\!-\!x)\log_2(1\!-\!x)$ is the binary entropy function.
For mixtures of Bell states, the fully entangled fraction $f(M)$ is simply the
largest eigenvalue of $M$.
 
We begin by considering the entanglement of a single pure state
$|\phi\rangle$.  It is convenient to write $|\phi\rangle$ in the
following orthogonal basis of completely entangled states:
\begin{equation}
\matrix{
|e_1\rangle = \hfill |\Phi^+\rangle \cr
|e_2\rangle = i|\Phi^-\rangle \cr
|e_3\rangle = i|\Psi^+\rangle \cr
|e_4\rangle = \hfill |\Psi^-\rangle \cr }
\label{es}
\end{equation}
Thus we write
\begin{equation}
|\phi\rangle = \sum_{j=1}^4\alpha_j|e_j\rangle.
\end{equation}
The entanglement of $|\phi\rangle$ can be computed directly as
the von Neumann entropy of the reduced density matrix of either
of the two particles.  On doing this calculation, one finds that the
entanglement of $|\phi\rangle$ is given by the simple formula
\begin{equation}
E = H[\half(1+\sqrt{1\!-\!C^2})], \label{pureent}
\end{equation}
where $C = |\sum_j\alpha_j^2|$.  (Note that one is squaring the
complex numbers $\alpha_j$, not their moduli.)  $E$ and $C$ both
range from 0 to 1, and $E$ is a monotonically increasing
function of $C$, so that
$C$ itself is a kind of measure of entanglement.  According to
Eq.~(\ref{pureent}), any {\em real} linear combination of the
states $|e_j\rangle$ is another completely entangled state
({\em i.e.}, $E=1$).  In fact, {\em every} completely entangled
state can
be written, up to an overall phase factor, as a real linear
combination of the $|e_j\rangle$'s.  (To see this, choose
$\alpha_1$ to be real without loss of generality.  Then if the
other $\alpha_j$'s are not all real, $C$ will be less than unity,
and thus so will $E$.)
 
Note that if one of the $\alpha_j$'s, say $\alpha_1$, is sufficiently
large in magnitude, then the other $\alpha_j$'s will not have
enough combined weight to make $C$ equal to zero, and thus the state
will have to have some entanglement.  This makes sense: if one
particular completely entangled state is sufficiently strongly
represented in
$|\phi\rangle$, then $|\phi\rangle$ itself must have some entanglement.
Specifically, if $|\alpha_1|^2 > \half$, then because the sum of
the squares of the three remaining $\alpha_j$'s cannot exceed
$1-|\alpha_1|^2$ in magnitude, $C$ must be at least
$|\alpha_1|^2 - (1 - |\alpha_1|^2)$, {\em i.e.},
$2|\alpha_1|^2 \!-\! 1$.  It follows from Eq.~(\ref{pureent}) that $E$
must be at least
$H[\half + \sqrt{|\alpha_1|^2(1-|\alpha_1|^2)}]$.  That is,
we have shown that
\begin{equation}
E(|\phi\rangle) \ge h(|\alpha_1|^2), \label{alpha_1}
\end{equation}
where $h$ is defined in Eq.~(\ref{h}).
This inequality will be very important in what follows.
 
As one might expect, the properties just described are not unique
to the basis $\{|e_j\rangle\}$.  Let $|e_j'\rangle =
\sum_k R_{jk}|e_k\rangle$, where $R$ is any real, orthogonal
matrix.  ({\em I.e.}, $R^TR = I$.)  We can expand $|\phi\rangle$
as $|\phi\rangle = \sum_j\alpha_j'|e_j'\rangle$, and the sum
$\sum_j{\alpha_j'}^2$ is guaranteed to be equal to
$\sum_j{\alpha_j}^2$ because
of the properties of orthogonal transformations.  Thus one can
use the components $\alpha_j'$ in Eq.~(\ref{pureent}) just as
well as the components $\alpha_j$.  In particular, the inequality
(\ref{alpha_1}) can be generalized by substituting for $\alpha_1$
the component of $|\phi\rangle$ along {\em any} completely entangled
state $|e\rangle$.  That is, if we define $w = |\langle e|\phi
\rangle |^2$ for some completely entangled $|e\rangle$, then
\begin{equation}
E(|\phi\rangle) \ge h(w).  \label{E>h}
\end{equation}
 
We now move from pure states to mixed states.
Consider an arbitrary mixed state $M$, and consider any
ensemble ${\cal E}={p_k,\phi_k}$ which is a decomposition of $M$ 
into pure states
\begin{equation}
M = \sum_k p_k |\phi_k\rangle\langle\phi_k|. \label{decomp}
\end{equation}
For an arbitrary completely entangled state $|e\rangle$, let
$w_k = |\langle e|\phi_k\rangle|^2$, and let
$w = \langle e|M|e \rangle = \sum_k p_k w_k$.  We can
bound the entanglement of the ensemble~(\ref{decomp})
as follows:
\begin{equation}
E({\cal E}) = \sum_k p_k E(|\phi_k\rangle)
\ge \sum_k p_k h(w_k) \ge
h\left[\sum_k p_k w_k\right] = h(w).  \label{string}
\end{equation}
This equation is true in particular for the minimal entanglement
ensemble realizing $M$ for which $E(M)=E(\cal E)$.
The second inequality follows from the convexity of the function
$h$.  Clearly we obtain the best bound of this form by
maximizing $w=\langle e|M|e \rangle$ over all completely
entangled states $|e\rangle$.  This maximum value of $w$ is
what we have called the fully entangled fraction $f(M)$.  We
have thus proved that
\begin{equation}
E(M) \ge h[f(M)], \label{bound}
\end{equation}
as promised.
 
To make the bound (\ref{bound}) more useful, we give the
following simple algorithm for finding the fully entangled
fraction $f$ of an arbitrary state $M$ of a pair
of qubits.  First, write $M$ in the basis $\{|e_j\rangle\}$
defined in Eq.~(\ref{es}).  In this basis, the completely
entangled states are represented by the real vectors,
so we are looking for the maximum value of
$\langle e|M|e \rangle$ over all real vectors $|e\rangle$.  But
this maximum value is simply the largest eigenvalue of the real
part of $M$. We have then: $f$ = the maximum eigenvalue of
Re~$M$, when $M$ is written in the basis of Eq.~(\ref{es}).
 
We now show that the bound (\ref{bound}) is actually achieved
for two cases of interest: (i)~pure states and (ii)~mixtures of
Bell states.  That is, in these cases, $E(M) = h[f(M)]$.
 
{\em(i) Pure states.} Any pure state can be changed by local rotations
into a state~\cite{Peres} of the form $|\phi\rangle =
\alpha|\uparrow\uparrow\rangle + \beta|\downarrow\downarrow\rangle$,
where $\alpha,\beta \ge 0$ and $\alpha^2 + \beta^2 = 1$.  Entanglement
is not changed by such rotations, so it is sufficient to show that the
bound is achieved for states of this form.  For $M=
|\phi\rangle\langle\phi|$, the completely entangled state maximizing
$\langle e|M|e \rangle$ is $|\Phi^+\rangle$, and the value of $f$ is
$|\langle\Phi^+|\phi\rangle|^2 = {(\alpha + \beta)^2 \over 2} = \half
+ \alpha\beta$.  By straightforward substitution one finds that
$h(\half + \alpha\beta) = H(\alpha^2)$, which we know to be the
entanglement of $|\phi\rangle$.  Thus $E(M) = h[f(M)]$, which is what
we wanted to show.
 
{\em(ii) Mixtures of Bell states.} Consider a mixed
state of the form
\begin{equation}
W = \sum_{j=1}^4 p_j|e_j\rangle\langle e_j|.\label{firstW}
\end{equation}
Suppose first that one of the eigenvalues $p_j$ is greater than
or equal to $\half$, and without loss of generality take this
eigenvalue to be $p_1$.  The following eight pure states, mixed
with equal probabilities, yield the state $W$:
\begin{equation}
\sqrt{p_1}|e_1\rangle + i(\pm \sqrt{p_2}|e_2\rangle
\pm \sqrt{p_3}|e_3\rangle \pm \sqrt{p_4}|e_4\rangle).
\end{equation}
Moreover, all of these pure states have the same entanglement,
namely,
\begin{equation}
E = h(p_1).
\end{equation}
(See Eq.~(\ref{pureent}).)  Therefore the average entanglement
of the mixture is also $\langle E \rangle = h(p_1)$.  But $p_1$
is equal to
$f(W)$ for this density matrix, so for this particular
mixture, we have $\langle E \rangle = h[f(W)]$.  Since
the right hand side is our lower bound on $E$, this mixture
must be a minimum-entanglement decomposition of $W$,
and thus $E(W) = h[f(W)]$.
 
If none of the eigenvalues $p_j$ is greater
than $\half$, then there exist phase factors $\theta_i$ such that
$\sum_j p_j e^{i\theta_j} = 0$.  In that case we can express
$W$ as an equal mixture of a different set of eight states:
\begin{equation}
\sqrt{p_1}e^{i\theta_1/2}|e_1\rangle
\pm \sqrt{p_2}e^{i\theta_2/2}|e_2\rangle
\pm \sqrt{p_3}e^{i\theta_3/2}|e_3\rangle
\pm \sqrt{p_4}e^{i\theta_4/2}|e_4\rangle.
\end{equation}
For each of these states, the quantity $C$ [Eq.~(\ref{pureent})]
is equal to zero, and thus the entanglement is zero.  It follows
that $E(W) = 0$, so that again the bound is achieved. (The
bound $h[f(W)]$ is zero in this case because $f$, the greatest
of the $p_j$'s, is less than $\half$.)
 
It is interesting to ask whether the bound $h[f(M)]$ is in fact
{\em always} equal to $E(M)$ for general mixed states $M$, not
necessarily Bell-diagonal.  It turns out that it is not.
Consider, for example, the mixed state
\begin{equation}
M = \half|\uparrow\uparrow\rangle\langle\uparrow\uparrow|
+ \half|\Psi^+\rangle\langle\Psi^+|. \label{counter}
\end{equation}
The value of $f$ for this state is $\half$, so that $h(f)=0$.
And yet, as we now show, it is impossible to build this
state out of unentangled pure states; hence $E(M)$ is
greater than zero and is not equal to $h(f)$.
 
To see this, let us try to construct the density
matrix of Eq.~(\ref{counter}) out of unentangled pure states.
That is, we want
\begin{equation}
M = \sum_k p_k |\phi_k\rangle\langle\phi_k|, \label{goal}
\end{equation}
where each $|\phi_k\rangle$ is unentangled.  That is, each
$|\phi_k\rangle$ is such that when we write it in the
basis of Eq.~(\ref{es}), {\em i.e.} as $|\phi_k\rangle =
\sum_{j=1}^4 \alpha_{k,j}|e_j\rangle$, the $\alpha$'s
satisfy the condition
\begin{equation}
\sum_{j=1}^4 \alpha_{k,j}^2 = 0.  \label{zero}
\end{equation}
Now the density matrix $M$, when written in the $|e_j\rangle$
basis, looks like this:
\begin{equation}
M = \left[ \matrix{{1\over 4} &{i\over 4} &0 &0 \cr
                      {-i\over 4} &{1\over 4} &0 &0 \cr
                      0 &0 &\half &0 \cr
                      0 &0 &0 &0}\right].
\end{equation}
Thus, in order for Eq.~(\ref{goal}) to be true, the $\alpha$'s
must be consistent with the following conditions:
\begin{equation}
\matrix{
\sum_k p_k |\alpha_{k,1}|^2 = {1\over 4} \cr
\sum_k p_k |\alpha_{k,2}|^2 = {1\over 4} \cr
\sum_k p_k |\alpha_{k,3}|^2 = {1\over 2} \cr
\sum_k p_k |\alpha_{k,4}|^2 = 0 \cr
\sum_k p_k \alpha_{k,1}\alpha_{k,2}^* = {i\over 4}. }
\label{sums}
\end{equation}
Evidently all the $\alpha_{k,4}$'s are equal to zero.
By Eq.~(\ref{zero}) the remaining $\alpha$'s satisfy
\begin{equation}
|\alpha_{k,1}|^2 + |\alpha_{k,2}|^2 \ge |\alpha_{k,3}|^2
\hskip0.5cm \hbox{for every }k.
\end{equation}
In fact, the ``$\ge$'' of this last relation must be
an equality, or else the sum conditions of Eq.~(\ref{sums})
would not work out.  That is,
\begin{equation}
|\alpha_{k,1}|^2 + |\alpha_{k,2}|^2 = |\alpha_{k,3}|^2
\hskip0.5cm \hbox{for every }k.
\end{equation}
Combining this last equation with Eq.~(\ref{zero}), we arrive
at the conclusion that for each $k$, the ratio of $\alpha_{k,1}$
to $\alpha_{k,2}$ is real.
But in that case there is no way to
generate the imaginary sum required by the
last of the conditions (\ref{sums}).  It is thus impossible
to build $M$ out of unentangled pure states; that
is, $E(M)>0$.
 
We conclude, then, that our bound is only a bound and not an
exact formula for $E$.  It turns out, in fact, that there are
two other ways to prove that the state $M$ has nonzero entanglement
of formation.  Peres~\cite{Peres2} and Horodecki 
{\em et al.}~\cite{horodecki} have recently developed a
general test for nonzero entanglement for states of two qubits and
has applied it explicitly to states like our $M$, showing that
$E(M)$ is nonzero.  Also, in Sec.~\ref{subsec:nonBell}
below, we show that one can distill some pure
entanglement from $M$, which would not be possible if $E(M)$
were zero.

\section{Purification}
\label{sec:D}
 
Suppose Alice and Bob have $n$ pairs of particles, each pair's state
described by a density matrix $M$.  Such a mixed state results if one or
both members of an initially pure Bell state is subjected to noise during
transmission or storage (cf. Fig.~\ref{f1}). Given these $n$ impure
pairs, how many pure Bell singlets can they distill by local actions;
indeed, can they distill any at all?  In other words, how much
entanglement can they ``purify'' out of their mixed state without
further use of a quantum channel to share more entanglement?
 
The complete answer is not yet known, but upper and lower bounds
are~\cite{purification}.  An upper bound is $E(M)$ per pair, because
if Alice and Bob could get more good singlets than that they could use
them to create more mixed states with density matrix $M$ than the
number with which they started thereby increasing their entanglement
by local operations, which we have proven impossible (Sec.~\ref{subsec:EMa}).
Lower bounds are given by construction.  We have found
specific procedures which Alice and Bob can use to purify certain
types of mixed states into a lesser number of pure singlets.  We call
these schemes {\em entanglement purification protocols} (EPP), which
should not be confused with the {\em purifications} of a mixed state
of~\cite{Jozsa}.
 
\subsection{Purification Basics}
\label{basics}
 
Our purification procedures all stem from a few simple ideas:
 
\begin{enumerate}
 
\item A general two-particle mixed state $M$ can be converted to a
Werner state $W_F$ (Eq.~(\ref{WF})) by an irreversible
preprocessing operation which increases the entropy ($S(W_F)>S(M)$),
perhaps wasting some of its recoverable entanglement, but rendering
the state easier deal with because it can thereafter be regarded as a
classical mixture of the four orthogonal Bell states
(Eqs.~(\ref{sing}), (\ref{bell1}), and (\ref{bell2}))
\cite{footfef}.  The simplest
such preprocessing operation, a {\it random bilateral
rotation}\cite{purification} or ``twirl'', consists of choosing an
independent, random SU(2) for each impure pair and applying it to both
members of the pair (cf. Fig. \ref{tommy}).  Because of the singlet
state's invariance under bilateral rotation, twirling has the effect
of removing off-diagonal terms in the two-particle density matrix in
the Bell basis, as well as equalizing the triplet eigenvalues.  
Actually, removing the off-diagonal terms is sufficient
as all of our EPP protocols operate successfully (with only 
minor modification) on a Bell-diagonal mixed state $W$ with, in general,
unequal triplet eigenvalues.  Equalization
of the triplet eigenvalues only adds unnecessary entropy to the mixture.
In Appendix~\ref{appx} it is shown that a continuum of rotations is
unnecessary: an arbitrary mixed state of two qubits can be converted
into a Werner $W_F$ or Bell-diagonal $W$ mixture by a ``discrete
twirl,'' consisting of a random choice among
an appropriate discrete set of bilateral rotations~\cite{foot2}.
We use $T$ to denote the nonunitary operation of performing either a
discrete or a continuous twirl.
\label{i1}
\begin{figure}[htbp]
\centerline{\psfig{figure=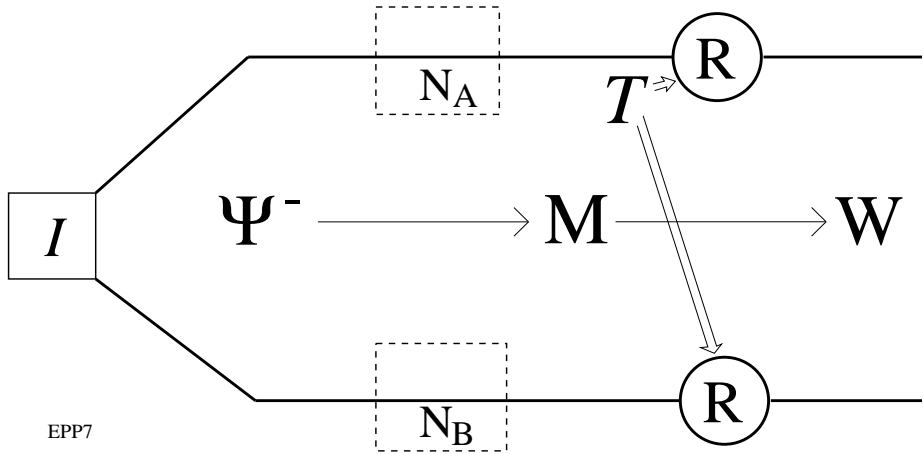,width=5in}}
\caption[Twirling]{The general mixed state $M$ of 
Fig.~\protect\ref{f1} can
be converted into one of the Werner form $W_F$ of
Eq.~(\protect\ref{WF}) if the particles on both Alice's and Bob's side
are subjected to the same random rotation $R$ (we refer to the act of
choosing a random SU(2) rotation and applying it to both particles as
a ``twirl'' $T$).}
\label{tommy}
\end{figure}
 
\item Once the initial mixed state $M$ has been rendered into
Bell-diagonal form $W$, it can be purified as if it were a classical
mixture of Bell states, without regard to the original mixed state $M$
or the noisy channel(s) that may have generated it~\cite{footrho}.
This is extremely convenient for the development of all
our protocols.  However, as we show in Appendix~\ref{twirlout}
all the purification protocols we will develop will also
work just as well on the original non Bell-diagonal mixtures $M$.
\label{i2}

\begin{table}[htbp]
\begin{tabular}{cr|llll|l}
 & &\multicolumn{4}{|c|}{source}& \\
 & &$\Psi^-$&$\Phi^-$&$\Phi^+$&$\Psi^+$&\\
\cline{2-7}
 &$I$&$\Psi^-$&$\Phi^-$&$\Phi^+$&$\Psi^+$\\
Unilateral $\pi$ Rotations:&$\sigma_x$&$\Phi^-$&$\Psi^-$&$\Psi^+$&$\Phi^+$&\\
 &$\sigma_y$&$\Phi^+$&$\Psi^+$&$\Psi^-$&$\Phi^-$&\\
 &$\sigma_z$&$\Psi^+$&$\Phi^+$&$\Phi^-$&$\Psi^-$&\\
\cline{2-7}
\vspace{.4 in}\\
 & &\multicolumn{4}{|c|}{source}&\\
 & &$\Psi^-$&$\Phi^-$&$\Phi^+$&$\Psi^+$&\\
\cline{2-7}
 &$I$&$\Psi^-$&$\Phi^-$&$\Phi^+$&$\Psi^+$&\\
Bilateral $\pi/2$ Rotations:&$B_x$&$\Psi^-$&$\Phi^-$&$\Psi^+$&$\Phi^+$&\\
 &$B_y$&$\Psi^-$&$\Psi^+$&$\Phi^+$&$\Phi^-$&\\
 &$B_z$&$\Psi^-$&$\Phi^+$&$\Phi^-$&$\Psi^+$&\\
\cline{2-7}
\vspace{.4 in}\\
 & &\multicolumn{4}{|c|}{source}& \\
 &target&$\Psi^-$&$\Phi^-$&$\Phi^+$&$\Psi^+$\\
\cline{2-7}
 & &$\Psi^+$&$\Phi^+$&$\Phi^-$&$\Psi^-$&(source)\\
 &$\Psi^-$&$\Phi^-$&$\Psi^-$&$\Psi^-$&$\Phi^-$&(target)\\
 \cline{2-7}
 & &$\Psi^+$&$\Phi^+$&$\Phi^-$&$\Psi^-$&(source)\\
Bilateral XOR:&$\Phi^-$&$\Psi^-$&$\Phi^-$&$\Phi^-$&$\Psi^-$&(target)\\
 \cline{2-7}
 & &$\Psi^-$&$\Phi^-$&$\Phi^+$&$\Psi^+$&(source)\\
 &$\Phi^+$&$\Psi^+$&$\Phi^+$&$\Phi^+$&$\Psi^+$&(target)\\
 \cline{2-7}
 & &$\Psi^-$&$\Phi^-$&$\Phi^+$&$\Psi^+$&(source)\\
 &$\Psi^+$&$\Phi^+$&$\Psi^+$&$\Psi^+$&$\Phi^+$&(target)\\
 
\end{tabular}
 
\caption{The unilateral and bilateral operations used by Alice
and Bob to map Bell states to Bell states.  Each entry of the BXOR
table has two lines, the first showing what happens to the source
state, the second showing what happens to the target state.
\label{bell_table}}
\end{table}
 
\item Bell states map onto one another under several kinds of local
unitary operations (cf. Table~\ref{bell_table}). These three sets of
operations are of two types: {\em unilateral\/} operations which are
performed by Bob or Alice but not both, and {\em bilateral\/}
operations which can be written as a tensor product of an Alice part
and a Bob part, each of which are the same.  The three types of
operations used are: 1) Unilateral rotations by $\pi$ radians,
corresponding to the three Pauli matrices $\sigma_x$, $\sigma_y$, and
$\sigma_z$; 2) Bilateral rotations by $\pi/2$ radians, henceforth
denoted $B_x$, $B_y$, and $B_z$; and 3) The bilateral application of
the two-bit quantum XOR (or controlled-NOT)\cite{G9,PT} hereafter
referred to as the BXOR operation (see Fig.~\ref{bxorfig}).
\begin{figure}[htbp]
\centerline{\psfig{figure=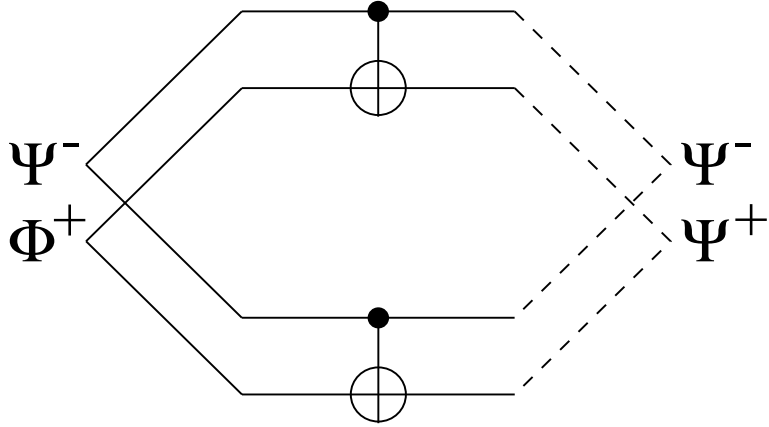,width=4in}}
\caption[BXOR operation]{The BXOR operation.  A solid dot indicates the source bit of
an XOR operation\protect\cite{G9} and a crossed circle indicates the target.
In this example a $\Psi^-$ state is the
source and a $\Phi^+$ is the target.  If the pairs are later brought
back together and measured in the Bell basis the source will remain
a $\Psi^-$ and the target will have become a $\Psi^+$, as per
Table~\protect\ref{bell_table}.
\label{bxorfig}}
\end{figure}
These operations and the Bell state mappings they implement, along with
individual particle measurements, are the basic tools Alice and Bob use
to purify singlets out of $W$.\label{i3}
 
\item Alice and Bob can distinguish $\Phi$ states from $\Psi$ states
by locally measuring their particles along the $z$ direction.  If they
get the same results they have a $\Phi$, if they get opposite results
they have a $\Psi$.  Note that if only one of the observers (say Bob)
needs to know whether the state was a $\Phi$ or a $\Psi$, the process
can be done without two-way communication.  Alice simply makes her
measurement and sends the result to Bob.  After Bob makes his
measurement, he can then determine whether the state had been a $\Phi$
or a $\Psi$ by comparing his measurement result with Alice's, without
any further communication.\label{i4}
 
\item For convenience we take $\ket{\Phi^+}$ as the standard state for
the rest of the paper.  This is because it is the state which, when
used as both source and target in a BXOR, remains unchanged.  It is
not necessary to use this convention but it is algebraically simpler.
We note that $\ket{\Phi^+}$ states can be converted to singlet
($\ket{\Psi^-}$) states using the unilateral $\sigma_y$ rotation, as
shown in Table~\ref{bell_table}.  The only complication is that the
nonunitary twirling operation $T$ of item \ref{i1} works only when
$\ket{\Psi^-}$ is taken as the standard state.  But a modified twirl
$T'$ which leaves $\ket{\Phi^+}$ invariant and randomizes the other
three Bell states may easily be constructed: the modified twirl would
consist of a unilateral $\sigma_y$ (which swaps the $\ket{\Phi^+}$'s
and $\ket{\Psi^-}$'s) followed by a conventional twirl $T$, followed
by another unilateral $\sigma_y$ (which swaps them back).\label{i5}
 
\item The preceding points all suggest a new notation for the Bell
states.  We use two classical bits to label each of the Bell states
and write
\begin{eqnarray}
&&\Phi^+=00\nonumber\\
&&\Psi^+=01\nonumber\\
&&\Phi^-=10\nonumber\\
&&\Psi^-=11.
\label{2bit}
\end{eqnarray}
The right, low-order or ``amplitude'' bit identifies the $\Phi/\Psi$
property of the Bell state, while the left, high-order or ``phase''
bit identifies the $+/-$ property.  Both properties could be
distinguished simultaneously by a nonlocal measurement, but local
measurements can only distinguish one of the properties at a time,
randomizing the other.  For example a bilateral $z$ spin measurement
distinguishes the amplitude while randomizing the phase.\label{i6}
\end{enumerate}
 
\begin{table}[htbp]
\begin{center}
\begin{tabular}{c|ll|ll|l}
 &\multicolumn{2}{|c|}{initial}&\multicolumn{2}{|c|}{after BXOR}&Test\\
Probability&S&T&S&T&result\\
\hline
$p_{00}^2$&00&00&00&00&P\\
$p_{00}p_{01}$&00&01&00&01&F\\
$p_{00}p_{10}$&00&10&10&10&P\\
$p_{00}p_{11}$&00&11&10&11&F\\
\hline
$p_{01}p_{00}$&01&00&01&01&F\\
$p_{01}^2$&01&01&01&00&P\\
$p_{01}p_{10}$&01&10&11&11&F\\
$p_{01}p_{11}$&01&11&11&10&P\\
\hline
$p_{10}p_{00}$&10&00&10&00&P\\
$p_{10}p_{01}$&10&01&10&01&F\\
$p_{10}^2$&10&10&00&10&P\\
$p_{10}p_{11}$&10&11&00&11&F\\
\hline
$p_{11}p_{00}$&11&00&11&01&F\\
$p_{11}p_{01}$&11&01&11&00&P\\
$p_{11}p_{10}$&11&10&01&11&F\\
$p_{11}^2$&11&11&01&10&P\\
\end{tabular}
\caption{Probabilities for each initial configuration of source and
target in a pair of Bell states drawn from the same ensemble, and the
resulting state configuration after a BXOR operation is applied.  The
final column shows whether the target state passes (P) or fails (F)
the test for being parallel along the $z$-axis (this is given by the
rightmost bit of the target state after the BXOR).  This table,
ignoring the probability column, is just the BXOR table of
Table~\protect\ref{bell_table} written in the bitwise notation of item
\protect\ref{i6} of Sec.~\protect\ref{basics}.
\label{recursion_table}}
\end{center}
\end{table}
 
\subsection{Purification Protocols}
\label{subsec:pure}
 
We now present several two- and one-way purification protocols.  All
begin with a large collection of $n$ impure pairs each in mixed state $M$,
use up $n\!-\!m$ of them (by measurement), while
maneuvering the remaining $m$ pairs into a collective state ${\boldmath{M'}}$ whose
fidelity $\bra{(\Phi^+)^m}$${\boldmath{M'}}$$\ket{(\Phi^+)^m}$ relative
to a product of $m$ standard $\Phi^+$ states approaches 1 in the limit of
large $n$.  The yield a purification protocol $P$ on input mixed
states $M$ is defined as
 \beq
D_P(M) = \lim_{n\rightarrow\infty}m/n.
 \eeq
If the original impure pairs $M$ arise from sharing pure EPR pairs
through a noisy channel $\chi$, then the yield $D_P(M)$, defines the
asymptotic number of qubits that can be reliably transmitted (via
teleportation) per use of the channel.  For one-way protocols the yield
is equal to the rate of a corresponding quantum error-correcting code
(cf. Section \ref{CandM}).  For two-way protocols, there is no
corresponding quantum error-correcting code. We will compare the yields
from our protocols with the rates of quantum error-correcting codes
introduced by other authors, and with known upper bounds on the one-way
and two-way distillable entanglements $D_1(W)$ and $D_2(W)$. These are
defined in the obvious way, e.g. $D_1(W)=\max\{D_P(W): \mbox{$P$ is a
1-EPP}\}$. No entanglement purification protocol has been proven optimal,
but all give lower bounds on the amount of entanglement that can be
distilled from various mixed states.
 
\subsubsection{Recurrence method}
\label{subsec:recur}
 
A purification procedure presented originally in~\cite{purification} is
the recurrence method.  This is an explicitly two-way protocol. Two
states are drawn from an ensemble which is a mixture of Bell states with
probabilities $p_i$ where $i$ labels the Bell states in our two-bit
notation. (As noted earlier, if the original impure state is not
Bell-diagonal, it can be made so by twirling).  The 00 state is again
taken to be the standard state and we take $p_{00}=F$.  The two states
are used as the source and target for the BXOR operation.  Their initial
states and probabilities, and states after the BXOR operation, are shown
in Table~\ref{recursion_table}.  Alice and Bob test the target states,
and then separate the source states into the ones whose target states
passed and the ones whose target state failed.  Each of these subsets is
a Bell state mixture, with new probabilities.  These {\em a posteriori}
probabilities for the `passed' subset are: \beq \begin{tabular}{ll}
$p_{00}'=(p_{00}^2+p_{10}^2)/p_{\rm pass}$&$p_{01}'=(p_{01}^2+
p_{11}^2)/p_{\rm pass}$\\ $p_{10}'=2p_{00}p_{10}/p_{\rm
pass}$&$p_{11}'=2p_{01}p_{11}/p_{\rm pass}$ \end{tabular} \eeq with \beq
p_{\rm
pass}=p_{00}^2+p_{01}^2+p_{10}^2+p_{11}^2+2p_{00}p_{10}+2p_{01}p_{11}.
\label{ppass} \eeq
 
Consider the situation where Alice and Bob begin with a large supply
of Werner states $W_F$.  They apply the preceding procedure and are
left with a subset of states which passed and a subset which failed.
For the members of the ``passed'' subset $p_{00}'>p_{00}$ for 
all $p_{00}>0.5$.
The members of the ``failed'' subset have
$p_{00}=p_{01}=p_{10}=p_{11}=1/4$.  Since the entanglement $E$ of this
mixture is 0, it will clearly not be possible to extract any
entanglement from the ``failed'' subset, so all members of this
subset are discarded.  Note that this is where the protocol explicitly
requires two-way communication.  Both Alice and Bob need to know the
results of the test in order to determine which pairs to discard.

\begin{figure}[p]
\psfig{figure=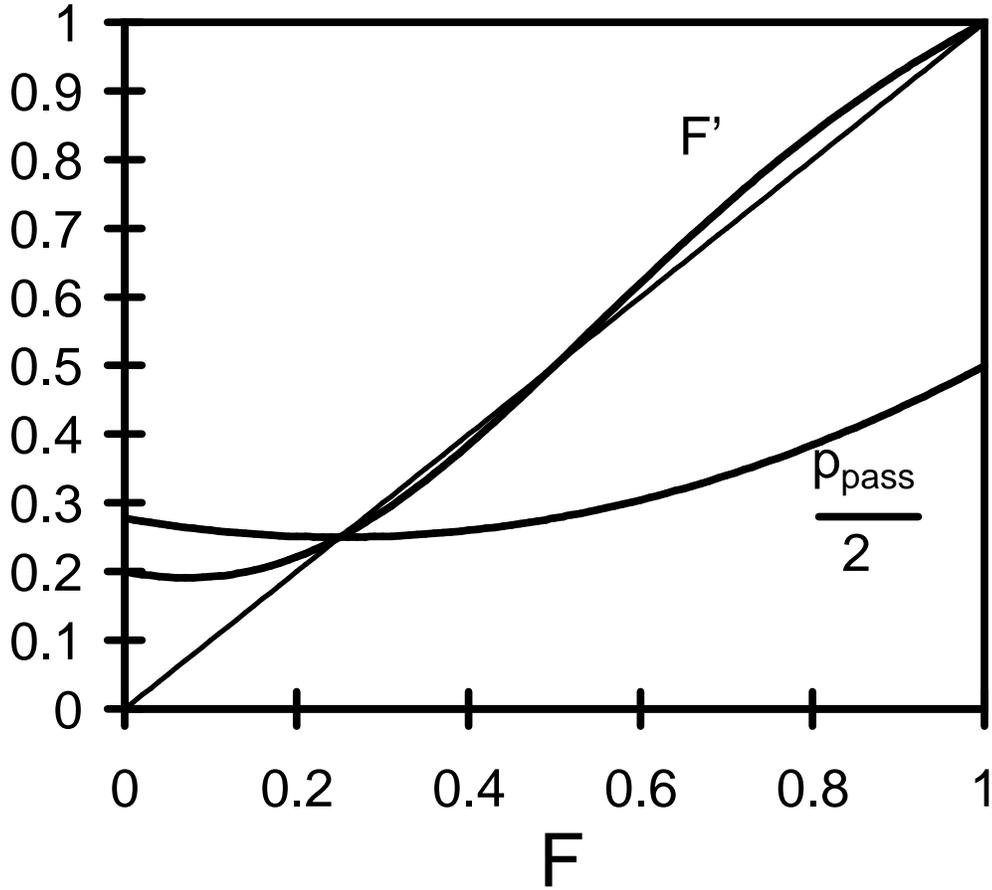}
\caption[Recurrence graph]
{Effect on the fidelity of Werner states of one step of purification,
using the recurrence protocol.  $F$ is the initial fidelity of the Werner
state (Eq.~(\protect\ref{WF})), $F'$ is the final fidelity of the
``passed'' pairs after one iteration.  Also shown is the fraction
$p_{pass}/2$ of pairs remaining after one iteration (cf.
Eq.~(\protect\ref{ppass})).}
\label{recurchart}
\end{figure}

\begin{figure}[p]
\psfig{figure=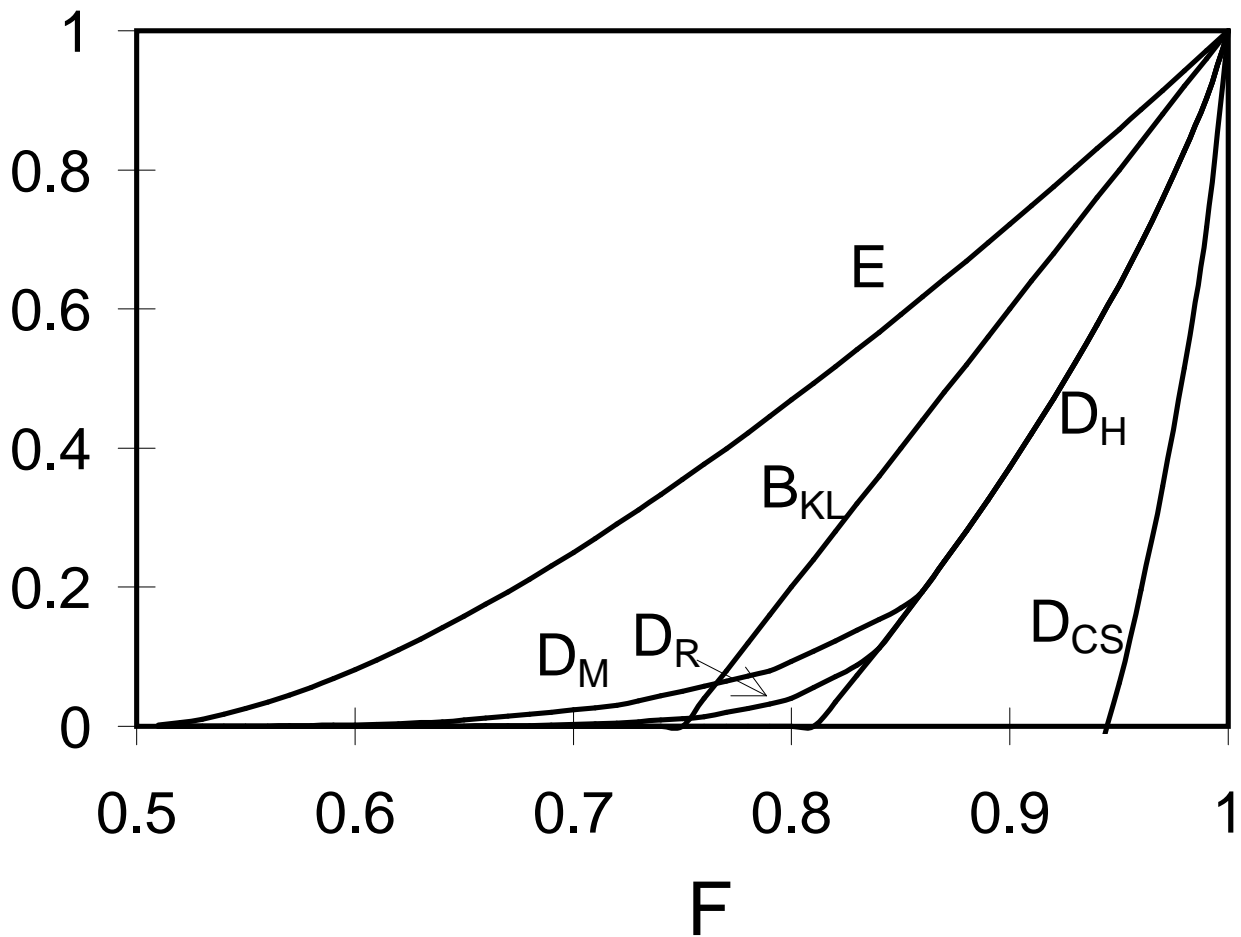}
\caption[Measures of entanglement versus
fidelity $F$ for Werner states $W_F$ of Eq.~(\protect\ref{WF})]
{Measures of entanglement versus
fidelity $F$ for Werner states $W_F$ of Eq.~(\protect\ref{WF}).  $E$
is the entanglement of formation, Eq.~(\protect\ref{string}).
$D_R$ is the
yield of the recurrence method of Sec.~\protect\ref{subsec:recur}
continued by the hashing method of
(Sec.~\protect\ref{breeding}). $D_M$ is the yield of the modified
recurrence method of C. Macchiavello\protect\cite{foot5}, continued by
hashing.  $D_H$ is the yield of the one-way hashing and breeding
protocols (Sec.~\protect\ref{breeding}) used alone. $D_{CS}$ is the
rate of the quantum error correcting codes proposed by Calderbank and
Shor\protect\cite{CS} and Steane\protect\cite{Steane}. 
$B_{KL}$ is
the upper bound for $D_1$ as shown in Sec.~\protect\ref{newupperbound} 
(following Knill and Laflamme~\protect\cite{lafknill}). 
\label{dchart}}
\end{figure}

\begin{figure}[p]
\psfig{figure=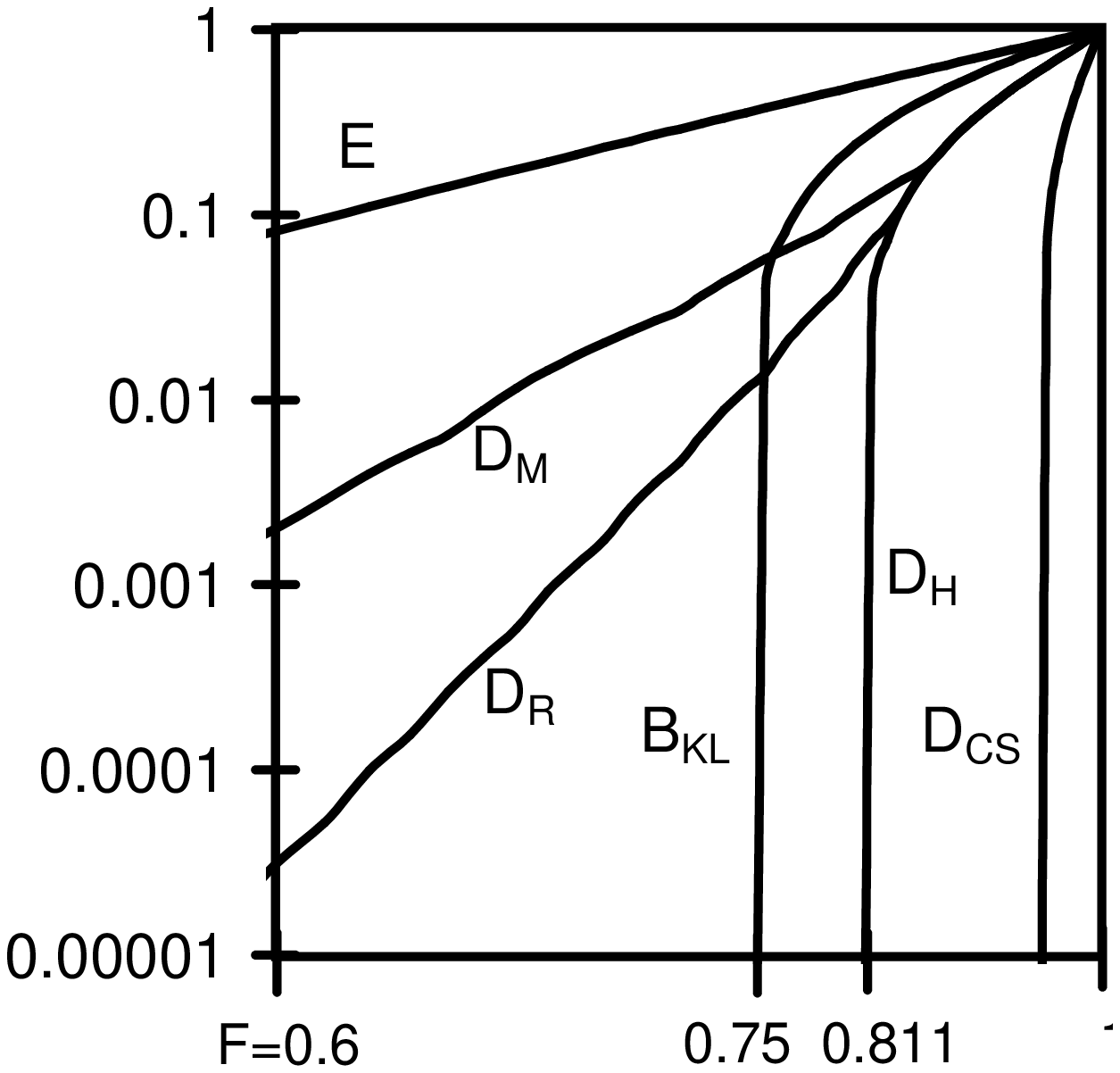}
\caption[logchart.eps]
{The same as Fig.~\protect\ref{dchart} exhibited on logarithmic
scales.  The value along the $x$-axis is proportional
to the logarithm of $(F-0.5)$.
In this form it is clear that $E$, $D_M$ and $D_R$ follow
power laws $(F\!-\!.5)^\alpha$.  The ripples in $D_M$ and $D_R$ are real,
and arise from the variable number of recurrence steps performed
before switching over to the hashing protocol\protect\cite{purification}.
}
\label{logchart}
\end{figure}
 
The members of the ``passed'' subset have a greater $p_{00}$ than
those in the original set of impure pairs.  The new density matrix is
still Bell diagonal, but is no longer a Werner state $W_F$.
Therefore, a twirl $T'$ is applied (Sec.~\ref{basics}, items~\ref{i1}
and~\ref{i5}), leaving the $p_{00}$ component alone and equalizing the
others~\cite{foot5}.  (It is appropriate in this situation to use the
modified twirl $T'$ which leaves $\Phi^+$ invariant, as explained in item
\ref{i5} of Sec.~\ref{basics}.)  We are left with a new situation
similar the the starting situation, but with a higher fidelity
$F'=p_{00}'$.  Figure~\ref{recurchart} shows the resulting $F'$ versus
$F$.  The process is then repeated; iterating the function of
Fig. \ref{recurchart} will continue to improve the fidelity.  This can
be continued until the fidelity is arbitrarily close to 1.
C.~Macchiavello~\cite{foot5} has found that faster convergence can be
achieved by substituting a deterministic bilateral $B_x$ rotation for
the twirl $T'$.  With this modification, the density matrix remains
Bell-diagonal, but no longer has the Werner form $W_F$ after the first
iteration; nevertheless its $p_{00}$ component increases more rapidly
with successive iterations.
 
Even with this improvement the recurrence method is rather
inefficient, approaching zero yield in the limit of high output
fidelity, since in each iteration at least half the pairs are lost
(one out of every two is measured, and the failures are discarded).
Figure~\ref{recurchart} shows the fraction of pairs lost on each
iteration.  A positive yield, $D_2$, even in the limit of perfect output
fidelity can be obtained by switching over from the recurrence method
to the hashing method, to be described in Section~\ref{hashing}, as
soon as so doing will produce more good singlets than doing another
step of recurrence.  The yield versus initial fidelity of these
combined recurrence-hashing protocols is shown in Figure~\ref{dchart}.
 
It is important to note that the recurrence-hashing method gives a
positive yield of purified singlets from all Werner states with
fidelity greater than 1/2.  Werner states of fidelity 1/2 or less have
$E=0$ and therefore can yield no singlets.  The pure hashing and
breeding protocols, described below, which are one-way protocols, work
only down to $F\approx .8107$, and even the best known one-way
protocol~\cite{jumpthegun} works only down to $F\approx .8096$.
 
\subsubsection{Direct purification of non-Bell-diagonal mixtures}
\label{subsec:nonBell}

Most of the purification strategies discussed in this paper assume
that the state to be purified is first brought to the Werner form,
or at least to Bell-diagonal form, by means of a twirling operation.
As we have said, though, this strategy
is somewhat wasteful and we use it only to make the analysis manageable.
In this subsection we give a simple example showing how a state can be
purified directly with no twirling.
For this particular example, it happens that the purification is
accomplished in a single step rather than in a series of steps that
gradually raise the fidelity.

Consider again the state $M$ of Eq.~(\ref{counter}):
\begin{equation}
M = \half|\uparrow\uparrow\rangle\langle\uparrow\uparrow|
+ \half|\Psi^+\rangle\langle\Psi^+|. \label{nonBell}
\end{equation}
Note that because the fully-entangled fraction (Eq.~(\ref{fef})) 
$f=1/2$  for this state, it cannot be purified by the recurrence
method.  However,
a collection of pairs in this state can be purified as 
using the following two-way protocol~\cite{richard}:
as in the recurrence method, Alice and Bob first perform the
BXOR operation
between pairs of pairs,
and then bilaterally measure each target pair in the
up-down basis.  One can show that if the outcome of
this measurement on a given target pair is ``down-down,'' then
the corresponding source pair is left in the completely
entangled state $\Psi^+$.  Alice and Bob therefore keep
the source pair only when they get this outcome, and discard
it otherwise.  The probability
of getting the outcome ``down-down''
is ${1\over{8}}$, and since each target pair had to be
sacrificed for the measurement, the yield from this
procedure is $D_2={1\over{16}}$.
The same strategy works for any state of the form
\begin{equation}
M = (1-p)|\uparrow\uparrow\rangle\langle\uparrow\uparrow|
+ p|\Psi^+\rangle\langle\Psi^+|,
\end{equation}
with yield $D_2=p^2/4$.

A recent paper by Horodecki {\em et al.}~\cite{horodecki2}
presents a more general approach to the purification of
mixed states which, like the above scheme, does not start
by bringing the states to Bell-diagonal form.  Their strategy
begins with a
filtering operation aimed at increasing the
fully entangled fraction $f$ (Eq.~(\ref{fef})) of the surviving pairs;
these pairs are then subjected to
the recurrence procedure described above.
These authors have shown that by this technique, one can distill
some amount of pure entanglement from any state of
two qubits having a nonzero entanglement of formation.
In other words, they have obtained for such systems the
very interesting result that
if $E(M)$ is nonzero, then so is $D_2(M)$.

\subsubsection{One-way hashing method}
\label{hashing}
 
This protocol uses methods analogous to those of universal hashing in
classical privacy amplification~\cite{privacyamplification}.  (We will
give a self-contained treatment of this hashing scheme here.)  Given a
large number $n$ of impure pairs drawn from a Bell-diagonal ensemble
of known density matrix $W$, this protocol allows Alice and Bob to
distill a smaller number $m\approx n(1\!-\!S(W))$ of purified pairs
(e.g. near-perfect $\Phi^+$ states) whenever $S(W)<1$. In the limit of
large $n$, the output pairs approach perfect purity, while the
asymptotic yield $m/n$ approaches $1\!-\!S(W)$.  This hashing protocol
supersedes our earlier breeding protocol~\cite{purification}, which we
will review briefly in Sec.~\ref{breeding}.
 
The hashing protocol works by having Alice and Bob each perform BXORs and
other local unitary operations (Table \ref{bell_table}) on
corresponding members of their pairs, after which they locally measure
some of the pairs to gain information about the Bell states of the
remaining unmeasured pairs.  By the correct choice of local
operations, each measurement can be made to reveal almost 1 bit about
the unmeasured pairs; therefore, by sacrificing slightly more than
$nS(W)$ pairs, where $S(W)$ is the von Neumann entropy (See
Eq. (\ref{tangy})) of the impure pairs, the Bell states of all the
remaining unmeasured pairs can, with high probability, be ascertained.
Then local unilateral Pauli rotations ($\sigma_{x,y,z}$) can be used
to restore each unmeasured pair to the standard $\Phi^+$ state.
 
The hashing protocol requires only one-way communication: after Alice
finishes her part of the protocol, in the process having measured
$n\!-\!m$ of her qubits, she is able to send Bob classical information
which, when combined with his measurement results, enables him to
transform his corresponding unmeasured qubits into near-perfect
$\Phi^+$ twins of Alice's unmeasured qubits, as shown in
Fig.~\ref{1way}.
 
Let $\delta$ be a small positive parameter that will later be allowed
to approach zero in the limit of large $n$. The initial sequence of
$n$ impure pairs can be conveniently represented by a $2n$-bit string
$x_0$ formed by concatenating the two-bit representations
(Eq. (\ref{2bit})) of the Bell states of the individual pairs, the
sequence $\Psi^-\Phi^+\Phi^-$ for example being represented 110010.
The {\em parity\/} of a bit string is the modulo-2 sum of its bits;
the parity of a subset {\sl s} of the bits in a string $x$ can be
expressed as a Boolean inner product ${\sl s}\cdot x$, i.e. the
modulo-2 sum of the bitwise AND of strings {\sl s} and $x$.  For
example ${\sl 1101}\cdot0111=0$ in accord with the fact that there are
an even number of ones in the subset consisting of the first, second
and fourth bit of the string 0111.  Although the inner product ${\sl
s}\cdot x$ is a symmetric function of its two arguments, we use a
slanted font for the first argument to emphasize its role as a subset
selection index, while the second argument (in Roman font) is the bit
string representing an unknown sequence of Bell states to be purified.
 
The hashing protocol takes advantage of the following facts:
\begin{itemize}
\item
the distribution $P_{X_0}$ of initial sequences $x_0$, being a product
of $n$ identical independent distributions, receives almost all its weight from
a set of $\approx2^{nS(W)}$ ``likely'' strings. If the likely
set ${\cal L}$ is defined as comprising the $2^{n(S(W)+\delta)}$ most
probable strings in $P_{X_0}$, then the probability that the initial
string $x$ falls outside ${\cal L}$ is
$O(\exp(-\delta^2n))$\cite{Schu}.
\item
As will be described in more detail later, the local Bell-preserving
unitary operations of Table \ref{bell_table} (bilateral $\pi/2$
rotations, unilateral Pauli rotations, and BXORs), followed by local
measurement of one of the pairs, can be used to learn the parity of an
arbitrary subset {\sl s} of the bits in the unknown Bell-state sequence
$x$, leaving the remaining unmeasured pairs in definite Bell states
characterized by a two-bits-shorter string $f_{\sl s}(x)$
determined by the initial sequence $x$ and the chosen subset {\sl
s}.
\item
For any two distinct strings $x\neq y$, the probability that they
agree on the parity of a random subset of their bit positions, i.e.,
that ${\sl s}\cdot x={\sl s}\cdot y$ for random {\sl s}, is
exactly 1/2. This is an elementary consequence of the distributive law
$({\sl s}\cdot x) \oplus ({\sl s}\cdot y) = {\sl
s}\cdot(x\oplus y)$.
\end{itemize}
 
The hashing protocol consists of $n\!-\!m$ rounds of the following
procedure.  At the beginning of the $(k+1)$'st round.
$k=0,1...n-m-1$, Alice and Bob have $n-k$ impure pairs whose unknown
Bell state is described by a $2(n-k)$-bit string $x_k$.  In
particular, before the first round, the Bell sequence $x_0$ is
distributed according to the simple {\it a priori} probability
distribution $P_{X_0}$ noted above.  Then in the $(k+1)$'st
round, Alice first chooses and tells Bob a random $2(n-k)$-bit
string ${\sl s}_k$.  Second, Alice and Bob perform local unitary
operations and measure one pair to determine the subset parity ${\sl
s}_k\cdot x_k$, leaving behind $n-k-1$ unmeasured pairs in a Bell state
described by the $(2(n-k)-2)$-bit string $x_{k+1}=f_{{\sl s}_k}(x_k)$.
 
Consider the trajectories of two arbitrary but distinct strings
$x_0\neq y_0$ under this procedure.  Let $x_k$ and $y_k$ denote
the images of $x_0$ and $y_0$ respectively after $k$ rounds,
where the same sequence of operations $f_{{\sl s}_0}, f_{{\sl
s}_1}...f_{{\sl s}_{n-m-1}}$, parameterized by the same random-subset
index strings ${\sl s}_0, {\sl s}_1...{\sl s}_{n-m-1}$, is used for
both trajectories.  It can readily be verified that for any $r<n$
the probability
\begin{equation}
P((x_r \neq y_r)\;\& \;\forall_{k=0}^{r-1} ({\sl s}_k\cdot x_k =
{\sl s}_k \cdot y_k))
\end{equation}
(i.e., the probability that $x_r$ and $y_r$ remain distinct while
nevertheless having agreed on all $r$ subset parities along the way,
${\sl s}_k\cdot x_k={\sl s}_k\cdot y_k$ for $k=0...r-1$) is at most
$2^{-r}$.  This follows from the fact that at each iteration the
probability that $x$ and $y$ remain distinct is $\leq 1$, while the
probability that, if they were distinct at the beginning of the
iteration they will give the same subset parity, is exactly 1/2.
Recalling that the likely set ${\cal L}$ of initial candidates has
only $2^{n(S(W)+\delta)}$ members, but with probability greater than
$1-O(\exp(-\delta^2n))$ includes the true initial sequence $x_0$, it
is evident that after $r=n\!-\!m$ rounds, the probability of failure,
i.e. of no candidate, or of more than one candidate, remaining at the
end for $x_m$, is at most
$2^{n(S(W)+\delta)-(n\!-\!m)}+O(\exp(-\delta^2n))$.  Here the first term
upper-bounds the probability of more than one candidate surviving,
while the second term upper-bounds the probability of the true $x_0$
having fallen outside the likely set.  Letting $n\!-\!m=n(S(M)+2\delta)$
and taking $\delta\approx n^{-1/4}$, we get the desired result, that
the error probability approaches 0 and the yield $m$ approaches
$n(1\!-\!S(M))$ in the limit of large $n$.
 
\begin{figure}[htbp]
\centerline{\psfig{figure=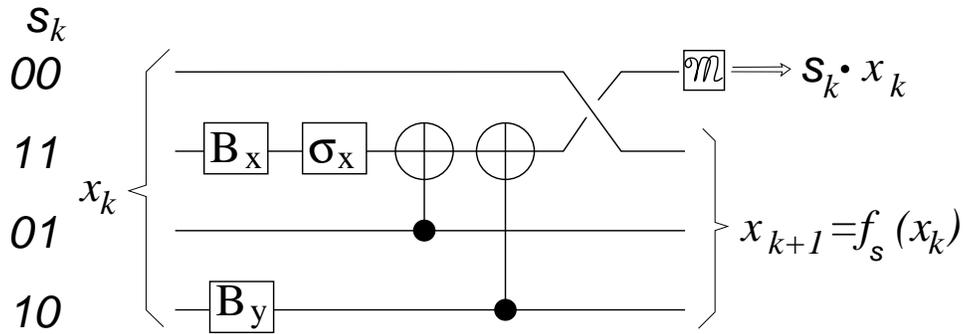,width=5in}}
\caption[Step k of one-way hashing]{Step $k$ of the one-way hashing protocol, used to determine
the parity ${\sl s}_k\cdot x_k$, for an arbitrary unknown set of four
Bell states represented by an unknown 8-bit string $x$ relative to a
known subset index string ${\sl s}={\sl 00,11,01,10}$.  If bilateral
measurement $\cal M$ yields a $\Psi$ state (i.e. if the
measurement result is 1), then half the candidates for $x$ are
excluded (e.g. $x$=00,00,00,00), but half are still allowed (e.g.
$x$=00,11,00,00). For each allowed $x$, the after-measurement Bell
states of the three remaining unmeasured pairs are a described by a
6-bit sequence $x_{k+1}=f_{\sl s}(x_k)$ deterministically computable
from $x$ and {\sl s}.
\label{figure10}}
\end{figure}

It remains to show how the local operations of Table \ref{bell_table}
can be used to collect the parity of an arbitrary subset of bits of
$x$ into the amplitude bit of a single pair.  We choose as the
destination pair, into which we wish to collect the parity ${\sl
s}\cdot x$, that pair corresponding to the first nonzero bit of $s$.
For example if ${\sl s}={\sl 00,11,01,10}$ (see Fig.~\ref{figure10}),
the destination will be the second pair of $x_k$.  Our goal will be to
make the amplitude bit of that pair after round $k$ equal to the
parity of: both bits of the second pair, the right bit of the third
pair, and the left bit of the fourth pair in the unknown input $x_k$.
Pairs such as the first, having {\sl 00} in the index string {\sl s},
have no effect on the desired subset parity, and accordingly are
bypassed by all the operations described below.
 
The first step in collecting the parity is to operate separately on
each of the pairs having a {\sl 01}, {\sl 10}, or {\sl 11} in
the index string, so as to collect the desired parity {\em for that
pair\/} into the amplitude (right) bit of the pair. This can be
achieved by doing nothing to pairs having {\sl 01} in the index
string, performing a $B_y$ on pairs having {\sl 10} (since $B_y$
has the effect of interchanging the phase and amplitude bits of a Bell
state), and performing the two rotations $B_x$ and $\sigma_x$ on pairs
with {\sl 11} in the index string ($B_x\sigma_x=\sigma_xB_x$ has
the effect of XORing a Bell state's phase bit into its amplitude bit).
 
The next step consists of BXORing all the pairs except those with {\sl
00} in the index string into the selected destination, in this case
the second pair.  The selected destination pair is used as the common
target for all these BXORs, causing its amplitude bit to accumulate
the desired subset parity ${\sl s}\cdot x$.  This follows from the
fact (cf. Table \ref{bell_table}) that the BXOR leaves the source's
amplitude bit unaffected while causing the target's amplitude bit to
become the XOR of the previous amplitude bits of source and target.
Recall that phase bits behave oppositely under BXOR: the target's
phase bit is unaffected while the source's phase bit becomes the XOR
of the previous values of source and target phase bits; this
``back-action'' must be accounted for in determining the function
$f_{\sl s}$.  Figure~\ref{figure10} illustrates this step of the
hashing method on an unknown 4-Bell-state sequence $x$ using the
subset index string ${\sl s}={\sl 00,11,01,10}$ mentioned before.
 
The hashing protocol distills a yield $D_H=1\!-\!S(W)$, which we have called
$D_0$ in our previous work\cite{purification}.  For the
Werner channel, parameterized completely by $F$,
\beq
S(W_F)=-F \log_2(F)-(1\!-\!F)\log_2((1\!-\!F)/3),
\eeq
giving a positive yield for Werner states with $F>0.8107$.  
Figures~\ref{dchart} and \ref{logchart} show $D_H(F)$, comparing it with
$E$ and with other purification protocols.
 
\subsubsection{Breeding method}
\label{breeding}
 
This protocol, introduced in Ref.~\cite{purification}, will not be
described here in detail, as it has been superseded by the one-way
hashing protocol described in the preceding section.  The breeding
protocol assumes that Alice and Bob have a shared pool of pure
$\ket{\Phi^+}=00$ states, previously prepared by some other method
(e.g. the recurrence method) and also a supply of Bell-diagonal impure
states which they wish to purify.  The protocol consumes the $\Phi^+$
states from the pool, but, if the impure states are not too impure,
produces more newly purified pairs than the number of pool states
consumed (in the manner of a breeder reactor).
 
\begin{figure}[htbp]
\centerline{\psfig{figure=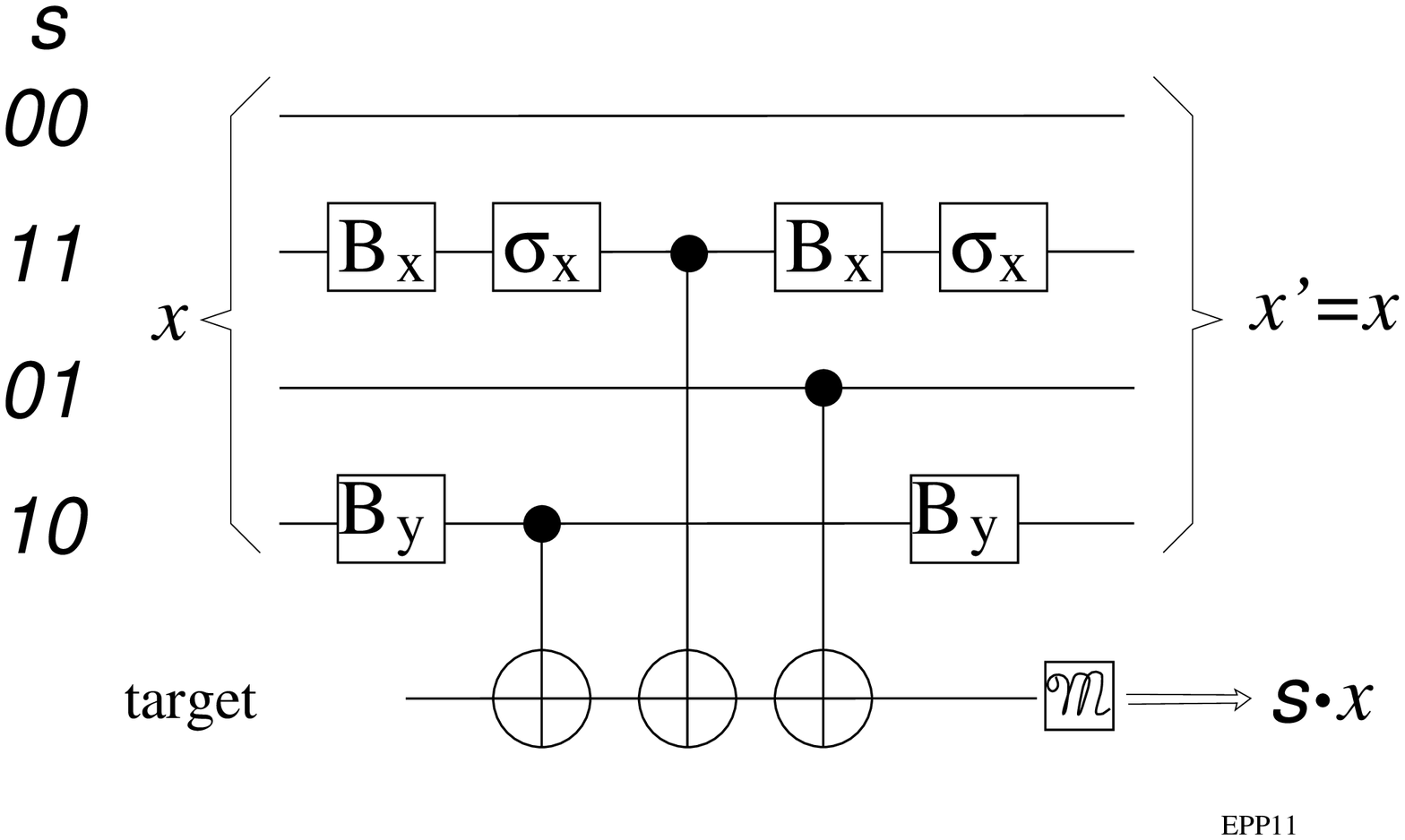,width=5in}}
\caption[EPP11.eps: Breeding]{Step $k$ of the one-way breeding protocol.
The scheme is
very similar to the hashing protocol of Fig.~\protect\ref{figure10},
except that the target for the BXORs is guaranteed to be a perfect
$\Phi^+$ state.  This allows the one-bit operations to be undone so
that there is no back-action on the string $x$.}
\label{breedfig}
\end{figure}

The basic step of breeding is very similar to that of hashing and is
shown in Fig. \ref{breedfig}.  Again a random subset {\sl s} of the
amplitude and phase bits of the Bell states is selected.  The parity
of this selected set is again gathered up in exactly the same way,
except that the target of the BXOR operations is one of the
pre-purified 00 states.  The use of the pure target simplifies the
action of the BXOR, in that the ``back action'' which changes the
state of the source bits is avoided in this scheme.  This means that
the input string $x$ can be restored to exactly its original value by
a simple undoing of the one-qubit local operations, as shown, This
offers the advantage that the (possibly very complicated) sequence of
boolean functions $f_{{\sl s}_0}, f_{{\sl s}_1}...f_{{\sl s}_{n-m-1}}$
do not have to be calculated in this case.  Once again, the result of
the parity measurement $\cal M$ is to reduce the number of
candidates for $x$ by almost exactly 1/2.  Thus, by the same argument
as before, after $n\!-\!m\approx nS(W)$ rounds of parity
measurements, it is probable that $x$ has been narrowed down to be
just one member of the likely set $\cal L$.  Thus, all $n$ of these
pairs can be turned into pure $\Phi^+$ states; however, since $n\!-\!m$
pure $\Phi^+$'s have been used up in the process, the net yield is
$m/n=D_H(F)$, exactly the same as in the hashing protocol.
 
\section{One-way $D$ and two-way $D$ are provably different}
\label{sec:D1D2}
 
It has already been noted that some of the entanglement purification
schemes use two-way communication between the two parties Alice and
Bob while others use only one-way communication.  The difference is
significant because one-way protocols can be used to protect quantum
states during storage in a noisy environment, as well as during
transmission through a noisy channel, while two-way protocols can only
be used for the latter purpose (cf. Section \ref{simplecode}).
Thus it is important to know whether there are mixed states for which
$D_1$ is properly less than $D_2$.  Here we show that there are, and
indeed that the original Werner state $W_{5/8}$, (i.e., the result of
sharing singlets through a 50\% depolarizing channel) cannot be
purified at all by one-way protocols, even though it has a positive
yield under two-way protocols.
 
\begin{figure}[htbf]
\centerline{\psfig{figure=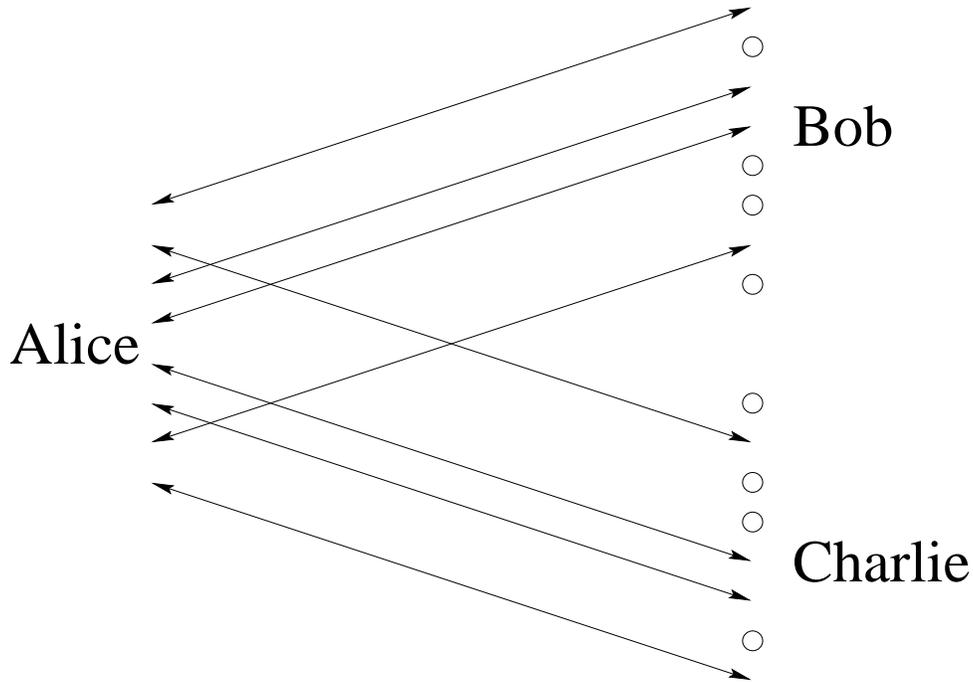,width=5in}}
\caption[A symmetric situation]
{A symmetric situation in which Bob and Charlie are
each equally entangled with Alice.  Two-headed arrows denote
maximally-entangled pairs, and open circles denote garbage states
(Eq.~(\protect\ref{garbage})).
\label{abc}}
\end{figure}

To show this, consider an ensemble where a state-preparer gives Alice
$n$ singlets, half shared with Bob and half shared with another person
(Charlie).  Alice is unaware of which pairs are shared with Bob
and which with Charlie.  Bob and Charlie are also given enough extra
garbage particles (either randomly selected qubits or any state
totally entangled with the environment but with no one else) so that
they each have a total of $n$ particles as well.  This situation is
diagrammed in Fig.~\ref{abc}.
From Alice and Bob's point of view,
each state has the density matrix $W_{5/8}$.
 
Alice, without hearing any information from Bob or Charlie, is
supposed to do her half of a purification protocol and then send on
classical data to the others.  Therefore, each particle Alice has looks
like a totally mixed state to her.  By symmetry, anything she could do
to assure herself that a particular particle is half of a good EPR
pair shared with Bob will also assure her that the same particle is
half of a good EPR pair shared with Charlie.  No such three-sided EPR
pair can exist.  If she used it to teleport a qubit to Bob she would
also have teleported it to Charlie, violating the no-cloning
theorem~\cite{WZ}.  Therefore, she cannot distill even one good EPR
pair from an arbitrarily large supply of $W_{5/8}$ states.  On the
other hand the combined recurrence-hashing method ($D_M$ in
Fig.~\ref{logchart}) gives a positive lower bound on the two-way yield
$D_2(W_{5/8}) > 0.00457$ so we can write
\beq
 D_1(W_{5/8}) =0 < 0.00457 \leq D_2(W_{5/8}).
\label{d10}
\eeq
It is also clear that any ensemble of Werner states can be
reduced to one of lower fidelity by local action (combining with
totally mixed states of Eq.~(\ref{garbage})).  Therefore $D_1(W_F)=0$
for all $F<5/8$.  Knill and Laflamme prove~\cite{lafknill} that
$D_1(W_F)=0$ for all $F<3/4$.  In Sec.~\ref{newupperbound} we 
explain their proof and, using the argument of Sec.~\ref{additivity},
obtain the bound
\beq
D_1 < 4f-3\ ,
\eeq 
as shown in Figs.~\ref{dchart} and \ref{logchart}.

A similar argument can be used to show that for some ensembles
$D_1$ is not symmetric depending on whether it is Alice or
Bob who starts the communication.  Suppose in the symmetric
situation of Fig.~\ref{abc} that Bob and Charlie {\em know} which
pairs are shared with Alice and which are garbage.  For this
ensemble the symmetry argument for Alice remains the same and
$D_{A\rightarrow B}=0$.  If the communication is from Bob to Alice,
though, it is easy to see he can use half of his particles, the
ones he knows are good pairs shared with Alice.  The other half
are useless since they have $E=0$ and could have been manufactured
locally.  Thus we have $D_{B\rightarrow A}=1/2$ and
$D_{A\rightarrow B}=0$.
 
Our no-cloning argument shows that Alice and Bob cannot generate good
EPR pairs by applying a 1-EPP to the mixed state $W_{5/8}$ generated
by sharing singlets through a 50\% depolarizing channel.  As a 
consequence, there is no quantum error-correcting code which can
transmit unknown quantum states reliably through
a 50\% depolarizing channel, as will be shown in the next
section. 
 
\section{Noisy Channels and Bipartite Mixed States}\label{CandM} In
preceding sections we have considered the preparation and purification
of bipartite mixed states, and we have shown that two-way entanglement
purification protocols can purify some mixed states that cannot be
purified by any one-way protocol.  When used in conjunction with
teleportation, purification protocols, whether one-way or two-way, offer
a means of transmitting quantum information faithfully via noisy
channels; and one-way protocols, by producing time-separated
entanglement, can additionally be used to protect quantum states during
storage in a noisy environment.   In this section we discuss the close
relation between one-way entanglement purification protocols and the
other well-known means of protecting quantum information from noise,
namely quantum error-correcting
codes (QECC)~\cite{ChuangLaf,shellgame,CS,Steane,Laflamme,EM,Sam,balance,newest}. 
 
A quantum channel $\chi$,
operating on states in an $N$-dimensional Hilbert space, may be defined
as (cf.~\cite{shellgame}) a unitary interaction of the input state with an
environment, in which the environment is supplied in a standard pure
initial state $\ket{\mbox{\boldmath{$0$}}}$ and is traced out 
(i.e. discarded) after
the interaction to yield the channel output,
generally a mixed state.  The quantum capacity $Q(\chi)$ of such a channel
is the maximum asymptotic rate of reliable transmission of unknown quantum states 
$\ket{\xi}$ in ${\cal H}_2$ through the channel that can be achieved by using
a QECC to encode the states before transmission and decode them afterward.

\begin{figure}[htbp]
\centerline{\psfig{figure=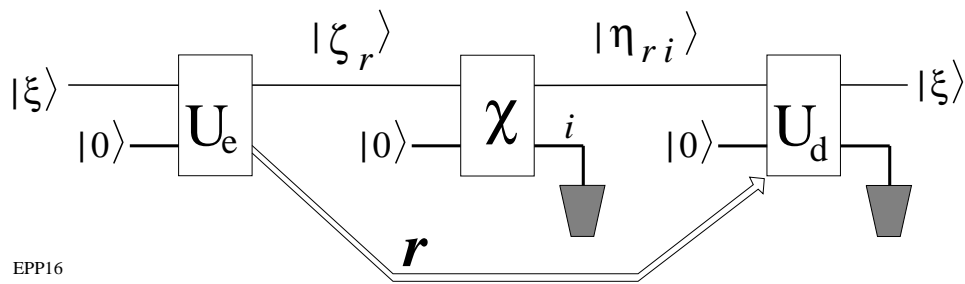,width=5in}}
\caption[General $Q_1$ style EPP]
{A general one-way QECC.  A classical side-channel from Alice
to Bob is allowed in addition to quantum channel $\chi$.}
\label{epp16}
\end{figure}

As in quantum teleportation~\cite{teleportation} we will also consider
the possibility that the quantum channel is supplemented with
classical communication.  This leads us to define the augmented quantum
capacities $Q_1(\chi)$ and $Q_2(\chi)$, of a channel supplemented
by unlimited one- and two-way classical communication. For example,
Fig.~\ref{epp16} shows a quantum error-correcting code, consisting
of encoding transformation $U_e$ and decoding transformation $U_d$, used
to transmit unknown quantum states $\ket{\xi}$ reliably through the
noisy quantum channel $\chi$, with the help of a one-way classical
side channel (operating in the same direction as the quantum channel).
Perhaps surprisingly, this one-way classical channel provides no
enhancement of quantum capacity:
\beq
Q_1=Q\ .\label{qequalsq1}
\eeq
This will be shown in Sec.~\ref{decodeproof}.
 
We consider also the case of a noisy quantum channel supplemented
by a {\em noiseless} quantum channel.  We will show 
in Sec.~\ref{additivity} that the capacity of $n$ uses of a
noisy channel supplemented by $m$ uses of a noiseless channel
of unit capacity is no greater than the sum of their
individual capacities, i.e. their quantum capacities are no 
more than additive.  We have no similar result for the case of 
two different imperfect channels.
 
In contrast to Eq.~(\ref{qequalsq1}) we will show that for may quantum 
channels two way
classical communication can be used to transmit quantum states through
the channel at a rate $Q_2(\chi)$ considerably exceeding the one-way capacity
$Q(\chi)$.  This is typically done by using the channel to share
EPR pairs between Alice and Bob, purifying the resulting
bipartite mixed states by a two-way entanglement purification protocol,
then using the resulting purified pairs to teleport
unknown quantum states $\ket{\xi}$ from Alice to Bob.
 
The analysis of $Q$ and $Q_2$ is considerably simplified by the fact that
an important class of noisy channels, including depolarizing
channels, can be mapped in a one-to-one fashion onto a corresponding
class of bipartite mixed states, with the consequence that the channel's
quantum capacity $Q_1=Q$ is given by the one-way distillable
entanglement $D_1$ of the mixed state, and vice versa. For example, a
depolarizing channel of depolarization probability $p=1-x$
(cf. Eq.~(\ref{garbage})) corresponds to a
Werner state $W_F$ of fidelity $F=1-(3p/4)$ and has $Q=D_1(W_F)$ and
$Q_2=D_2(W_F)$.
 
The correspondence between channels and mixed states is established by
two functions, $\hat{M}(\chi)$ defining the bipartite mixed state
obtained from channel $\chi$ and $\hat{\chi}(M)$ defining the channel
obtained from bipartite mixed state $M$. 
The bipartite mixed state $\hat{M}(\chi)$ is
obtained by preparing a standard maximally entangled state of two
$N$-state subsystems,
 \beq \Upsilon = N^{-1/2}\sum_{i=1}^N
\ket{e_j}\otimes\ket{e_j}
 \eeq  and transmitting Bob's part
through the channel $\chi$. For example a Werner state $W_F,$ with
$F=1\!-\!3p/4$ results when half a standard EPR pair is transmitted through
a $p$-depolarizing channel. 
 
The mapping in the other direction, from
mixed states to channels, is obtained by teleportation.  Given a
bipartite mixed state $M$ of two subsystems, each having Hilbert space
of dimension $N$, the channel $\hat{\chi}(M)$ is defined by using mixed
state $M$, instead of the standard maximally entangled state
$\proj{\Upsilon}$, in a teleportation~\cite{teleportation} 
channel (see Fig.~\ref{qecc}).  It can be readily 
shown that for Bell-diagonal mixed states the two mappings are mutually
inverse $\hat{M}(\hat{\chi}(M))=M$; we shall call the channels
corresponding to such mixed states ``generalized depolarizing
channels''.
 
For more general channels and mixed states, the two mappings are not
generally mutually inverse.  For example, $\hat{\chi}(M)$, for
the bipartite state $M=\outerprod{\!\uparrow\uparrow}{\uparrow\uparrow\!}$, 
is the $p=1$ depolarizing channel, and $\hat{M}(\hat{\chi}(M))=G$ of
Eq.~(\ref{garbage}).
 
Nevertheless, two quite general inequalities will be demonstrated in
Sections~\ref{ineq1} and~\ref{ineq2}:
\beq
\forall_M  \;\;\; D_1(M) \geq Q(\hat{\chi}(M))
\label{qecc1epp}\eeq
and
\beq
\forall_{\chi} \;\;\; D_1(\hat{M}(\chi)) \leq Q(\chi).
\label{timereveq}\eeq
 
If (as in the case of a Bell diagonal state and its corresponding
generalized depolarizing channel) the mapping is reversible,
so that $ M=\hat{M}(\chi)$ and $\chi=\hat{\chi}(M)$, the two inequalities
are both satisfied, resulting in the equality mentioned earlier, viz.
 \beq
D_1(M)=Q(\chi).
\label{bigeq}
 \eeq
Equation~(\ref{qecc1epp}) follows from the ability, to be demonstrated
in the Sec.~\ref{ineq1}, to transform a QECC on
$\hat{\chi}(M)$ into a 1-EPP on $M$; Eq. (\ref{timereveq}) follows, as
shown in Sec.~\ref{ineq2}, from the fact that any 1-EPP on $\hat{M}(\chi)$,
followed by quantum teleportation, results in a QECC on $\chi$ 
with a classical side channel.
 
A trivial extension of these arguments also shows that the corresponding
results for two-way classical communication are true, namely:
\beq
\forall_M  \;\;\; D_2(M) \geq Q_2(\hat{\chi}(M))
\eeq
and
\beq
\forall_{\chi} \;\;\; D_2(\hat{M}(\chi)) \leq Q_2(\chi)\ ,
\eeq
and if $\hat{M}(\hat{\chi}(M))=M$ then
\beq
D_2(M)=Q_2(\chi).
\eeq

\subsection{A forward classical side channel does not increase
quantum capacity}
\label{decodeproof}

To demonstrate Eq.~(\ref{qequalsq1}), we note that
any one-way protocol for transmitting $\ket{\xi}$ through channel 
$\chi$ can be described as in Fig~\ref{epp16}.  The sender 
Alice codes $\ket{\xi}$ and an ancillary state 
$\ket{\mbox{\boldmath{$0$}}}$ using 
unitary transformation $U_e$.  She then performs an incomplete
measurement on the coded system giving classical results
$r$ which she sends on to Bob, the receiver. 
(if $r$ contains any information about 
the quantum input $\ket{\xi}$ the strong no-cloning theorem~\cite{BBM} 
would prevent the original state from being recovered perfectly,
even if the channel were noiseless.  However, $r$ might contain 
information on how the input $\ket{\xi}$ is coded.)  She also sends 
the remaining quantum state through $\chi$ as encoded state $\ket{\zeta_r}$.  
The channel maps $\ket{\zeta_r}$ onto $\ket{\eta_{ri}}$ for a noise 
syndrome $i$.  

Consider the unitary transformation Bob uses for decoding 
in the case of some value of the classical data $r$ for
for which the decoding is successful 
and without loss of generality name this case $r\!=\!0$.
(For a code which corrects with asymtotically perfect fidelity there may
be some cases of $r$ for which the correction doesn't work.) 
We also consider error syndrome $i$ which is successfully corrected
by $U_d$.  We have
\beq
U_d(r\!=\!0)(\ket{\eta_{0i}} \otimes \ket{\mbox{\boldmath{$0$}}})=\ket{\xi}
\otimes\ket{a_i}\ .
\eeq
(For our choice of $i$ the final $\ket{a_i}$ state can 
without loss of generality be taken to be $\ket{\mbox{\boldmath{$0$}}}$ 
in an appropriately sized Hilbert space.)  Applying $U_d^{-1}(r\!=\!0)$ gives
\beq
U_d^{-1}(r\!=\!0) (\ket{\xi}\otimes\ket{\mbox{\boldmath{$0$}}})=
\ket{\eta_{0i}}\otimes\ket{\mbox{\boldmath{$0$}}}\ .
\eeq
There must exist another unitary operation $U_s$ which rotates
$\ket{\eta_{0i}}$ into the noiseless coded vector
$\ket{\zeta_0}$. Thus, 
\beq
U_sU_d^{-1}(r\!=\!0) (\ket{\xi}\otimes\ket{\mbox{\boldmath{$0$}}})=
\ket{\zeta_0}\otimes\ket{\mbox{\boldmath{$0$}}}\ .
\eeq
In other words, $U_sU_d^{-1}(r\!=\!0)$ takes $\ket{\xi}$ into 
$\ket{\zeta_0}$ along with some ancillary inputs and outputs
always in a standard 
$\ket{\mbox{\boldmath{$0$}}}$ state.
Therefore $U_sU_d^{-1}(r\!=\!0)$ is a good encoder.
Since this encoder always results in the correct code vector
corresponding to classical data $r\!=\!0$ this data need not be sent 
to Bob at all, as he will have anticipated it. 
Thus, $U_sU_d^{-1}(r\!=\!0)$ and $U_d$ 
form a code needing no classical side-channel.

It may happen that for a large block code which only error-corrects
to some high fidelity ($|\langle \xi | \xi_f \rangle| > 1-\epsilon$ 
where $\ket{\xi_f}$ is the final output of the decoder)
that {\em no} case is corrected perfectly.
Then the coded states produced by $U_sU_d^{-1}(r\!=\!0)$ will
be imperfect.  After transmission through the noisy
channel and correction by $U_d$ the final output will then be
less perfect than in the original code.  Nevertheless, because
of unitarity it is clear that as $\epsilon \rightarrow 0$ 
the fidelity of this code will also approach unity.

Thus any protocol using classical one-way data transmission
to supplement a quantum channel can be converted into a protocol
in which the classical transmission is unnecessary and with the
same capacity $Q=Q_1$.  We have also now shown that the
encoding stage is unitary, in the sense that no extra classical 
or quantum results accumulate in Alice's lab.  

If the error syndrome $i=0$, corresponding to no error, is 
decoded with high fidelity by $U_d$ then $U_s$ can be taken to be the 
identity.  Thus, the encoding and decoding transformations can 
in this case be written in a form where $U_e=U_d^{-1}$, a 
fact independently shown by Knill and Laflamme~\cite{lafknill}.  
If the $i=0$ error syndrome is not decoded with high fidelity by
$U_d$~\cite{footnonoise} then the encoder cannot be the inverse of 
the decoder.  The proof is simple:
$U_e(\ket{\xi}\otimes\ket{\mbox{\boldmath{$0$}}})=\ket{\zeta}$ (where
we have dropped the $r$ subscripts since it has been proven the classical
data is never needed) and therefore 
$U_e^{-1}\ket{\zeta}=(\ket{\xi}\otimes\ket{\mbox{\boldmath{$0$}}}).$
Thus $U_e^{-1}$ decodes the noiseless coded vectors $\ket{\zeta}$ which
is exactly what $U_d$ has been assumed not to do.

\subsection{Additivity of perfect and imperfect quantum channel capacities}
\label{additivity}


Consider a channel of capacity $Q>0$ supplemented by a perfect
channel of capacity 1.  Suppose the imperfect channel is used
$n$ times and the perfect channel is used $m$ times.  We will call  
the maximum number of bits transmitted through the channels in this
case $T$.
If the capacity of this joint channel is additive then 
$T=T_a=Qn+m$.

Suppose the number of bits transmitted is superadditive, i.e. $T>T_a$.  
From the definition of noisy channel capacity we know that we can 
use an imperfect channel $t$ times to simulate
a perfect channel being used $m$ times where $Qt=m$.
We now use the imperfect channel a total $n+t$ 
times and we can transmit $T$ qubits through this two-part
use of the imperfect channel.  But $T>T_a=Qn+m$ so 
\begin{equation}
T>Qn+Qt\ . 
\label{tg}
\end{equation}
The capacity of this channel is $Q'=\frac{T}{n+t}$.  Using Eq.(\ref{tg})
we can write
\begin{equation}
Q'=\frac{T}{n+t} > \frac{Qn+Qt}{n+t}=Q\ . 
\end{equation}
A capacity of $Q'> Q$ has been achieved using only the original
imperfect channel whose capacity was $Q$.  This cannot be so.

\subsection{QECC $\rightarrow$ 1-EPP proving
$\boldmath{\forall_M \; D_1(M) \geq Q(\hat{\chi}(M))}$}

\label{ineq1} 
 
\begin{figure}[htbp]
\centerline{\psfig{figure=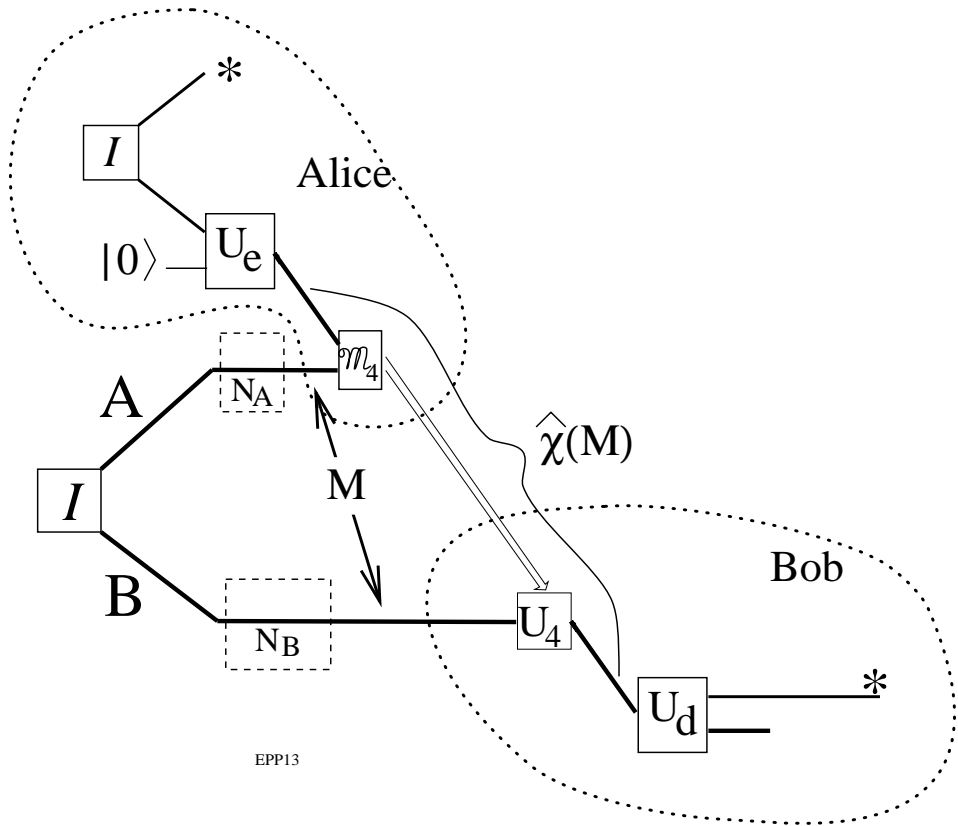,width=5in}}
\caption[A QECC can be transformed into a 1-EPP]
{A QECC can be transformed into a 1-EPP.
Teleporting ($M_4, U_4$) via a mixed state $M$ defines the noisy channel $\hat{\chi}(M)$.
If a quantum error-correcting code $\{U_e, U_d\}$ can correct the errors
in this channel, the code and channel can be used to share pure 
entanglement between Alice and Bob (*).  This establishes 
inequality~(\ref{qecc1epp}), viz.
$\forall_M \;\;\; D_1(M) \geq Q(\hat{\chi}(M))$.}
\label{ineq1fig}
\end{figure}

To demonstrate this inequality (cf. Fig.~\ref{ineq1fig}) we use
bipartite mixed states
$M$ in place of the standard maximally entangled states
$(\Phi^+)$ to teleport $n$ qubits from Alice to Bob.  This teleportation
defines a certain noisy channel $\hat{\chi}(M)$, so designated
on the center right of the figure. Alice
prepares $n$ qubits to be teleported through this channel by applying the
encoding transformation
$U_e$ of a QECC to $m$ halves of EPR pairs which she generates in her lab
(upper left) at $I$ and to
$n-m$ ancillas in the standard $\ket{0}$ state.  The resulting quantum-encoded
$n$ qubits are teleported to Bob at lower right through the noisy channel.
There Bob applies the decoding transformation $U_d$.  If the code can
successfully correct the errors introduced by the noisy teleportation,
then the result is that Alice and Bob share $m$ time-separated EPR pairs
(*).  Indeed the whole figure can be regarded as a one-way
purification protocol whereby Alice and Bob prepare $m$ good EPR pairs from $n$
of the initial mixed states $M$, using a QECC of rate $Q=m/n$ able to correct
errors in the noisy quantum channel $\hat{\chi}(M)$.  Thus $D_1(M)$ must be
at least as great as the rate $Q(\hat{\chi}(M))$ of the best QECC able
to achieve reliable quantum transmission through $\hat{\chi}(M)$.
 
\subsection{
1-EPP $\rightarrow$ QECC proving
$\boldmath{\forall_{\chi} \; D_1(\hat{M}(\chi)) \leq Q(\chi)}$} 
\label{ineq2}
 
In the same style as the last section, we establish the second
inequality by exhibiting an explicit protocol.  The object is to show
that, given the existence of a 1-EPP acting on the mixed state $\hat{M}(\chi)$
obtained from quantum channel $\chi$, Alice can successfully transmit
arbitrary quantum states $|\xi\rangle$ to Bob.  The capacity $Q$ of
this quantum channel is the same as $D_1$ for the 1-EPP; this
establishes that the capacity of $\chi$ is at least as good as the $D_1$
of the corresponding 1-EPP.
\begin{figure}[htbp]
\centerline{\psfig{figure=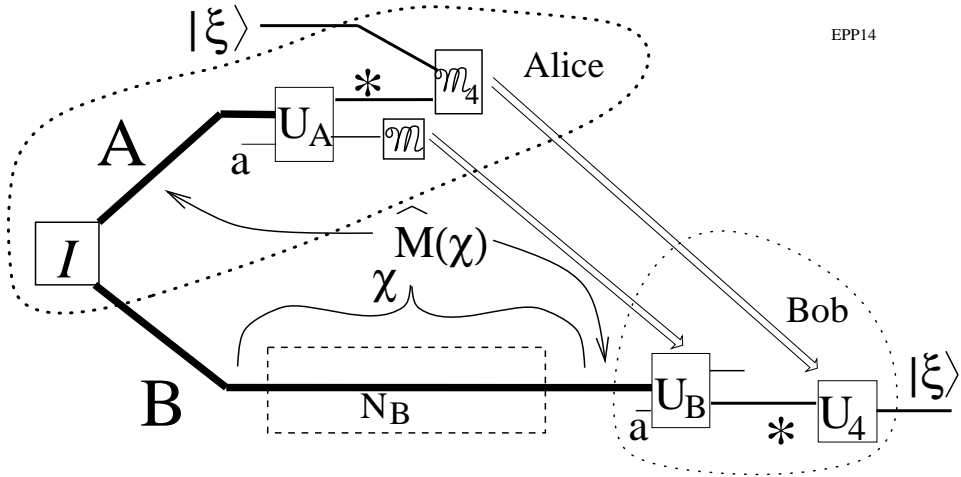,width=5in}}
\caption[A 1-EPP can be transformed into a QECC]
{A 1-EPP can be transformed into a QECC.  Given $\chi$,
Alice creates mixed states $\hat{M}(\chi)$ by passing halves
of entangled states $\Phi^+$ from source $I$ through the
channel.  Alice and Bob perform a 1-EPP resulting in 
perfectly entangled states (*) which are then used to teleport
$\ket{\xi}$ safely to Bob, completing a QECC.}
\label{ineq2fig}
\end{figure}
 
In fact, this protocol just involves the application of quantum
teleportation~\cite{teleportation}
mentioned in the introduction.  In Fig.~\ref{ineq2fig}
we show more explicitly the necessary construction, which has already
been touched on in Figs.~\ref{1way} and \ref{qecc}.  Alice and Bob
are connected by channel $\chi$.  Alice arranges to share the bipartite
mixed state $\hat{M}(\chi)$ with Bob by passing halves (the $B$ particles)
of maximally entangled
states ($\Phi^+$) from source $I$ through $\chi$ to Bob.
Then Alice and Bob
partake in the 1-EPP protocol.  We have represented this procedure
somewhat more generally than is necessary for the hashing-type
procedures shown earlier, or for the finite-block protocols to be
derived below.  We simply indicate that they must preform two
operations $U_A$ and $U_B$, and that Alice will perform some
measurements $\cal M$ and pass the results to Bob.  The
measurements which Bob would perform in the hashing protocol are
understood to be incorporated in $U_B$.  Also, we have accounted
for the possibility that either Alice or Bob might employ an
ancilla $a$ for some of their processing operations.
 
By hypothesis, this protocol leaves Alice and Bob with $nD_1$ maximally
entangled states (*).  They then may use this resource to teleport
$nD_1$ unknown quantum bits in the state $|\xi\rangle$.  Thus, the
net effect is that Alice and Bob, using channel $\chi$ supplemented
by one-way classical communication, have a means
of reliably transmitting quantum data, with capacity $D_1(\hat{M}(\chi))$.
This is exactly a QECC on $\chi$ with a one-way classical side-channel.
However Eq.~(\ref{qequalsq1}) (proven in Sec.~\ref{decodeproof})
states that the same capacity can
be obtained without the use of classical communication. 
Thus, the ultimate capacity $Q$ of channel $\chi$ must be at least
as great.  This establishes the inequality.
 
\section{Simple quantum error-correcting codes}
\label{simplecode}
 
For most of the remainder of this paper, we will exploit the
equivalence which we have established between 1-EPP on $\hat{M}(\chi)$
and a QECC on $\chi$.  
 
We note that when the 1-EPP has the property that
the unitary transformations
$U_B$ and $U_4$ performed by Bob can be done ``in place''
(i.e. no ancilla qubits need to be introduced, see Fig.~\ref{1way}),
the 1-EPP can be transformed into a particularly simple style
of QECC, exactly like the schemes which have been introduced by
Shor~\cite{shellgame} and have now been extended by many
others~\cite{CS,Steane,Laflamme,EM,Sam,balance,newest}, which are also
all done ``in place.'' 
As we have seen in Figs.~\ref{ineq1fig}
and \ref{ineq2fig}, some versions of 1-EPP and QECC may require 
ancilla $a$ for their implementation. 
 
The proof of the correspondence between the in-place 1-EPP
and in-place QECC is immediate, following Sec.~\ref{ineq2}.
The 1-EPP is used to make a QECC as in Fig.~\ref{ineq2fig}.  The unitary
transformations $U_B$ and $U_4$ performed by Bob are combined as 
a $U_d$ and $U_d$ is performed in place by assumption.  
Thus $U_e=U_s U_d^{-1}$ (see Sec.~\ref{decodeproof}) can also 
be done in place.  
 
As a simple consequence of this result, the one-way hashing protocol
of Sec.~\ref{hashing} can be reinterpreted as an explicit error
correction code, and indeed it does the same kind of job as the recent
quantum error correction schemes based on linear-code theory of
Calderbank and Shor~\cite{CS} and Steane~\cite{Steane}: in the limit of
large qubit block size $n$, it protects an arbitrary state in a
$2^m$-dimensional Hilbert space from noise.  
We note that the hashing protocol actually does somewhat better
than the linear-code schemes.  $D_1(\hat{M}(\chi))$, and therefore
$Q(\chi)$ (see Eq.~(\ref{bigeq})),
is higher for hashing than for the linear-code scheme,
as shown in Figs.~\ref{dchart} and \ref{logchart}.
 
We will make further contact with this other work on error-correction coding
in finite blocks by showing how finite blocks
of EPR pairs can be purified in the presence of noise which only
affects a finite number of the Bell states.  When transformed into an
error correcting code, this becomes a procedure for recovering from a
finite number of qubit errors, as in Shor's procedure in which one
qubit, coded into nine qubits, is safe from any error on a single
qubit.  We develop efficient numerical strategies based on the
Bell-state approach which look for new coding schemes of this type,
and in fact we find a code which does the same job as Shor's using
only five EPR pairs.
 
\subsection{Another derivation of a QECC from a restricted 1-EPP}
\label{subsec:equ}
 
Another way to derive the in-place QECC from the in-place 1-EPP is
to exploit the
symmetry between measurement and preparation in quantum mechanics.
Here we will restrict our attention to noise models which are one-sided
(i.e., $N_A$ absent in Fig.~\ref{1way}), or {\it effectively} one-sided.
An important case where the noise is effectively one-sided
is when the mixed state $M$ obtained in Fig.~\ref{tommy} is Bell-diagonal,
i.e., has the form of $W$ (Eq.~(\ref{firstW})).  We can say that,
subjected to this noise, the pure Bell state is taken to an
ensemble of each of the four Bell states, with some probabilities.  Using
the notation of Sec.~\ref{subsec:recur} these are
$p_{00}$, $p_{01}$, $p_{10}$ and $p_{11}$:
\begin{equation}
|\Phi^+\rangle\ \ \rightarrow\ \ \{\sqrt{p_{00}}|\Phi^+\rangle,
\sqrt{p_{10}}|\Phi^-\rangle,\sqrt{p_{01}}|\Psi^+\rangle,\sqrt{p_{11}}
|\Psi^-\rangle\}=\{R_{mn}|\Phi^+\rangle\}.\label{belmod}
\end{equation}
(Here $R_{mn}$ are proportional to the operators
$\{I,\sigma_x,\sigma_y,\sigma_z\}$ of Table~\ref{bell_table}.)  It is
easy to show that the same mixed state could be obtained if the $B$
particles were subjected to a generalized depolarizing channel, and
$N_A$ were absent.  More generally, we require that $N_{A,B}$ be such
that the resulting $M$ could be obtainable from some channel $\chi$;
$M=\hat{M}(\chi)$ for some $\chi$.  This is a fairly 
obvious restriction to make,
since we are planning on defining a QECC on this effective quantum
channel $\chi$.  Note also that, since the twirling of Sec.~\ref{basics} 
(item~\ref{i1}) converts any bipartite mixed state into a Werner state,
for some purposes {\em any} noise can be made effectively 
one-sided.
 
We will now show that under these conditions, the operations performed
by Alice in Fig.~\ref{ineq2fig} can be greatly simplified.  Consider
the joint state of the $A$ and $B$ particles
after Alice has applied the unitary
transformation $U_1$ of Fig.~\ref{1way} as part of the purification 
protocol, but before
the one-sided noise $N_B$ has acted on the $B$ particles.  The joint state is
still a pure, maximally entangled state.  For convenience, we assume
that the source $I$ produces $\Phi^+$
Bell states.  (If it produced another type of Bell state, some
additional simple rotations can be inserted in the derivation we are
about to give.)  The initial product of $n$ Bell states may
be written
\begin{equation}
|{\bf \Phi}\rangle_i=\frac{1}{\sqrt{2^n}}\sum_{x=0}^{2^n-1}|x\rangle_A
|x\rangle_B.
\end{equation}
After the application of the unitary transformation $U_1$ to Alice's
particles, the new state of the system is
\begin{equation}
|{\bf \Phi}\rangle_f=\frac{1}{\sqrt{2^n}}\sum_{x=0}^{2^n-1}
\sum_{y=0}^{2^n-1}(U_1)_{x,y}|y\rangle_A|x\rangle_B.
\end{equation}
But notice that by a simple change of the dummy indices, this state
can be rewritten
\begin{equation}
|{\bf \Phi}\rangle_f=\frac{1}{\sqrt{2^n}}\sum_{x=0}^{2^n-1}
\sum_{y=0}^{2^n-1}|x\rangle_A(U_1^T)_{x,y}|y\rangle_B.
\end{equation}
That is, the unitary transformation applied to the $A$ particles is
completely equivalent to the same operation (transposed) applied to
the $B$ particles.
 
Alice's tasks in the 1-EPP protocol are thus reduced to making
one-particle measurements $\cal M$ on $n\!-\!m$ of the $A$
particles, making Bell measurements ${\cal M}_4$ between the $m$
qubits $|\xi\rangle$ to be protected and her remaining $m$ particles
(as in quantum teleportation~\cite{teleportation}), and applying
$U_1^T$ to the $B$ particles before sending them, along with her
classical measurement results, to Bob.  (Recall from the Introduction
that $m$ is the yield of good singlets from the purification
protocol.)  
 
However, the $n\!-\!m$ one-particle measurements
$\cal M$ can be eliminated entirely.  We use the
property of $\Phi^+$ states that if one of the particles is
measured to be $|0\rangle$ or $|1\rangle$ in the $z$ basis, 
then the other particle is ``collapsed'' into the same
state~\cite{EPR,Bell}.  So, rather than creating $n\!-\!m$ entangled states
at $I$, Alice simply prepares $n\!-\!m$ qubits in a definite state and
sends them directly into the $U_1^T$ operation.  To mimic the
randomness of the measurement $\cal M$, Alice might do $n\!-\!m$
coin flips to decide what the prepared state of these $B$ particles
will be, and send this classical data on to Bob.  But this is
unnecessary, since by hypothesis, the 1-EPP always yields perfect
entangled pairs (*), no matter what the values of the $\cal M$ 
measurements were.  So, Alice and Bob may as well pre-agree on some
particular definite set of values (e.g., all 0's), and Alice will
always pre-set those $B$ particles to that state.\cite{footas}
 
The only $A$ particles remaining in the protocol at this point are the
$m$ particles forming the halves of perfect EPR pairs with Bob, and
which are immediately used for teleportation to Bob.  But we note
that, following the usual rules of teleportation, the measurement
${\cal M}_4$ causes the corresponding $B$ particles, immediately after
their creation at source $I$, to be in the state $|\xi\rangle$ (if the
measurement outcome were 00), or a rotated version,
$\sigma_{x,y,z}|\xi\rangle$ (for the other measurement outcomes).
Again, the protocol should succeed no matter what the value of this
measurement; therefore, if Alice and Bob pre-agree that this classical
data should be taken to have the value 00, then Alice can eliminate
the $A$ particles entirely, eliminate the preparation $I$ of entangled
states, and simply feed in the $|\xi\rangle$ states directly as $B$
particles into the $U_1^T$ transformation.  (Bob also does the $U_4$
operation of Fig.~\ref{1way} appropriate for 00, namely, a no-op.)
 
\begin{figure}[htbp]
\centerline{\psfig{figure=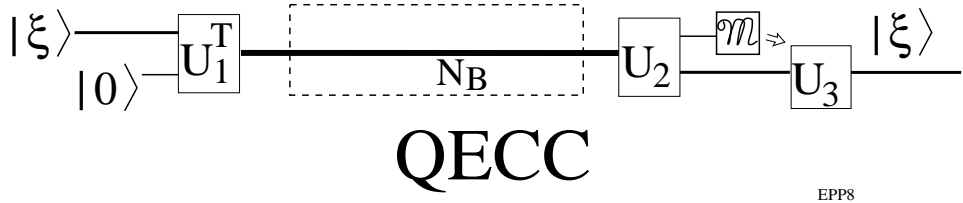,width=5in}}
\caption[EPP8.eps: 1-EPP into QECC]{The one-way purification protocol of Fig.~\protect\ref{qecc}
may be transformed into the quantum-error-correcting-code protocol
shown here.  In a QECC, an arbitrary quantum state $|\xi\rangle$,
along with some qubits which are originally set to $|0\rangle$, are
encoded in such a way by $U_1^T$ that, after being subjected to errors
$N_B$, decoding $U_2$ followed by measurement $\cal M$,
followed by final rotation $U_3$, permits an exact reconstruction of
the original state $|\xi\rangle$.}
\label{qeccf}
\end{figure}
 
Finally we step back to see the effect that this series of
transformations has produced, as summarized in Fig.~\ref{qeccf}.  All
use of bipartite states $I$, and the corresponding $A$ particles, has
been eliminated, along with all the measurement results transmitted to
Bob.  The net effect is that Alice has taken the $m$-qubit unknown
quantum state $|\xi\rangle$ along with $n\!-\!m$ ``blank'' qubits,
processed them with $U_1^T$, and sent them on channel $\chi$ to Bob.  He
is able to use his half of the protocol, without any additional
classical messages, to reconstruct $|\xi\rangle$.  This, of course, is
precisely the in-place QECC that we want.
 
\subsection{Finite block-size purification and error correcting codes}
\label{subsec:MC}
 
We have now shown that Bell-state purification procedures can be
mapped directly into quantum error correcting codes.  This gives an
alternative way to look for quantum error correction procedures within
the purification approach.  This can be both analytically and
computationally useful.  In fact, we can take over everything which we
obtained via the hashing protocol of Sec.~\ref{hashing}, in which Alice
and Bob perform a sequence of unilateral and bilateral unitary
operations to transform their bipartite state from one collection of
Bell states to another, in order to gain information about the errors
to which their particles have been subjected.
 
In this section we will show that this approach can also be used to do
purification, and thus error correction, in small, finite blocks of
qubits, in the spirit of much of the other recent work on
QECC~\cite{ChuangLaf,shellgame,Steane,Laflamme,EM,Sam, balance,newest}.  In these
procedures the object is slightly different than in the protocols
which employ asymptotically large block sizes: Here, we wish purify a
finite block of $n$ EPR pairs, of which no more than $t$ have
interacted with the environment (i.e., been subjected to noise).  The
end result is to be $m<n$ maximally entangled pairs, for which $F=1$
exactly.  The explicit result we present below will be for $n=5$,
$m=1$, and $t=1$.  This protocol thus has the same capability as the
one recently reported by Laflamme {\it et al.}\cite{Laflamme},
although the quantum network which we derive below is simpler in some
respects.  We are still investigating the extent to which our two
protocols are equivalent.
 
The general approach will be the same as in Sec.~\ref{sec:D},
however, our earlier emphasis was on error correction in
asymptotically large blocks of states.  To deal with the finite-block
case, we will need a few small but important modifications:
\begin{itemize}
\item There will again be a set ${\cal L}$ of possible collections
of Bell states after the action of the noise $N_B$; but rather
than being a ``likely set'' defined by the fidelity of the channel,
we will characterize the noise by a promise that the number of
errors cannot
exceed a certain number $t$.  Cases with $t+1$ errors are not just
deemed to have low probability; they are declared to be
disallowed, following Shor~\cite{shellgame}.
\item The set ${\cal L}$ will have a definite, finite size; if the
size of the Bell state block is $n$ and the number of erroneous Bell
states to be corrected is $t$, then the size of the set is~\cite{EM}
\begin{equation}
S=\sum_{p=0}^t3^p\choose{n}{p}.\label{syn}
\end{equation}
Borrowing the traditional language of error
correction, each member of the set, indexed by $i$, $1\le i\le S$,
defines an {\it error syndrome}.
The ``3'' in Eq.~(\ref{syn}) corresponds to the number of
possible incorrect Bell states occurring in the evolution of
Eq. (\ref{belmod}): there is either a phase error
($\Phi^+\rightarrow\Phi^-$), an
amplitude error ($\Phi^+\rightarrow\Psi^+$) or both
($\Phi^+\rightarrow\Psi^-$)\cite{Steane,EM}.  It has been noted\cite{CS,EM}
that correcting these three types of error is sufficient to
correct any arbitrary noise to which the quantum state is subjected
which we prove in Appendix~\ref{twirlout}.
\item The object of the error correction is slightly different than in
Sec.~\ref{sec:D}; in the earlier case it was to find a protocol where
the fidelity of the remaining EPR pairs approached unity
asymptotically as $n\rightarrow\infty$.  In the finite-block case,
the object is to find a protocol such that the fidelity attains
exactly 100\%, that is, $m$ good EPR pairs are guaranteed to be
recoverable from the original set of $n$ Bell states for every single
one of the $S$ error syndromes.
\end{itemize}
 
Let us emphasize again that, in the purification language which we
have developed, the quantum error correction problem has been turned
into an entirely classical exercise: given a set of $n$ Bell states,
we use the operations of item~\ref{i2} in Sec.~\ref{basics} to create a
classical Boolean function which maps these Bell states onto others
such that, for all $S$ of the error syndromes, the first $m$ Bell
states are always the same when the measurement results on the
remaining $n\!-\!m$ Bell states are the same.
 
We will develop this informal statement of the problem in a more
formal mathematical language.  First, recall the code which we
introduced for the Bell states in item~\ref{i5} of Sec.~\ref{basics} in
which, for example, the collection of Bell states $\Phi^+\Phi^-\Phi^+$
is coded as the 6-bit word $001000$.  As in our hashing-protocol
discussion (Sec.~\ref{hashing}), we denote such words by $x^{(i)}$,
where the superscript $i$ denotes the word appropriate for the
$i^{th}$ error syndrome.  These words have $2n$ bits, and we will
sometimes denote by $x_k^{(i)}$ the $k^{th}$ bit of the word.
 
\begin{figure}[htbp]
\centerline{\psfig{figure=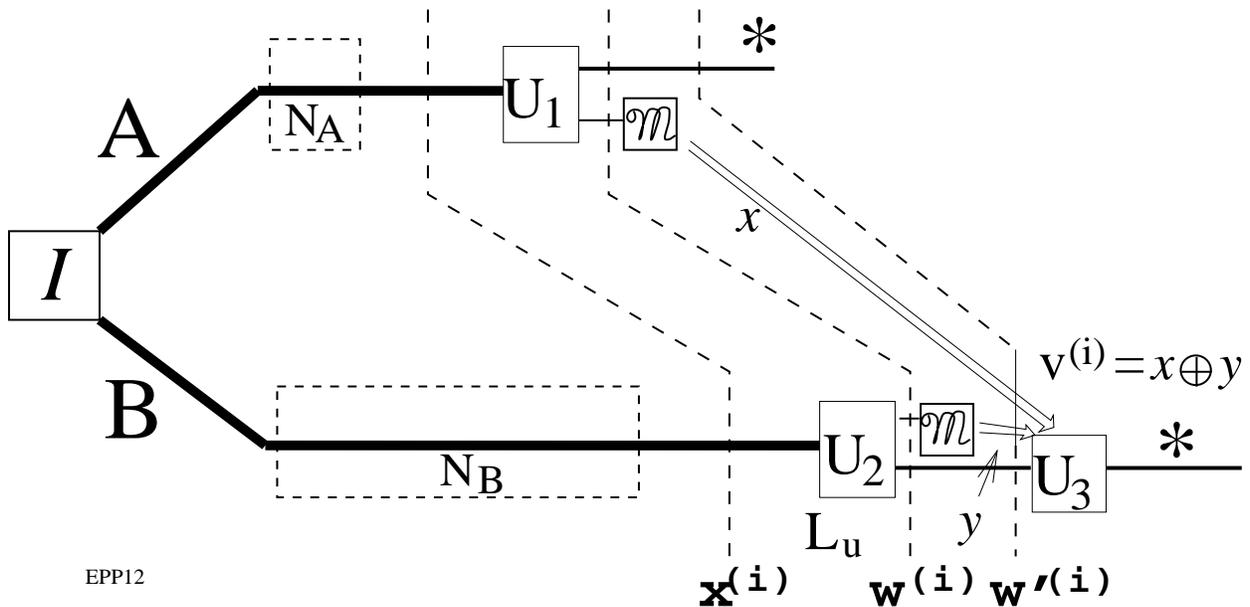,width=6.5in}}
\caption[EPP12.eps: 1-EPP marked with new notation]
{The 1-EPP of Fig.~\protect\ref{1way} marked with the notation
used in this section.}
\label{1eppmark}
\end{figure}
 
Alice and Bob subject $x^{(i)}$ to the unitary transformations $U_1$
and $U_2$.  They are confined to performing sequences of the
unilateral and bilateral operations introduced in Table
\ref{bell_table}.  In particular, they can do either:
\begin{enumerate}
\item a bilateral XOR, which flips the
low (right) bit of the target iff the low bit of the source is 1, and
flips the high (left) bit of the source iff the high bit of the target
is 1;
\item a bilateral $\pi/2$ rotation $B_y$ of both spins in a pair
about the $y$-axis, which interchanges the high and low bits;
\item a unilateral (by either Alice or Bob) $\pi$ rotation $\sigma_z$
of one spin about the $z$-axis, which complements the low bit; or
\item a composite operation $\sigma_x B_x$, where
the $\sigma_x$ operation is unilateral and the $B_x$ is bilateral; the
simple net effect of this sequence of operations is to flip the low bit
iff the high bit is one.
\end{enumerate}
It is easy to show that with these four operations, Alice and Bob can
do anything which they can do with the full set of operations in Table
\ref{bell_table}.  In our classical representation, the effect of such
a sequence of operations is to apply a classical Boolean function
$L_u$ to $x^{(i)}$, yielding a string $w^{(i)}$:
\begin{equation}
w^{(i)}=L_u(x^{(i)}).
\end{equation}
We use the symbol $L_u$ for this function because, with the operations
that Alice and Bob have at their disposal, $L_u$ is constrained to be
a linear, reversible Boolean function.  This is easy to show for the
sequences of the four operations given above.  Note, however, that
not all linear reversible Boolean functions are obtainable with
this repertoire.
A linear Boolean function\cite{Cop2} can be
written as a matrix equation
\begin{equation}
w^{(i)}=M_ux^{(i)}+b.\label{boolin}
\end{equation}
Here the matrix $M$ and the vector $b$ are boolean-valued ($\in\{0,1\})$,
and addition is defined modulo 2.  Reversibility adds an
additional constraint: $\det(M)=1$ (modulo 2).  In a moment we will
write down the condition which the set of $w^{(i)}$ must satisfy in
order for purification to succeed.
 
The next step of purification is a measurement $\cal M$
of $n\!-\!m$ of the Bell states.  As discussed in item~\ref{i5} of
Sec.~\ref{basics}, after learning Alice's measurement result, Bob can
deduce the low bit of each of the measured Bell states.  If we write
these measurement results for error syndrome $i$ as another boolean
word $v^{(i)}$ (of length $n\!-\!m$), the measurement can be expressed as
another linear boolean function:
\begin{equation}
v^{(i)}=M_mw^{(i)}.
\end{equation}
The matrix elements of $M_m$ are
\begin{equation}
(M_m)_{kl}=\delta_{k,2(m+k)}.
\end{equation}
The state of the remaining unmeasured Bell states is coded in a truncated
word $w'$ of length $2m$:
\begin{equation}
w'^{(i)}=(w_1w_2...w_{2m})^{(i)}.
\end{equation}
 
We now have all the machinery to state the condition for a successful
purification.  The object is to perform a final rotation $U_3$ on the
state coded by $w'$ and restore it, for every error syndrome $i$, to
the state $00...0$.  Whatever $w'$ is, such a restoring $U_3$ is always
available to Bob; for each Bell state, he does the Pauli rotations:
\begin{equation}
\begin{array}{cl}
\mbox{Bell state} & U_3 \mbox{ transformation} \\
00 & I \mbox{ (do nothing)} \\
01 & \sigma_z \\
10 & \sigma_x \\
11 & \sigma_y. \\ \end{array}
\end{equation}
But Bob must know which of these four rotations to apply to each
of the remaining $m$ Bell states.  The only information he has on which
of them to perform are the bits of the measurement vector $v^{(i)}$.
This information will be sufficient, if for every error syndrome which
produces a distinct $w'$, $v$ is distinct; in this case, Bob will
know exactly which final rotation $U_3$ to apply.
 
This, then, is our final condition for successful purification.  In
more mathematical language, we require an operation $L_u$ for which
\begin{equation}
\forall_{i,j}\ w'^{(i)}\ne w'^{(j)}\ \Longrightarrow\ v^{(i)}\ne v^{(j)}.
\label{goodcon}
\end{equation}
We will shortly show the results of a search for $L_u$ which satisfy
Eq. (\ref{goodcon}).
 
But first, we touch a point which has been raised in the recent
literature:\cite{Steane,CS,EM,Laflamme} Bob will obviously know which
rotation $U_3$ to apply if from the measurement he learns the precise
error syndrome, that is if for each error syndrome the measurement
outcome is distinct.  This ``condition for learning all the errors''
may be stated mathematically in a way parallel to Eq.
(\ref{goodcon}):
\begin{equation}
\forall_{i,j}\ i\ne j\ \Longrightarrow\ v^{(i)}\ne v^{(j)}.
\label{badcon}
\end{equation}
This condition is obviously {\it sufficient} for successful error
correction; however, it is more restrictive than Eq. (\ref{goodcon}),
and it is not a {\it necessary} condition.  If Eq. (\ref{badcon}) {\it
were} a necessary condition for error correction, then a comparison of
the number of possible distinct measurements $v^{(i)}$ with the number
of error syndromes $S$ leads\cite{EM,Laflamme} to a restriction on the
block size in which a certain number of errors can be corrected:
\begin{equation}
S=\sum_{p=0}^t3^p\choose{n}{p}\le 2^{n\!-\!m}.\label{D0}
\end{equation}
It is this bound which is attained, asymptotically, by the hashing and
breeding protocols above.  However, Eq. (\ref{goodcon}) puts no
obvious restriction on the block size in which error correction can
succeed, suggesting that the bound Eq. (\ref{D0}) can actually be
exceeded.  For example, if the transformation $L_u$ were permitted to
be any arbitrary boolean function, then it would be capable of
setting $w'=00...0$ for every syndrome $i$, in which case {\it no}
error correction measurements $v$ would be needed.
 
However, $L_u$ is very strongly constrained in addition to being a
linear, reversible boolean function, and we are left uncertain to what
degree the bound Eq. (\ref{D0}) may be violated.  For the small cases
which we have explored below, in which one Bell state is restored from
single-qubit errors ($m=1$, $t=1$), we find that the bound of
Eq. (\ref{D0}) is {\it not} exceeded.  All solutions which we find
which satisfy Eq. (\ref{goodcon}) also happen to identify every error
syndrome uniquely (Eq. (\ref{badcon})).  The present work, therefore,
does not demonstrate that Eq. (\ref{goodcon}) actually leads to more
power error-correction schemes than Eq. (\ref{badcon}).  However, Shor
and Smolin\cite{jumpthegun} have recently exhibited a family of new
protocols which, at least asymptotically for large $n$, exceed the
bound Eq. (\ref{D0}) by a small but finite amount.
 
\subsection{Monte Carlo results for finite-block purification protocols}
\label{subsec:MCres}
 
For the single-error ($t=1$), single-purified-state ($m=1$) case, we
have performed a Monte-Carlo computer search for unitary
transformations $U_1$ and $U_2$.  The program first tabulates the
$x^{(i)}$ for all the allowed error syndromes $i$, as shown in Table
\ref{syndrome}.  (For the case of $t=1$ there are $S=3n+1$ error
syndromes, since either of the $n$ Bell states could suffer three
types of error, plus one for the no-error case.)  The program then
randomly selects one of the four basic operations enumerated above,
and randomly selects a Bell state or pair of Bell states to which to
apply the operation.  The program then checks whether the resulting
set of states $w^{(i)}$ satisfies the error-correction condition
of Eq. (\ref{goodcon}).  If the answer is no, then the program
repeats the procedure, adding another random operation.  If the answer
is yes, the programs saves the list of operations, and starts over,
seeking a shorter solution.  Two ``shortness'' criteria were explored:
fewest total operations, and fewest total BXOR's (since two-bit operations
could be the more difficult ones to implement in a physical apparatus
\cite{G9}).
 
A simple argument akin to the one of Sec.~\ref{sec:D1D2} shows that
error correction in a block of 2 ($t=1$, $m=1$, $n=2$) is impossible.
We performed an extensive search for $n=3$ and $n=4$ codes; it would
not be possible to detect the complete error syndrome for these cases
(Eq.  (\ref{D0})), but it would appear {\it a priori} possible to
satisfy Eq. (\ref{goodcon}).  Nevertheless, no solutions were found,
strongly suggesting that, for this case, $n=5$ is the best block code
possible\cite{Laflamme}.  Knill and Laflamme have recently proved
this~\cite{lafknill}.
 
\begin{figure}[p]
\centerline{\psfig{figure=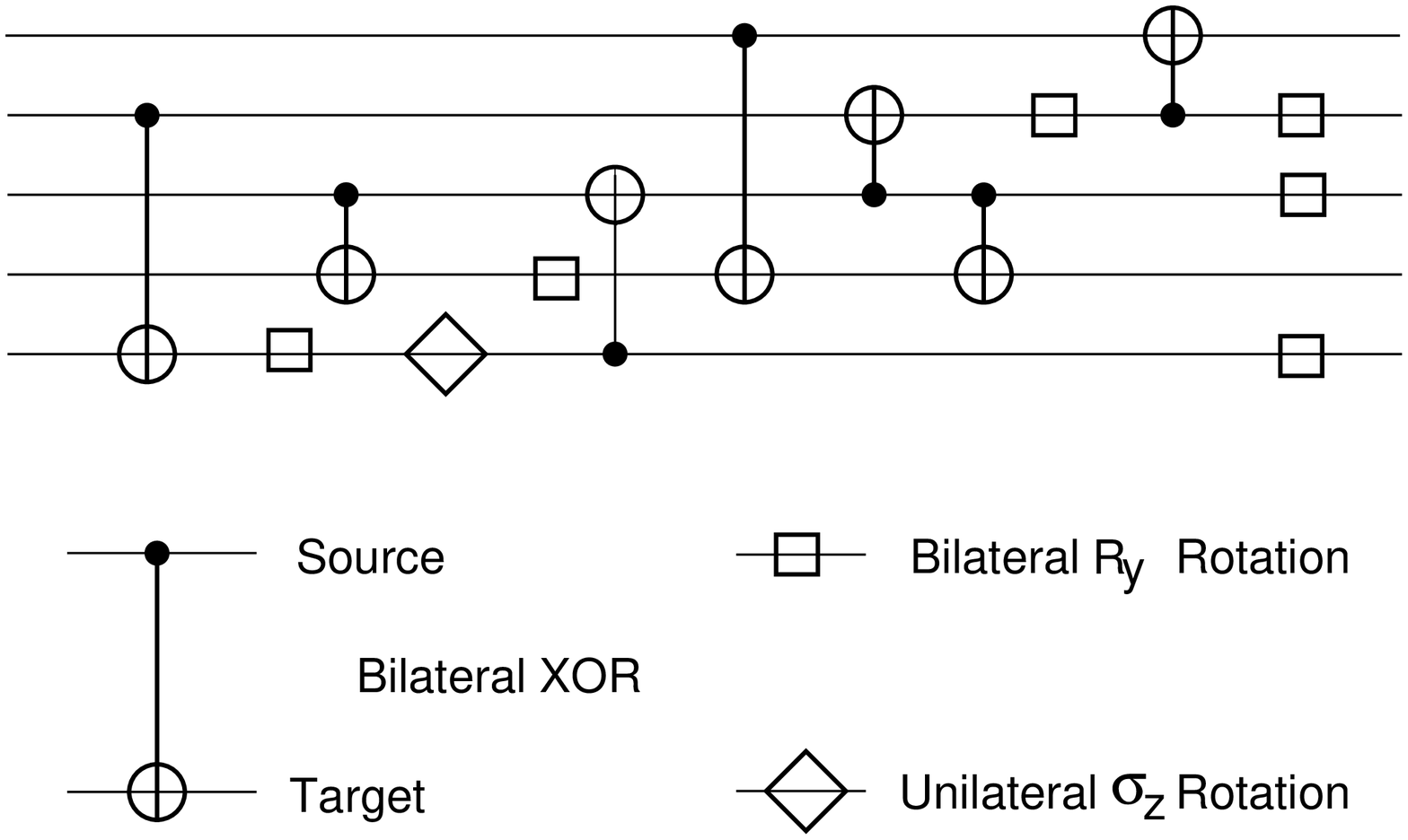,width=6.5in}}
\caption[music.eps]{The quantum gate array, determined by our computer search,
which protects one qubit from single-bit errors in a block of five.
``Bilateral'' and ``unilateral'' refer to whether both Alice and Bob,
or only Alice (or Bob), perform the indicated steps in the 2-EPP; in
the QECC version, it corresponds to whether the operation is done in
both coding and decoding, or in just the coding (or decoding)
operations.
\label{music}}
\end{figure}

Our search found many solutions for $n=5$ with similar numbers of
quantum gate operations.  The minimal network which was eventually
found was one with 11 operations, 6 of which were BXORs.  Here we
present a complete analysis of a slightly different solution, which
involves 12 operations, 7 of which are BXORs.  The gate array for
this solution is shown in Fig. \ref{music}.  The complete action
of $U_1$ and $U_2$ produced by this quantum network is given in
Table \ref{syndrome}.
 
\begin{table}[htbp]
\begin{tabular}{r|lllll|lllll|llll}
\cline{1-15}
&\multicolumn{5}{|c|}{Initial state}&\multicolumn{5}{c|}{Final state}&\multicolumn{4}{c}{Measurement}\\
{\it i$\:$}&\multicolumn{5}{|c|}{$x^{(i)}$}&\multicolumn{5}{c|}{$w^{(i)}$}&\multicolumn{4}{c}{result $v^{(i)}$}\\
\cline{1-15}
 1&00&00&00&00&00&00&00&00&00&01&0&0&0&1\\
 2&01&00&00&00&00&01&00&00&01&01&0&0&1&1\\
 3&10&00&00&00&00&10&01&00&00&01&1&0&0&1\\
 4&11&00&00&00&00&11&01&00&01&01&1&0&1&1\\
 5&00&01&00&00&00&00&01&00&00&00&1&0&0&0\\
 6&00&10&00&00&00&01&10&01&00&01&0&1&0&1\\
 7&00&11&00&00&00&01&11&01&00&00&1&1&0&0\\
 8&00&00&01&00&00&10&00&11&11&01&0&1&1&1\\
 9&00&00&10&00&00&00&00&01&00&00&0&1&0&0\\
10&00&00&11&00&00&10&00&10&11&00&0&0&1&0\\
11&00&00&00&01&00&10&01&01&10&01&1&1&0&1\\
12&00&00&00&10&00&00&00&01&01&00&0&1&1&0\\
13&00&00&00&11&00&10&01&00&11&00&1&0&1&0\\
14&00&00&00&00&01&00&00&00&00&00&0&0&0&0\\
15&00&00&00&00&10&01&11&11&01&11&1&1&1&1\\
16&00&00&00&00&11&01&11&11&01&10&1&1&1&0
\end{tabular}
\caption{Possible initial Bell states and the resulting final
state after the gate array of Fig.~\protect\ref{music} has been
applied.
\label{syndrome}}
\end{table}
 
Note that, as indicated above, this code not only satisfies the actual
error-correction criterion Eq. (\ref{goodcon}), but it also satisfies
the stronger condition Eq. (\ref{badcon}); all the error syndromes
are distinguished by the measurement results $v^{(i)}$.
 
It is interesting to note, as a check, that the tabulated transformation
is indeed a reversible, linear boolean operation.  The reader may
readily confirm that the results of Table \ref{syndrome} are obtained
from the linear transformation Eq. (\ref{boolin}), with
 
\begin{equation}
M_u=\left(\begin{array}{llllllllll}
1&0&0&0&0&1&0&1&0&0\\
0&1&1&0&0&0&0&0&1&0\\
0&0&1&0&0&0&0&0&1&0\\
1&0&0&1&0&0&0&1&1&0\\
0&0&0&0&0&1&0&0&1&0\\
0&0&1&0&1&1&1&1&1&0\\
0&0&0&0&0&1&0&1&0&0\\
0&0&0&0&0&1&1&0&1&0\\
0&1&0&0&0&0&0&0&1&0\\
0&0&0&1&1&0&1&0&0&1\end{array}
\right)
\end{equation}
and
\begin{equation}
b=(0\ 0\ 0\ 0\ 0\ 0\ 0\ 0\ 0\ 1).
\end{equation}

\subsection{Alternative conditions for successful quantum error correction code}
\label{subsec:linal}
 
While all of our work has involved deriving QECCs using the 1-EPP
construction, it is possible, and instructive, to formulate the
conditions for a good error correcting code directly in the QECC
language.  As Shor first showed\cite{shellgame}, in this language the
requirements become a set of constraints which the subspace into which
the quantum bits are encoded must satisfy.  In the course of our work
we derived a set of general conditions for the case of
error-correcting a single bit ($m=1$).  They are quite similar to
conditions which other workers have formulated recently\cite{EM,MZ}.
Knill and Laflamme have recently obtained the same 
condition~\cite{lafknill}.
 
We will assume that only one qubit is to be protected, but the
generalization to multiple qubits is straightforward.  Suppose a qubit
is encoded (by $U_1^T$ in Fig.~\ref{qeccf}) as a state
\begin{equation}
|\xi\rangle=\alpha|v_0\rangle+\beta|v_1\rangle,\label{Psistart}
\end{equation}
where $\alpha$ and $\beta$ are arbitrary except for the normalization
condition
\begin{equation}
|\alpha|^2+|\beta|^2=1,
\end{equation}
and $|v_0\rangle$ and $|v_1\rangle$ are two basis vectors in the
high-dimensional Hilbert space of the quantum memory block.
Can $|v_0\rangle$ and $|v_1\rangle$
be chosen such that, after the quantum state is subjected to Werner-type
errors,
the original quantum state can still be perfectly
reconstituted as the state of a single qubit,
\begin{equation}
|\xi_f\rangle=\alpha|0\rangle+\beta|1\rangle?\label{final}
\end{equation}
We shall derive the conditions which $|v_0\rangle$ and $|v_1\rangle$
must satisfy in order for this to be true.
 
We specify the action of the noise as a mapping of the original
quantum state into an ensemble of unnormalized state vectors given by
applying the linear operators $R_i$ to the original state vector:
\begin{equation}
|\xi\rangle\ \ \ \ \rightarrow\ \ \ \ \{R_i|\xi\rangle\}.\label{altspec}
\end{equation}
For each error syndrome $i$ there is an (unnormalized)
operator $R_i$ specifying the effect of the noise, as in Eq.~(\ref{belmod}).
For single-bit errors, the
$R_i$'s are just proportional to a $\sigma_x$, $\sigma_y$, or $\sigma_z$
operator applied to one of the quantum-memory qubits, as discussed below.
Two-bit errors would involve operators like
$R_i=\sigma_{x,y,z}^\alpha\sigma_{x,y,z}^\beta$ applied to two different
qubits $\alpha$
and $\beta$, and so forth.  Equivalently to Eq. (\ref{altspec}), the
effect of the noise $N_B$ in Fig.~\ref{qeccf} can be expressed as a
ensemble of normalized state vectors $|\xi_i\rangle$ with their
associated probabilities $p_i$:
\begin{equation}
|\xi\rangle\ \ \ \ \rightarrow\ \ \ \ \{p_i,|\xi_i\rangle \}=
\{ \langle\xi|R_i^\dagger R_i|\xi\rangle,\frac{R_i|\xi\rangle}
{\sqrt{\langle\xi|R_i^\dagger R_i|\xi\rangle}} \} .\label{POVMnorm}
\end{equation}
 
The Werner noise can be set up so that the $p_i$'s are the
probabilities that the environment ``measures'' the $i^{th}$ outcome
of a pointer or ancilla space.  We can evaluate the probability $p_i$
(for the $i^{th}$ outcome of these measurements) for the state
Eq. (\ref{Psistart}) using the expression in Eq. (\ref{POVMnorm}):
\begin{equation}
p_i=(\alpha^\ast,\beta^\ast)\times\left (\begin{array}{cc}
\langle v_0|R_i^\dagger R_i|v_0\rangle&\langle v_0|R_i^\dagger R_i|v_1\rangle\\
\langle v_1|R_i^\dagger R_i|v_0\rangle&\langle v_1|R_i^\dagger R_i|v_1\rangle
\end{array}\right )\times\left (\begin{array}{c}\alpha\\ \beta
\end{array}\right ).\label{bigprob}
\end{equation}
We have used the linearity of the operators $R_i$.  The matrix notation
used in Eq. (\ref{bigprob}) will prove useful in a moment.
 
The first, necessary condition which must be satisfied in order that
the state may be reconstituted as in Eq.~(\ref{final}) is that the
environment producing the Werner noise can acquire no information
about the initial quantum state by doing this ancilla measurement.
This will be true so long as $p_i$ in Eq. (\ref{bigprob}) is not a
function of the state vector coefficients $\alpha$ and $\beta$.  It
may be noted that the right hand side of Eq. (\ref{bigprob}) has the
form of the expectation value of a $2\times 2$ Hermitian operator in
the state $(\alpha, \beta)^T$.  It is a well-known theorem of linear
algebra that such an operator can only have an expectation value
independent of the state vector $(\alpha,\beta)^T$ iff the Hermitian
operator is proportional to the identity operator.  This gives us the
first two conditions that the state vector may be recovered exactly:
$\forall_i$,
\begin{eqnarray}
&&\langle v_0|R_i^\dagger R_i|v_0\rangle=\langle v_1|R_i^\dagger R_i|v_1
\rangle=p_i,\nonumber\\
&&\langle v_1|R_i^\dagger R_i|v_0\rangle=0.
\end{eqnarray}
 
If this condition is satisfied, then the ensemble of state vectors
in Eq. (\ref{altspec}) can be written in the simplified form:
\begin{equation}
\alpha|v_0\rangle+\beta|v_1\rangle\ \ \ \ \rightarrow\ \ \ \ \{p_i,
\frac{\alpha R_i|v_0\rangle+\beta R_i|v_1\rangle}
{\sqrt{p_i}}\}.\label{simplified}
\end{equation}
 
Now, given that the environment learns nothing from the measurement, a
further, sufficient condition is that there exist a unitary
transformation ($U_2$) which takes each of the state vectors of
Eq. (\ref{simplified}) to a vector of the form:
\begin{equation}
\frac{1}{\sqrt{\langle v_0|R_i^\dagger R_i|v_0\rangle}}
(\alpha R_i|v_0\rangle+\beta R_i|v_1\rangle)\ \ \ \ \rightarrow\ \ \ \
(\alpha|0\rangle+\beta|1\rangle)|a_i\rangle.
\end{equation}
Here $|a_i\rangle$ is a normalized state vector of all the qubits
excluding the one which will contain the final state
Eq. (\ref{final}).  Because of unitarity, the angle between any two
state vectors must be preserved.  Taking the dot product of the state
vectors resulting from two different syndromes $i$ and $j$,
and equating the result before and after the
unitary operation gives:
\begin{eqnarray}
\frac{1}{\sqrt{\langle v_0|R_i^\dagger R_i|v_0\rangle}
\sqrt{\langle v_0|R_j^\dagger R_j|v_0\rangle}}\times
\;\;\;\;\;\;\;\;\;\;\;\;\;\;\;\;\;\;\;\;\;\;\;\;\; && \nonumber \\
(\alpha^\ast,\beta^\ast)\times\left (\begin{array}{cc}
\langle v_0|R_i^\dagger R_j|v_0\rangle&\langle v_0|R_i^\dagger R_j|v_1\rangle\\
\langle v_1|R_i^\dagger R_j|v_0\rangle&\langle v_1|R_i^\dagger R_j|v_1\rangle
\end{array}\right )\times\left (\begin{array}{c}\alpha\\ \beta
\end{array}\right )= && \nonumber \\
|\alpha|^2\langle a_i|a_j\rangle+
|\beta|^2\langle a_i|a_j\rangle=\langle a_i|a_j\rangle.&&\label{mongo}
\end{eqnarray}
In the last part we have used the normalization condition to
eliminate $\alpha$ and $\beta$.  Now, since the right hand side
of Eq. (\ref{mongo}), and the prefactor of the left hand side, are
independent of $\alpha$ and $\beta$, so must be the expectation
value of the $2\times 2$ Hermitian operator.  We again conclude
that this Hermitian operator must be proportional to the identity
operator, and this gives the final necessary and sufficient
conditions\cite{footus}
for successful storage of the quantum data: $\forall_{i,j}$,
\begin{eqnarray}
&&\langle v_0|R_i^\dagger R_j|v_0\rangle=\langle v_1|R_i^\dagger R_j|v_1
\rangle,\label{thelasta}\\
&&\langle v_1|R_i^\dagger R_j|v_0\rangle=0.\label{thelastb}
\end{eqnarray}
\newline$\Box$
 
For the specific 5-qubit code described above, we found (by another,
simple computer calculation) that the two basis vectors of
Eq. (\ref{Psistart}) are:
\begin{eqnarray}
\ket{v_0} \propto (-&\!\!\ket{00000}\!\!& -
\ket{11000} -
\ket{01100} -
\ket{00110} -
\ket{00011} - \\
&\!\!\ket{10001}\!\!& +
\ket{10010} +
\ket{10100} +
\ket{01001} +
\ket{01010} + \nonumber\\
\ket{00101} +
&\!\!\ket{11110}\!\!& +
\ket{11101} +
\ket{11011} +
\ket{10111} +
\ket{01111} )\nonumber\label{vc1}
\end{eqnarray}
i.e. a superposition of all even-parity kets, with particular
signs, and
\begin{equation}
\ket{v_1} = \mbox{the corresponding vector with 0 and
1 interchanged.}\label{vc2}
\end{equation}
It is easy to confirm that this
pair of vectors satisfies the conditions Eqs. (\ref{thelasta}) and 
(\ref{thelastb}).  It is
interesting to note that these two vectors do {\it not} span the same
two-dimensional subspace as the ones recently reported by Laflamme
{\it et al.}\cite{Laflamme}; but it has recently been shown
that they are related to one another by one bit rotations~\cite{DS}.

\subsection{Implications of error-correction conditions on 
channel capacity}
\label{newupperbound}
Knill and Laflamme\cite{lafknill} have used the error correction 
conditions (Eqs. (\ref{thelasta}) and (\ref{thelastb})) to provide 
a stronger upper bound for $Q$ and $D_1$ than the one of Sec.~\ref{sec:D1D2}
by showing that $D_1=0$ when $F=0.75$. 
We indicate this on Figs.~\ref{dchart} and \ref{logchart} using our 
channel-additivity result of Sec.~\ref{additivity} to extend this to the linear
bound shown.
Their proof is as follows: write the coded qubit basis states (cf. 
Eqs.~(\ref{vc1}) and (\ref{vc2})) as 
\begin{equation}
|v_i\rangle=\sum_x\alpha^i_x|x\rangle=\sum_{y:z}\alpha^i_{y:z}|y:z\rangle.
\end{equation}
Here $x$ stands for an $n$ bit binary number, and $y:z$ stands for a partitioning
of $x$ into a $2t$-bit substring $y$ and an $(n-2t)$-bit substring $z$.  (The 
partitioning may be arbitrary, and need not be into the least significant and
most significant bits.)  Knill and Laflamme then consider the reduced
density matrices on the $y$ and the $z$ spaces:
\begin{equation}
\rho^i_{n-2t}=\sum_{y,z_1,z_2}\alpha^i_{y:z_1}\alpha^{i*}_{y:z_2}|z_1\rangle
\langle z_2|
\end{equation}
\begin{equation}
\rho^i_{2t}=\sum_{y_1,y_2,z}\alpha^i_{y_1:z}\alpha^{i*}_{y_2:z}|y_1\rangle
\langle y_2|
\end{equation}
Knill and Laflamme then prove two operator equations.  First:
\begin{equation}
\rho^0_{n-2t}\rho^1_{n-2t}=0.
\end{equation}
This is proved by using the condition for a successful error-correction code
(Eq. (\ref{thelastb})),
where the linear operator $R_i$ operates on a set of $t$ bits, and $R_j$
operates on a different set of $t$ bits.  (These $R$'s should be taken
as projection operators in this proof.)  Likewise, by applying Eq. 
(\ref{thelasta}) with the same operators $R_i$ and $R_j$,
they prove
\begin{equation}
\rho^0_{2t}=\rho^1_{2t}.
\end{equation}
These two equations give a contradiction when the two substrings are of the
same size, because it says that reduced matrices are simultaneously orthogonal
and identical.  This says that no code can exist if $2t=n-2t$, which 
corresponds to $F=1-t/n=0.75$.  As a bonus, these results give an interesting 
insight into the behavior of coded states: no measurement on $2t$ qubits can 
reveal anything about whether a 0 or a 1 is encoded, while there exists a 
measurement on $n-2t$
qubits which will distinguish with certainty a coded 0 from a coded 1.

This result shows that the lowest fidelity Werner channel with
finite capacity must have $F>0.75$.  Call that fidelity $F_0$.
Consider a channel with fidelity $F$ between $F_0$ and 1.  The capacity
of this channel is no greater than that of a composite 
channel consisting of a perfect channel used a fraction
$\frac{F-F_0}{1-F_0}$ of the time and a channel with fidelity $F_0$
used $\frac{1-F}{1-F_0}$ of the time because the first channel
is the same as the composite channel provided one is unaware of
whether the fidelity is 1 or $F_0$ on any particular use
of the channel.  (This construction is akin to that of Sec.~\ref{sec:D1D2}.)
By the channel additivity argument of Sec.~\ref{additivity} the
capacity of the composite channel, which bounds the capacity of the
fidelity $F$ channel, cannot exceed $\frac{F-F_0}{1-F_0}.$  Since
$F_0$ cannot be below 0.75 we obtain the straight-line bound 
\beq
Q=D_1\le 4F-3\ ,
\eeq
as shown in Figs. \ref{dchart} and \ref{logchart}.

\section{Discussion and Conclusions}
\label{sec:theend}
 
There has been an immense amount of recent activity and progress
in the theory of quantum error-correcting codes, including
block codes with some error-correction capacities in blocks of
two\cite{newest} three\cite{EM,Sam}, and four\cite{newest}.  Codes
which completely correct single-bit errors have now been reported for
block sizes of five as in the present work\cite{Laflamme},
seven\cite{Steane}, eight\cite{balance}, and nine\cite{shellgame};
this is in addition to the work using linear-code theory of families
of codes which work up to arbitrarily large block
sizes\cite{CS,Steane}.  A variety of subsidiary criteria have been
introduced, such as correcting only phase errors,
maintaining constant energy in the coded state, and correction by a
generalized watchdogging process.  Much of this work can be
expressed in entanglement purification language, in some cases
more simply.
 
Our results highlight the different uses to which a quantum channel may
be put.  When a noisy quantum channel is used for classical
communication, the goal---by optimal choice of preparations at the
sending end, measurements at the receiving end, and classical
error-correction techniques---is to maximize the throughput of reliable
classical information.  When used for this purpose, a simple
depolarizing channel from Alice to Bob has a positive classical capacity
$C>0$ provided it is less than 100\% depolarizing.  Adding a parallel classical
side channel to the depolarizing quantum channel would increase the
classical capacity of the combination by exactly the capacity of the
classical side channel.
 
When the same depolarizing channel is used in connection with a QECC or
EPP to transmit unknown quantum states or share entanglement, its
quantum capacity $Q$ is positive only if the depolarization
probability is sufficiently small ($<1/3$), and this capacity is not
increased at all by adjoining a parallel classical side channel. On the
other hand, a classical back channel, from Bob to Alice, does enhance
the quantum capacity, making it positive for all depolarization
probabilities less than $2/3$.

It is instructive to compare our results to the simpler theory of
noiseless quantum channels and pure maximally-entangled
states.  There
the transmission of an intact two-state quantum system or qubit (say
from Alice to Bob) is a very strong primitive, which can be
used to accomplish other weaker actions, in particular the undirected
sharing of an ebit of entanglement between Alice and Bob, or the
directed transmisson of a bit of classical information from Alice to
Bob. (These two weaker uses to which a qubit can be put are mutually
exclusive, in the sense that $k$ qubits cannot be used simultaneously to
share $\ell$ ebits between Alice and Bob {\em and\/} to transmit $m$
classical bits from Alice to Bob if \mbox{$\ell+m>k$}.~\cite{exclusive})

A noisy quantum channel $\chi$, if it is not too noisy, can similarly be
used, in conjunction with QECCs, for the reliable transmission of
unknown quantum states, the reliable sharing of entanglement, or the
reliable transmission of classical information.  Its capacity for the
first two tasks, which we call the quantum capacity $Q(\chi)$, is a
lower bound on its capacity $C(\chi)$ for the third task, which is the
channel's conventional classical capacity.
 
Most error-correction protocols are designed to deal with error
processes that act independently on each qubit, or affect only a bounded
number of qubits within a block. 
A quite different error model arises in quantum cryptography, where the
goal is to transmit qubits, or share pure ebits, in such a way as to
shield them from entanglement with a malicious adversary.  Traditionally
one grants this adversary the ability to listen to all classical
communications between the protagonists Alice and Bob, and to interact
with the quantum data in a highly correlated way designed to defeat
their error-correction or entanglement-purification protocol.
It is not yet known whether protocols can be developed to deal
successfully with such an adversarial environment.
 
Even for the simple error models which introduce no entanglement
between the message qubits, there are still a wide range of open
questions.  As Fig.~\ref{dchart} has shown, we still do not know
what the attainable yield is for a given channel fidelity; but we
are hopeful that the upper and lower bounds we have presented can
be moved towards one another, for both one-way and two-way protocols.
 
Improving the lower bounds is relatively straightforward, as it simply
involves construction of protocols with higher yields.  An
important step towards this has been the realization that it is not
necessary to identify the entire error syndrome to successfully purify. 
This has permitted the lower bound for one-way protocols (and thus for
QECCs) to be raised slightly above the $D_H$ curve of
Fig.~\ref{dchart} (see Ref.~\cite{jumpthegun}). 
 
Improvement of the upper bounds is more problematical.  For two-way
protocols, we presently have no insight into how this bound can be
lowered below $E$.   Characterizing $D_1$, $D_2$ and $E$ for all mixed
states would be a great achievement~\cite{H3distill}, but even that would not necessarily
provide a complete theory of mixed state entanglement. Such a theory
ought to describe, for any two bipartite states $M$ and $M'$, the
asymptotic yield with which state $M'$ can be prepared from state $M$ by
local operations, with or without classical communication.  In general,
the most efficient preparation would probably not proceed by distilling
pure entanglement out of $M'$, then using it to prepare $M$; it is even
conceivable that there might be incomparable pairs of states, $M$ and
$M'$ such that neither could be prepared from the other with positive
yield.

Surprisingly, basic questions about even the classical capacity of
quantum channels remain open.  For example, it is not known whether
the classical capacity of two parallel quantum
channels can be increased by entangling their inputs.

For us, all of this suggests that, even 70 years after its
establishment,
we still are only beginning to understand the full
implications of the quantum theory.  Its capacity to store,
transmit, and manipulate information is clearly different from
anything which was envisioned in the classical world.  It still
remains to be seen whether the present surge of interest in
quantum error correction will enable the great potential power of
quantum computation to be realized, but it is clearly a step in this
direction.

\bigskip
Acknowledgments: We are especially grateful to Peter Shor for
many important
insights which he has imparted to us during the course of this
work.  We thank G. Brassard, R. Cleve,
A. Ekert, R. Jozsa, M. Knill, R. Laflamme, R. Landauer, 
C. Macchiavello, S. Popescu, and B. Schumacher for helpful
discussions.
 
\appendix
 
\section{Appendix: Implementation of Random Bilateral Rotation}
\label{appx}
 
In this appendix we show how an arbitrary density matrix of two
particles can be brought into the Werner form by making a random
selection, with uniform probabilities, from a set of 12 operations
$\{U_i\}$ which involve identical rotations on each of the two particles.
(Thus, the rotations $U_i$ are members of a particular SU(2) subset of
SU(4).)  After such a set of rotations the density matrix is
transformed into an arithmetic average of the rotated matrices:
\begin{equation}
M_T=\frac{1}{N}\sum_{i=1}^NU_i^\dagger M U_i.\label{avgg}
\end{equation}
$N$ will be 12 in the example we are about to give.  The $4\times 4$
density matrix $M$, expressed in the Bell basis, has three parts which
behave in different ways under rotation: 1) the diagonal singlet
($\Psi^-$) matrix element, which transforms as a scalar; 2) three
singlet-triplet matrix elements, which transform as a vector under
rotation; and 3) the $3\times 3$ triplet block, which transforms as a
second-rank symmetric tensor.  In the desired Werner form the vector
part of the density matrix is zero, and the symmetric second-rank
tensor part is proportional to the identity.
 
The mathematics of this problem is the same as that which describes
the tensor properties of a large collection of molecules as would
occur in a liquid, glass, or solid.  In the case of a liquid, all
possible orientations of the molecules occur.  Because of the
orientational averaging (mathematically equivalent to
Eq. (\ref{avgg}), where the sum runs over all SU(2) operations),
vector quantities become zero (e.g., the net electric dipole moment of
the liquid is zero), while second-rank tensor quantities become
proportional to the identity (e.g., the liquid's dielectric response
is isotropic)\cite{Tink}.
 
But following the molecular-physics analogy further, we know that
crystals, in which the molecular units only assume a discrete set of
orientations, can also be optically isotropic and non-polar.  It is
also well known that only cubic crystals have sufficiently high
symmetry to be isotropic.  This suggests that if the sum in
Eq. (\ref{avgg}) is over the discrete subgroup of SU(2) corresponding
to the symmetry operations of a tetrahedron (the simplest object with
cubic symmetry), then the desired Werner state will result; and this
turns out to be the case.
 
The bilateral rotations $B_{x,y,z}$ introduced in Sec.~\ref{hashing}
are the appropriate starting point for building up the desired set of
operations.  In fact they correspond to 4-fold rotations of a cube
about the $x$-, $y$-, and $z$-axes.  This is not evident from their
action on Bell states as shown in Table~\ref{bell_table}
where they appear to correspond to 2-fold operations.  This is because
this table does not show the effect of the $B$ rotations on the phase
of the Bell states.  Phases are not required in the purification
protocols described in the text, because the density matrix in all
these cases is already assumed to be diagonal, so that the phases do
not appear.  But for the present analysis they do, so we repeat the
table with phases in Table~\ref{Bwithphase}.
 
\begin{table}[htbp]
\begin{tabular}{cr|rrrr|l}
 & &\multicolumn{4}{|c|}{source}&\\
 & &$\Psi^-$&$\Phi^-$&$\Phi^+$&$\Psi^+$&\\
\cline{2-7}
 &$I$&$\Psi^-$&$\Phi^-$&$\Phi^+$&$\Psi^+$&\\
Bilateral $\pi/2$ Rotations:&$B_x$&$\Psi^-$&$\Phi^-$&$i\Psi^+$&$i\Phi^+$&\\
 &$B_y$&$\Psi^-$&$-\Psi^+$&$\Phi^+$&$\Phi^-$&\\
 &$B_z$&$\Psi^-$&$i\Phi^+$&$i\Phi^-$&$\Psi^+$&\\
\cline{2-7}
\end{tabular}
\caption[B rotations with phases]{Modification of part of Table~\protect\ref{bell_table},
including the phase-changes of the Bell states.}
\label{Bwithphase}
\end{table}
 
When presented in this way, it is evident that these operations are
4-fold (that is, $B_i^4=I$) , and indeed, they are the generators of
the 24-element group of rotations of a cube, known as the group O in
crystallography\cite{Tink}.  (It is also isomorphic to $S_4$, the
permutation group of 4 objects.)
 
Now, as mentioned above, only the rotations which leave a tetrahedron
invariant are necessary to make the density matrix isotropic.  This is
a 12-element subgroup of O know as T (which is isomorphic to
$A_4$, the group of all even permutations of 4 objects).  Written in
terms of the $B_i$'s, these twelve operations are
\begin{equation}
\{U_i\}=
\begin{array}{l}
I  \mbox{(identity)}\\
B_xB_x\\
B_yB_y\\
B_zB_z\\
B_xB_y\\
B_yB_z\\
B_zB_x\\
B_yB_x\\
B_xB_yB_xB_y\\
B_yB_zB_yB_z\\
B_zB_xB_zB_x\\
B_yB_xB_yB_x.\end{array}M\rightarrow W_F
\end{equation}
It is easily confirmed by direct calculation, using Table
\ref{Bwithphase}, that this set of 12 $\{U_i\}$, when applied to a
general density matrix $M$ in Eq. (\ref{avgg}), results in a Werner
density matrix $W_F$ of Eq.~(\ref{WF}).
 
There are a couple of special cases in which the set of rotations
can be made simpler.  If it is only required that the state $M$ be
taken to some Bell-diagonal state $W$ (Eq.~(\ref{firstW})), then a
smaller subset, corresponding to the orthorhombic crystal group
$D_2$ (an abelian four-element group) may be used:
\begin{equation}
\{U_i\}=
\begin{array}{l}
I\\
B_xB_x\\
B_yB_y\\
B_zB_z.\end{array}\;\;\;M\rightarrow W
\end{equation}
Finally there is another special case, which arises in some of our
purification protocols, in which the density matrix $W$ is already
diagonal in the Bell basis, but is not isotropic (i.e., the triplet
matrix elements are different from one another).  To carry $W$ into
$W_F$, the discrete group in Eq. (\ref{avgg}) can be again be reduced,
in this case to the three-element group with the elements
\begin{equation}
\{U_i\}=
\begin{array}{l}
I\\
B_xB_xB_xB_y\\
B_xB_xB_xB_z.
\end{array}\;\;\;W\rightarrow W_F
\label{triple}
\end{equation}
One further feature of any set $\{U_i\}$ that takes the density matrix
to the isotropic form $W_F$, which can be used to simplify the set, is
that the modified set $\{RU_i\}$, for any bilateral rotation $R$, also
results in a Werner density matrix $W_F$ in Eq. (\ref{avgg}).  Since
the density matrix is already isotropic, any additional rotation $R$
leaves it isotropic.  (A cubic crystal has the same dielectric
properties no matter how it is rotated.)  For example, if we take
$R=B_x$, the three operations of Eq. (\ref{triple}) take the form
\begin{equation}
\{U_i\}=
\begin{array}{l}
B_x\\
B_y\\
B_z.
\end{array}\;\;\;W\rightarrow W_F
\label{easytriple}
\end{equation}
 
\section{Appendix: General-noise error correction}
\label{twirlout}
In this appendix we present an argument, based on
twirling, that correcting amplitude and phase errors corrects
every possible error.
We have derived finite-block purifications under the assumption that
the pairs which are affected by the environment are subject to errors
of the Werner type, in which the Bell state evolves into a classical
mixture of Bell states (see Eq.~(\ref{belmod})).  But the most general
effect which noise can have on a Bell state appears very different
from the Werner noise model, and is characterized by the $4\times 4$
density matrix $M$ into which a standard Bell state $\Phi^+$ evolves
(see Fig. \ref{tommy}).  Many additional parameters besides the
fidelity $F=\langle\Phi^+|M|\Phi^+\rangle$ are required for the
specification of this general error model.  A general $4\times 4$
density matrix of course requires 15 real parameters for its
specification.  However, not all of these parameters define distinct
errors, since any change of basis by Alice or Bob cannot essentially
change the situation (in particular, the ability to purify EPR pairs
cannot be changed).  This says that 6 parameters, those involved in
two different SU(2) changes of basis, are irrelevant.  But this still
leaves 9 parameters which are required to fully specify the most
general independent-error model\cite{foot3}.  How then does correction
of just amplitude, phase, and both, deal with all of these possible
noise conditions, characterized by 9 continuous parameters?
\begin{figure}[htbp]
\centerline{\psfig{figure=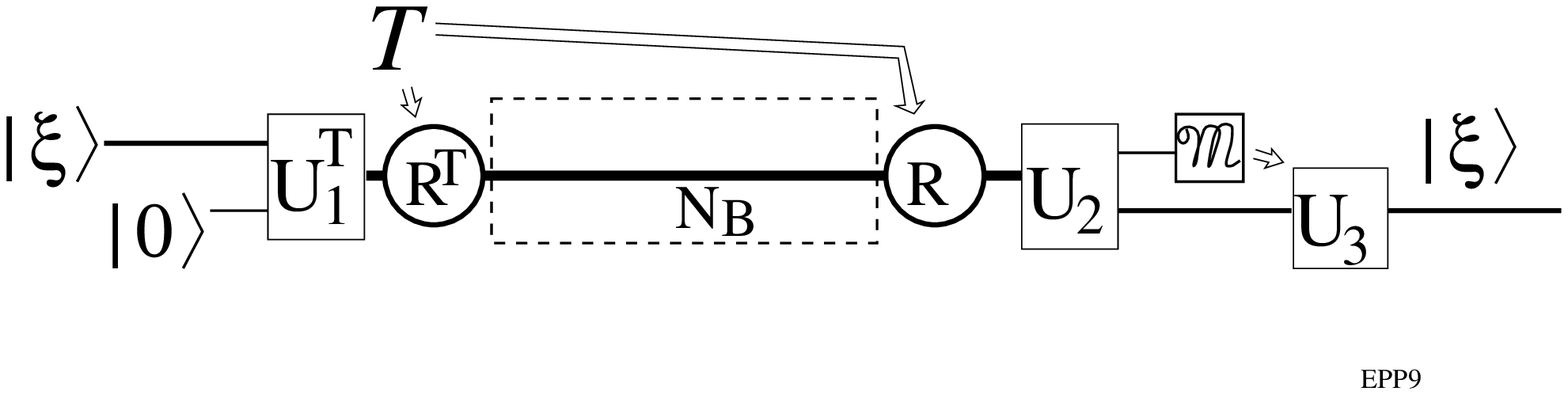,width=5in}}
\caption[EPP9.eps Twirl need not be performed]
{If the state is subject to the initial and final rotations
$R^T$ and $R$ (the ``twirl'' $T$) in the QECC of Fig.~\protect\ref{qeccf},
then the action of the noise $N_B$ is guaranteed to be of a simple form
in which only three types of errors, amplitude, phase, or
amplitude-and-phase, can occur on each qubit\protect\cite{EM}; this
corresponds to the Werner mixed state $W_F$ in the purification picture.
As described in the text, for finite-block error correction the QECC
protocol will succeed even if the twirl $T$ is not performed.}
\label{newt}
\end{figure}
 
To show this we will again introduce the ``twirl'' of
Fig.~\ref{tommy}, although in the end it will be removed again.
Recall that any density matrix is transformed into one of the Werner
type by the random twirl.  (See item~\ref{i5} of Sec.~\ref{basics} for
the method of twirling the $\Phi^+$ state.)  Thus, if twirling is
inserted as shown in Fig.~\ref{newt}, or in the corresponding places
in Fig. \ref{1way}, then the channel is converted to the Werner type,
and the error correction criteria we will describe in the next section
will work.
 
But let us consider the action of the twirl in more detail.  Let us
personify the twirl action {\it T} in Fig.~\ref{newt} (or in the
corresponding purification protocol of Fig.~\ref{1way}, as in
Fig.~\ref{tommy}) by saying that an agent (``Tom'') performs the twirl
for the $n$ bits by randomly choosing $n$ times from among one of 12
bilateral rotations tabulated in Appendix~\ref{appx}.  Tom makes a record of
which of these $12^n$ actions he has taken; he does not, however,
reveal this record to Alice or Bob.  Without this record, but with a
knowledge that Tom has performed this action, Alice and Bob conclude
that the density matrix of the degraded pairs has the Werner form.
They proceed to use the protocol they have developed to purify $m$ EPR
pairs perfectly.  Now, suppose that after this has been done, Tom
reveals to Alice and Bob the twirl record which he has heretofore kept
secret.  At this point, Alice and Bob now have a revised knowledge of
the state of the particle pairs which entered their purification
protocol; in fact, they now know that the density matrix is just some
particular rotated version of the non-Werner density matrix in which
the environment leaves the EPR pairs.  Nevertheless, this does not
change the fact that the purification protocol has succeeded.  Indeed,
we must conclude that it succeeds for each of the $12^n$ possible
values of Tom's record, and in particular it succeeds even in the case
that each of Tom's $n$ rotations was the identity operation.  Thus,
the purification protocol works on the original non-Werner errors,
even if Tom and his twirling is completely removed.  This completes
the desired proof, and we will thus develop protocols for correcting
Werner type errors, Eq. (\ref{belmod}), keeping in mind their
applicability to the more general case.

A slight extension of the above arguments shows that asymptotic
large-block purification schemes such as our hashing protocol
of Sec.~\ref{hashing} are also
capable of correcting for non-Werner error.  Consider a non
Bell-diagonal product density matrix of $n$ particles, 
${\bf M}=(M)^n$, whose fidelity is such that, after twirling, 
it can be successfully purified, resulting in entangled
states whose final fidelity with respect to perfect singlets
approaches 1 in the limit $n\rightarrow\infty$.  The hashing
protocol produces truly perfect singlets of unit fidelity
for a likely set $\cal L$ of error syndromes containing nearly all
the probability.
This means that we
can write 
${\bf M}=(1-\epsilon) {\bf M'}+\epsilon\mbox{\boldmath{$\delta$}}{\bf M}$, 
where
${\bf M'}$ can be purified with exactly 100\% final fidelity.  By the
above arguments, ${\bf M'}$ can be successfully purified even if
twirling is not performed.  Since $\epsilon\rightarrow 0$ as
$n\rightarrow\infty$, the original state ${\bf M}$ will also be
purified to fidelity approaching 1, even without twirling.


\begin{thebibliography}{99}
\bibitem{foot1} Systems need not interact directly in order to become
entangled.  For example two systems $A$ and $B$ that have never
interacted can become entangled indirectly, by a sequence of
interactions in which $A$ becomes entangled with $C$ and $B$ with $D$,
followed by a joint measurement on $C$ and $D$, followed by unitary
operation on $A$ or $B$ depending on the result of the joint
measurement.  This is an example of quantum
teleportation~\protect{\cite{teleportation}}.
\bibitem{footfef} Alice and Bob should choose a basis for
the Bell states which maximizes the fully entangled fraction
$f$ Eq.~(\protect\ref{fef}).
\bibitem{foot2} Ref. \protect{\cite{purification}} states incorrectly
that a set of 4 discrete rotations is sufficient to simulate a random
bilateral rotation; in fact 12 are both necessary and sufficient.
\bibitem{footrho}This follows from
the fact that ensembles with the same density matrix are
indistinguishable by measurements performed on members of the
ensembles.  (See, e.g., Ref.~\protect\cite{Peres}, p. 75.)  (Such measurements
could include executing a purification protocol followed by testing
the quality of the resulting purified pairs.)
\bibitem{foot5}  C.~Macchiavello (private communication) pointed
out that the yield of the recurrence protocol could be increased by
substituting a deterministic $B_x$ operation for the random twirl $T'$.
\bibitem{richard} This purification
strategy was worked out in a discussion with Richard Jozsa.
\bibitem{footnonoise}For a large-block code which
only error-corrects to some high fidelity, the code may not need
to properly treat the no-error case because this case is highly
improbable.  Such a case may also arise for certain peculiar block 
error models from which the no error case is excluded.
\bibitem{footas}If the 1-EPP is not perfect, but only produces entangled
states with some high fidelity $1-\epsilon$, then it is obvious that there
is at least one set of values for the $\cal M$ measurements for
which the final fidelity is at least $F=1-\epsilon$, since $F$ is
the fidelity averaged over all possible measurement results.  It is
this set which Alice and Bob should pre-agree to use for the state
preparation of these $B$ particles.
\bibitem{footus} We note that the conditions for the ``error prevention''
(or ``error watchdogging'') protocol of Ref.~\protect\cite{newest} are
a very simple subset of the conditions which we have derived for error
correction: ``error prevention'' requires that 
Eqs. (\protect\ref{thelasta}) and (\protect\ref{thelastb})
be satisfied not for all $i$ and $j$, but just for those $i$ and $j$
which refer to errors involving the same qubit.  Thus it is obvious
that every QECC is an ``error prevention'' scheme, but not vice versa.
\bibitem{foot3}In the Bell basis, this restriction to 9 parameters is
achieved by making the matrix elements
$\langle\Phi^+|M|\Phi^-\rangle$, $\langle\Phi^+|M|\Psi^-\rangle$,
$\langle\Phi^-|M|\Psi^+\rangle$, and $\langle\Psi^+|M|\Psi^-\rangle$
purely real, and making $\langle\Phi^+|M|\Psi^+\rangle$ and
$\langle\Phi^-|M|\Psi^-\rangle$ purely imaginary.  This corresponds to
making the reduced density matrices $\rho_A$ and $\rho_B$ diagonal,
and making additional phase adjustments ($z$-axis rotations: see
Ref.~\protect\cite{G9}) to the A and the B particles.
\end{thebibliography}
\end{document}